\documentclass[13pt
]{ceurart}

\usepackage{color}
\usepackage{float}
\usepackage[figuresright]{rotating}
\usepackage{tabularx}
\usepackage{subfigure} 
\usepackage{algorithm}
\usepackage{algpseudocode}
\usepackage{appendix}
\usepackage{graphicx}
\usepackage{lineno}
\usepackage{pdflscape}
\usepackage{geometry}

\usepackage{listings}
\begin{document}

\conference{}

\title{A Mobile Data Driven Hierarchical Deep Reinforcement Learning Approach for Real-time Demand-Responsive Railway Rescheduling and Station Overcrowding Mitigation}


%
\author[1,2]{Enze Liu}[%
email=liuenze@chd.edu.cn,
]
\address[1]{College of Transportation Engineering, Chang'an University, Xi'an Shaanxi, China, 710064}
\address[2]{Institute for Transport Studies, University of Leeds, 34-40 University Road, Leeds, LS2 9JT, United Kingdom}
\address[3]{School of Civil Engineering, University of Leeds, Woodhouse Lane, Leeds, LS2 9JT, United Kingdom}

\author[2]{Zhiyuan Lin}[%
email=Z.Lin@leeds.ac.uk,
]
\cormark[1]

\author[2,3]{Judith Y.~T. Wang}[%
]

\author[1]{Hong Chen}[%
]

\cortext[1]{Corresponding author.}

\begin{abstract}
Real-time railway rescheduling is an important technique to enable operational recovery in response to unexpected and dynamic conditions in a timely and flexible manner. Current research relies mostly on origin-destination (OD) based data and model-based methods for estimating train passenger demands. These approaches primarily focus on averaged disruption patterns, often overlooking the immediate uneven distribution of demand over time. In reality, passenger demand frequently deviates significantly from OD data and model-based predictions, especially during a disaster, where passenger behavior is influenced by a variety of factors. Disastrous situations such as flooding in Zhengzhou, China in 2022 has created not only unprecedented effect on Zhengzhou railway station itself, which is a major railway hub in China, but also other major hubs connected to Zhengzhou, e.g., Xi'an, the closest railway hub west of Zhengzhou.  Real-time rescheduling for rail services to and from Xi'an has become a challenge with added complexity.  In this present study, we define a real-time demand-responsive (RTDR) railway rescheduling problem focusing two specific aspects of this complex situation, namely, volatility of the demand, and management of station crowdedness and its potential impact on passenger behaviour. For the first time, we propose a data-driven approach utilising real-time passenger mobile data (MD) to deal with this RTDR problem. A hierarchical deep reinforcement learning (HDRL) framework is designed to perform real-time rescheduling of railway services in a demand-responsive manner.  The use of MD in this new HDRL framework has enabled the modelling of passenger dynamics in response to train delays and station crowdedness, and a real-time optimisation for rescheduling of train services in view of the change in demand as a result of passengers' behavioural response to the situation.  Results show that the agent can steadily satisfy over 62\% of the demand with only 61\% of the original rolling stock, ensuring continuous operations without overcrowding during the disruption. Moreover, the agent exhibits adaptability when transferred to a new environment with increased demand, highlighting its effectiveness in addressing unforeseen disruptions in real-time settings.
\end{abstract}


\maketitle

\cfoot{\thepage}

\section{Introduction}
With the development of intelligent sensing and computing technologies, the railway system provides potential to automatically adjust train schedules in response to unexpected disruptions. Real-time rescheduling is an advanced technique capable of making prompt adaptations to unforeseen events. During a long-term disruption, the railway system usually goes through several rounds of rescheduling, interacting with the evolving condition, which aims to mitigate the consequence with limited rolling stock and the disrupted network. 

The challenges for managing a long-term disruption come from the uncertainty of disruption duration, demand propagation, and resource availability. State-of-art technologies can be classified into phase-based optimization approaches \citep{ghaemi_macroscopic_2018} and rolling horizon based methods \citep{RN131}. 
Phase-based approaches divide the post-disruption period into three phases: survivability (transition), response, and recovery. The first two phases aim to improve the adaptation of the disrupted railway system to the unexpected emergency condition. Most emergency-responsive strategies assume that disruption duration, passenger demand and backup resources are determined at the beginning, which enables to reschedule the operation plan based on optimization models. This, however, prevents such methods from dealing with real-time problems with uncertainty.
As opposed to the phase-based approaches, the rolling horizon approaches can effectively handle the uncertainty of disruptions while improving computational efficiency by solving subproblems in a step-by-step manner.  It is an online technology that interacts with real-time information, such as passenger demand, available rolling stock, and disruption information. It renews the operating plan when new information is available. However, it is hard to determine the granularity of the horizon as overly long intervals can reduce the accuracy and efficiency of the solution while excessively short intervals can result in redundant computations and lose the foresight of long-term benefit.
In summary, when dealing with long-term disruptions that require frequent schedule updating, rescheduling strategies are often divided into subproblems using two key principles. The first principle uses the estimated duration of the disruption, which is commonly applied in phase-based optimization approaches. The second principle involves dividing the open-end time span into fixed time intervals, which facilitates iterative and adaptive rescheduling through rolling horizon algorithms. 

Incorporating the consideration of passenger behavior into an optimization model would increase the model's complexity and the computational time, which may limit its practical application in real-time \citep{binder_2017}. Passengers' unpredictable response to the interrupted service would increase the uncertainty for demand estimation during disruption. Real-time in-station demands for trains are usually estimated by model-based methods using static origin-destination (OD) demand data. For example, \cite{RN150} used a sigmoid function to represent passengers' arriving process before the departure time, \cite{VEELENTURF2017133} set a deadline of waiting tolerance to capture passengers' returning process due to extended delays, \cite{BINDER2021103368} uses a stochastic model to split the demands on routes, and \cite{Liu2022AME} address passengers' participating willingness considering the in-station overcrowding and quality of alternative service. Model-based methods typically prioritize adherence to the original timetable and use passenger demand to assess the importance of a train. Rescheduling plans are designed to accommodate high-demand OD pairs. However, in practical post-disruption conditions, passengers' response behaviour can lead to numerous unforeseeable circumstances, which may deviate from the theoretical estimations. Especially in certain transport hub cities, passengers may transfer to other intercity modes due to the limited competency of the disrupted railway service. As a result, the real-time in-station demand may be over-estimated using OD demand if passengers' arriving and returning are not taken into consideration.

During a disruption,  the number of passengers entering a railway station steadily increases with time, resulting in congestion as trains are delayed. This escalating passenger density poses a significant challenge, as passengers might find themselves somehow trapped within the station once they entered. In-station overcrowding is associated with various secondary risks, including pedestrian panic \citep{Wang2015ModelingAS}, stampede accidents \citep{Zhang2020RailwaySR}, and the spreading of diseases \citep{Aghabayk2021EffectsOC}. Although \cite{Liu2022AME} described railway station overcrowding as a risk that is not immediately life-threatening, the severity of this risk tends to escalate with time, highlighting the importance of timely reactions to mitigate potential hazards. A significant challenge in addressing in-station crowdedness during disruptions arises from the limited availability of real-time data on in-station passenger dynamics. This lack of real-world records makes it difficult to monitor and quantify the level of crowdedness within stations, thereby impeding the ability to take proactive actions to mitigate potential risks with a data-driven approach. 


Several data-driven approaches have been studied to estimate the real-time in-station passenger demand. For instance, in an urban rail system, passengers' arrival, leaving and transfer processes during a disruption can be inferred from smart card data \citep{Yin2019HybridDA,MO2022102628,LI202330}. However, in the intercity railway system, though the seat reservation system provides the upper bound of a train demand at a station, passengers' entering and exiting times are often not recorded. With the emergence of mobile data (MD), it has been possible to capture anonymous passenger movement in real time in a grid network. The use of MD has facilitated the implementation of demand-responsive services in various public transit systems \citep{RN129,RN116,RN119}. Therefore, this paper attempts to analyse the MD of a railway station area to infer real-time in-station demand by monitoring passengers' entry, waiting, and exit movements, enabling more efficient real-time rescheduling in a demand-responsive manner.


Due to the need of prompt schedule updates during disruption, ensuring high computational efficiency is an important aspect in real-time rescheduling \citep{CACCHIANI201415}.  \citet{CORMAN201515} define a closed-loop control for real-time rescheduling, which distinguishes the real-time algorithm from real-time control. The open-loop tool uses a real-time algorithm only once at the beginning of the disruption, while the closed-loop tool applies multiple rapid algorithms at variable suitable time intervals based on conditions and predictions until a target stability is achieved. The main difference between multiple open-loop and closed-loop controls is the ability to modify the parameters of an action based on the results from previous actions. 


Building upon the nature of dynamic decision-making process for a long-term disruption, this paper introduces a data-driven and closed-loop approach, named deep reinforcement learning (DRL). Reinforcement learning may be applied to map the current situation to appropriate actions so as to maximize the long-term accumulative rewards \cite{RN157}. One possible way is to train a DRL framework with historical data to learn the foresight value of an action in a long-term decision-making process. The DRL approach uses the deep learning method with multi-layer Artificial Neutral Network (ANN) as the agent, which learns to predict a proper action by referring to numerous variables of current state. As a value-based model-free method, the Deep Q-Network (DQN) approach is an improved version of Q-learning, to tackle the ``curse of dimensionality'' of input variables, while the output decisions are countable \citep{mnih_playing_2013}. By monitoring the real-time value of the input variables, such as demand, backup resource, crowdedness, etc., the DRL agent is capable of making a decision in milliseconds. In practice, the demand and timetable patterns in the railway system are regulated. During normal days, the agent has ample time to undergo adequate offline training for understanding the regular patterns. This enables the agent to incorporate complex factors, such as passenger behavior, into the real-time rescheduling. It only needs to update its value function (i.e., ANN) using the near-time data to handle unexpected disruptions and improving the performance by the real-world data during long-term abnormal conditions. 


However, due to the limited availability of rolling stock, the decision-making instances are significantly restricted. Most of the time, the agent is in a monitoring phase, receiving minimal rewards from the environment. Only when dispatching a train, the agent has the potential to obtain a reward based on the efficacy of its decision. This scenario poses a challenge in DRL known as the ``sparse reward'' dilemma. To address this, an enhanced approach employing a \emph{feudal hierarchy structure}, termed Hierarchical Deep Reinforcement Learning (HDRL) \citep{pateria_hierarchical_2022,Dayan1992FeudalRL}, is adopted. Two sub-agents operates at different temporal levels. An upper-level agent constantly monitors the in-station conditions in real-time to determine whether dispatching a train is warranted, and when the dispatching decision is received, a lower-level agent formulates the specific rescheduling plan. Accordingly, the real-time demand-responsive rescheduling is realized by an HDRL method which is trained to recognize the value of dispatching a train facing various post-disrupted conditions,  enabling it to operate in real-life online scenarios. 

Our study builds a time-variant environment of passenger-railway system with MD and train timetable, which contains full-day data to simulate the open-end disrupted period. An HDRL approach is proposed to map the in-station condition, backup resource and on-route headway to the dispatching and rescheduling decision. The agent is in a feudal hierarchy structure with two levels where the upper level is to learn in what condition to dispatch a train referring to the dynamic in-station conditions, and the lower level is to determine the specific rescheduling plan referring to the real-time OD demands and on-route headway. The agent is trained to maximize the reward of the demand satisfaction, capacity utilization, and minimize the penalty of denying passengers, in-station overcrowding, time delay and unsafe headway for the whole period of disruption. Our proposed HDRL model was tested in a real-world disaster scenario occurred on July 20 2022 in Zhengzhou, a major central rail hub in China. Xi'an, an upstream heavy-demand station on the west of Zhengzhou, is selected as the focused station. The experiments conducted in this study encompassed several aspects, including the efficiency of offline training, capability analysis of the agent, benchmark with meta-heuristics and the online applicability during the Zhengzhou flood period. Notably, the agent satisfied up to 62\% demand on the 9 routes with only 61\% of the original rolling stock, and avoided in-station overcrowding during the whole period of disruption. After sufficient training, our HDRL agent outperformed our customized meta-heuristics in terms of demand satisfaction, rolling stock allocation and overall rewards. When the agent was transferred to a real-world disrupting environment with higher demand, it stably maintained the solving capability similar to the offline training results, which shows its adaptability in handling unforeseen disruptions in real-time settings.


The remainder of this paper is organized as follows. 
Section~\ref{sec:liter_rev}  reviews the state-of-art real-time rescheduling strategies, demand-response (re)scheduling and reinforcement learning approaches in railway system to clarify the contribution of this paper. 
Section~\ref{sec:problem_statement}  describes the problem of real-time demand-responsive rescheduling and the features of MD.
Section~\ref{sec:DRL_framework}  proposes the hierarchical deep reinforcement learning framework for rescheduling trains in real time.
Section~\ref{sec:experiments}  reports the computational experiments, results and discussions of applying the real-time demand-responsive rescheduling approach to the instance based on data from the Zhengzhou flood period, 
and Section~\ref{sec:conclusions} concludes the paper and presents areas of future research. 

\section{Literature review}\label{sec:liter_rev}

\subsection{Real-time train rescheduling approach}

When a natural or man-made disaster occurs in the railway system, the railway network topology is changed due to disruptions and blockages so that trains need to be rescheduled to save the system performance. Recovery from such disruptions is more complex and time-consuming than dealing with daily disturbances. During the disrupting period, it is necessary to adjust the original timetable to an abnormal one and rearrange the trains until the physical disruption is removed. The dynamic performance of a railway system in post-disruption condition can be described as a multi-phase process, also known as the bathtub model \citep{RN131}. Therefore, the real-time rescheduling that aims to stabilize the exacerbation of the disruption's effects can be classified as a static optimization for a specific period or a dynamic control problem when the phases are not clearly identified.

\paragraph{Phase-based approaches} During the transition and response phases that occur immediately after the disruption, it is critical to adjust the operation schedule facing the disrupted conditions. Strategies such as rolling stock rescheduling, rerouting, short-turning are implemented to prevent the exacerbation of delay and cancellation.
\cite{Kroon2015} proposed a rolling stock rescheduling model to reduce the passenger delay, considering dynamic passenger flows. An iterative heuristic was proposed to solve the model and a lower bound is calculated to assess the solution quality.
\cite{VEELENTURF2017133} proposes an extension to the iterative heuristic approach of  \cite{Kroon2015} for rescheduling the rolling stock and timetable, considering passengers behavior, and adapting the stop patterns to minimize the delays and the number of denied passengers. 
\cite{DOLLEVOET2017203} proposes an iterative framework to reschedule the timetable, rolling stock, and crew simultaneously to obtain an overall feasible solution in an acceptable solving time.
\cite{RN105} presents a mixed integer programming to determine the waiting station during disruption and reordering trains after recovery in case of a complete blockage. The objective is to minimize the total weighted train delay and the number of canceled trains, adhering to the original timetable.
\cite{ghaemi_macroscopic_2018} optimizes short-turning solutions for complete blockages in the multi-phase framework. Rescheduling measures of short-turning, partial cancellation, and re-timing are considered on both sides of the blockage.  
\cite{RN132} reschedules a disrupted timetable that integrates flexible stopping and flexible short-turning with reordering, retiming, and cancelling. Passengers' planned paths are used to estimate decision weights that minimize passenger delays. The proposed model considers realistic infrastructure characteristics, multiple headways, rolling stock circulations, and all phases of a disruption.
\cite{HONG2021103025} consider rescheduling of trains and reassigning passengers after disruptions. The model includes train retiming, rerouting, reordering and reserving with the objective of minimizing the total train delay while maximizing the number of passengers to be served. 
\cite{RN93} presents a real-time train rescheduling model on double-track high-speed railways. This approach allows trains in both directions to share sidings at stations and uses an integer linear programming based on a space-time network to minimize arrival deviation. 
The above real-time optimization models aim to rapidly provide optimal solution for the given conditions, which can be applied multiple times for updated conditions. 

\paragraph{Rolling horizon approaches} Accounting for the uncertainty of disrupting duration, rolling horizon approach is widely used in real-time rescheduling  to capture the long-term benefit and modify the schedule during the situation progress. \cite{RN76} proposes a rolling horizon approach for rolling stock rescheduling, which is presented as an online combinatorial decision problem with uncertainty modeled by a sequence of information updates. \cite{zhan_2016} uses mixed integer programming to reschedule train services in partially blocked situations and adopts a rolling horizon approach to handle the uncertainty of the disruption duration.
\cite{zhu_2020}  proposes a rolling horizon two-stage stochastic timetable rescheduling model with short-turning strategy to manage uncertain disruptions in the railways. Compared to deterministic rescheduling models, the stochastic method generates better rescheduling solutions with less train cancellations and delays.
\cite{RN131} proposes two approaches, the sequential approach and the combined approach, to handle multiple connected disruptions that occur unexpectedly in railway networks. A rolling horizon approach is also introduced, which considers delaying, reordering, cancelling, flexible short-turning, and flexible stopping to minimize cancellations and deviations from the planned timetable.
\cite{zhang_2023} proposes a multistage decision optimization approach and a rolling horizon algorithm for real-time train timetable rescheduling in a high-speed railway network with seat reservation. Passengers transfer is allowed and the inconvenience is minimized by the model.
Among these strategies, the length of horizon and the frequency of updates have been shown to have an impact on both the computational efficiency and the quality of solutions. Therefore, striking a balance between computational efficiency and solution accuracy is a challenging task during practical implementation.

Table \ref{Tab-review-RT} summarizes the relevant studies on real-time rescheduling for disruptions. The table provides information about the function of the model (i.e. objectives, and strategies), assumptions (i.e. demand and disrupting duration) and solution methods.

The table highlights the fact that only a limited number of studies incorporate the dynamic nature of passenger demand in their strategies, with the majority employing static demand to assess the importance of trains. The objectives of the various strategies mainly aim to minimize delay, cancellations, and deviations from the original timetable. However, the studies have overlooked the real-time in-station demand and the risk of in-station overcrowding during disruptions.

\begin{table}[htbp]
  \caption{State-of-art real-time rescheduling strategies for railway disruption}
  \label{Tab-review-RT}
  \begin{tabularx}{\linewidth}{lXXccc}
    \toprule
    Paper & Addressed problem & Objective & Demand & Duration & Algorithm \\
    \midrule
    \cite{Kroon2015} & RS, PR, CA & Operating cost, Delay & Dynamic & De & IH \\
    \cite{VEELENTURF2017133} & RS, PR, RT, SS, CA & Demand, Delay, Cancellation, Operating cost & Dynamic & De & IH \\
    \cite{DOLLEVOET2017203} & RS, PR, RT, SS, CA & Demand, Delay, Cancellation, Operating cost& Dynamic & De & IH \\
    \cite{RN105} & RT, CA & Delay, Cancellation & - & De & CS\\
    \cite{ghaemi_macroscopic_2018} & ST, CA & Delay, Cancellation & - & De & CS \\
    \cite{RN132} & RT, ST, CA, SS & Delay, Cancellation & Static & De & CS \\
    \cite{HONG2021103025} & RT, SS, PR & Delay, Demand & Static & De & CS \\
    \cite{RN93} & RT & Deviation & - & De & ADMM \\
    \cite{RN76} & RS, CA & Operating cost & - & Un & RH \\
    \cite{zhan_2016} & RT, CA & Deviation, Cancellation & - & Un & RH, CS\\
    \cite{zhu_2020} & RT, ST, CA, SS & Delay, Cancellation & - & Un & RH, CS\\
    \cite{RN131} & RT, ST, CA, SS & Delay, Cancellation & - & Un & RH, CS\\
    \cite{zhang_2023} & RT, SS, PR & Delay, Passenger inconvenience & Static & Un & RH, CS\\       
    This paper & RT, RS, RR, SS, CA, PR & Demand, Delay, Overcrowding & Dynamic & Un & DRL \\
    \bottomrule
    \multicolumn{6}{p{\linewidth}}{Denotation: RS: rolling stock rescheduling, PR: passenger reassignment, RT: retiming, RR: rerouting, CA: cancellation, SS: stop pattern, ST: short-turning, De: determined, Un: undetermined, IH: iterative heuristic, CS: commercial solver, ADMM: alternating direction method of multipliers, RH: rolling horizon}\\
  \end{tabularx}
\end{table}

As indicated in the table, most rescheduling problems are addressed, including retiming, rerouting, cancellations, stop scheduling, and rolling stock rescheduling. Some papers also consider passenger reassignment to align passenger demand with the rescheduling plan. The primary objectives focus on minimizing the consequences of disruptions, encompassing operating costs, time delays, timetable deviations, unsatisfied demand, train cancellations, and passenger inconveniences. Among the 13 papers, 6 papers explicitly claim to incorporate passenger demand in their rescheduling strategies, with 3 of them treating passenger demand as a dynamic factor. Additionally, 8 papers assume predefined disruption durations, while 5 papers assume an open-ended time span. Among the various solving approaches, 8 papers use commercial solvers to obtain exact solutions. However, when passenger demand is considered as a dynamic factor, only heuristic approaches are feasible due to the complexity of passenger mobility.

This paper integrates rescheduling strategies of retiming, rerouting, cancellations, stop scheduling, rolling stock rescheduling, and passenger reassignment. The objectives are to minimize unsatisfied demand, reduce passenger time delays, and introduce a novel objective of minimizing in-station overcrowding at the target station. The paper proposes an iterative offline-training approach, DRL, which effectively manages the uncertainty of disruption durations and dynamic passenger demand.

\subsection{Demand-responsive strategy}

There are plenty of studies facilitating the demand-responsive (re)scheduling for general railway system, including urban rail and intercity railway. Table \ref{Tab-review-DR} summarizes the state-of-art demand-responsive (re)scheduling approaches accounting for the required data, passengers' mobility, objectives, supplying capacities and solution methods. In demand-responsive strategies, the satisfaction of passenger demand, in-station waiting time, and crowdedness are crucial factors to address the consequence due to the imbalance between demand and supply, such as surging demand, overcrowded station, and lack of rolling stock. Numerous approaches have been researched to improve demand satisfaction in regular operating conditions and manage headway during abnormal demand conditions. However, little research focuses on demand propagation and the allocation of limited rolling stock in the post-disruption scenario. In such conditions, a complete network blockage often results in a significantly reduced number of available rolling stock, causing significant delays. This situation then leads to stranded passengers and overcrowded stations. However, in most studies, the static number of in-station passengers is assumed to be the demand waiting for the service, while the reduction of passenger quantity due to their independent leaving is not considered.

In the intercity railway system, OD demand and seat reservation data adhering to the timetable are typically used to evaluate the necessity of dispatching a train at a given time. Many studies consider dynamic passenger flow, including passenger demand and route split, while rescheduling the timetable to cater for the original timetable and high-demand trains and cancelling trains with lower demand. However, passengers' independent behavior can cause bias in demand estimation. Due to a lack of real-time demand data, model-based methods are used to simulate passenger mobility during disruption. The cancellation and deviation between the rescheduled and original timetable serve as important indicators in evaluating the effectiveness of rescheduling methods, while the feasibility and convenience of demand recomposition and passenger reassignment captures the passengers' profit on the alternative lines.

\begin{table}[htbp]
  \caption{Demand-responsive train (re)scheduling in general railway system}
  \label{Tab-review-DR}
  \begin{tabularx}{\linewidth}{lcccXcc}
    \toprule
    Paper & System & Data & Mobility & Objective & Capacity & Solution \\
    \midrule
    \cite{sun_demand-driven_2014} & U & FS & AW & Waiting time & SU & CS\\
    \cite{niu_optimizing_2013} & U & FS & AW  & Demand Satisfaction, Waiting time & SU & H\\
    \cite{zhu_bi-level_2017} & U & OD & AW & Waiting time, Arriving deviation, Overcrowding & LI & H\\
    \cite{zhao_integrated_2021} & U & FS & AW & Waiting time, Operating cost & LI & H\\
    \cite{besinovic_matheuristic_2022} & U & FS & AW & Operating cost, Waiting time & LI &H \\
    \cite{RN150} & I & OD & AW & Waiting time & SU & CS \\
    \cite{VEELENTURF2017133} & I & OD & AW \& L & Demand Satisfaction, Delay, Operating cost & LI & H \\
    \cite{corman_integrating_2017} & I & OD & AW & Delay & SU & H \& RH  \\
    \cite{meng_integrated_2019} & I & OD & AW & Demand Satisfaction, Capacity utilization, Operating cost & LI & H \& DP \\
    \cite{yin_hybrid_2019} & I & OD & AW & Waiting time, Operating cost & LI & H\\
    This paper & I & MD & AW \& L & Demand Satisfaction, Delay, Overcrowding, Capacity utilization & LI & DRL \\
    \bottomrule
    \multicolumn{7}{p{\linewidth}}{Denotation: U: urban rail transit, I: intercity railway, FS: fare system, OD: historical OD demand, MD: mobile data, AW: arriving and waiting, L: independently leaving, SU: sufficient, LI: limited, H: heuristic, CS: commercial solver, RH: rolling horizon, DP: dynamic programming, DRL: deep reinforcement learning}\\
  \end{tabularx}
\end{table}

\subsection{Research gap and potential contributions}

There exists a research gap between real-time rescheduling and demand-responsive strategy due to data availability and solution efficiency. Specifically, most strategies are studied based on the emergence of fare collection systems in urban transit, which record passengers' arrival and departure times as well as OD locations. However, the seat reservation dataset in intercity railway systems does not record exact entry and exit times, limiting the development of demand-responsive services in these systems. Additionally, demand-responsive strategies that consider time-varying passenger quantities and route choices are complex and often solved using heuristic algorithms, which may not be suitable for real-time application requiring high quality results.

Recently, certain research has been employing reinforcement learning approaches for real-time rescheduling in operational control \citep{SEMROV2016250,zhu_re_2020,wu_deep_2021,zhu_reinforcement_2023} and disturbance management \citep{ning_deep_2019,Liao_deep_2021, wang_policy-based_2021}.
Thanks to the offline training process, an agent can effectively adapt to complex environments during regular operational days. However, there has not been a reinforcement learning approach for real-time rescheduling considering the time-variant demand in response to a complete disruption. Therefore, the contributions of this paper can be summarized as follows:

\begin{enumerate}[(i)]
    \item This paper proposes a real-time rescheduling approach for a long-term disruption considering dynamic in-station passenger demand and crowdedness. 
    The scenario of disruption addresses the challenges of passenger independent leaving, in-station overcrowding, reduced number of rolling stock, open-ended disruption duration, integrated rescheduling on multiple routes, and delays due to detours.
    The approach incorporates strategies of rerouting, retiming, reordering, cancellation, stop scheduling, and passenger reassignment to maximize the demand satisfaction and minimize the in-station overcrowding and arrival delay. \\

    \item To our best knowledge, this paper is the first study that bridges the gap between real-time rescheduling and demand-responsive requirements by incorporating MD as the demand-estimating approach. The MD is analyzed to reveal the real-world passengers mobility within the station during disruption and to evaluate the actual efficiency of dispatching a train. The agent is trained based on the historical MD to understand the regular pattern of demand propagation and to map the real-time information and the value of dispatching in disrupted condition. Due to the cost of data possessing, this study only select the MD of a high-demand station in the upstream to illustrate the applicability of this research. \\

    \item A novel data-driven approach, HDRL, is proposed to address the real-world passenger demand iteratively, effectively adapting to long-term disruptions with uncertain end times. To overcome the challenge of sparse rewards, a hierarchical structure is designed inspired by feudal hierarchy. In this structure, the upper-level agent is responsible for monitoring in-station demand and making dispatching decisions. Meanwhile, the lower-level agent is tasked with formulating specific rescheduling plans based on information from different routes. The agent undergoes offline training, and is transferred to a new online environment. 
\end{enumerate}

\section{Problem statement}\label{sec:problem_statement}

This study is motivated by a real-world flood disaster on 20th July 2021 in Zhengzhou, a central railway hub city in China. A large number of railway routes in China were severely disrupted or totally suspended for weeks and trains had to engage in emergency braking maneuvers near many stations. Because of the complete blockage of the central hub, the circulation of rolling stock was broken, causing significant reduction on the number of rolling stock in routes related to Zhengzhou. Facing prevalent train cancellations, thousands of passengers were stranded in affected up- and down-stream stations around Zhengzhou. Among the upstream stations, those hubs with high passenger volumes are particularly sensitive to passenger overcrowding. Xi'an, a high-demand hub station west to Zhengzhou, is selected as the focused station to reschedule the routes and timetable for available trains.

\begin{figure*}[htbp]
  \centering
  \includegraphics[width=\linewidth]{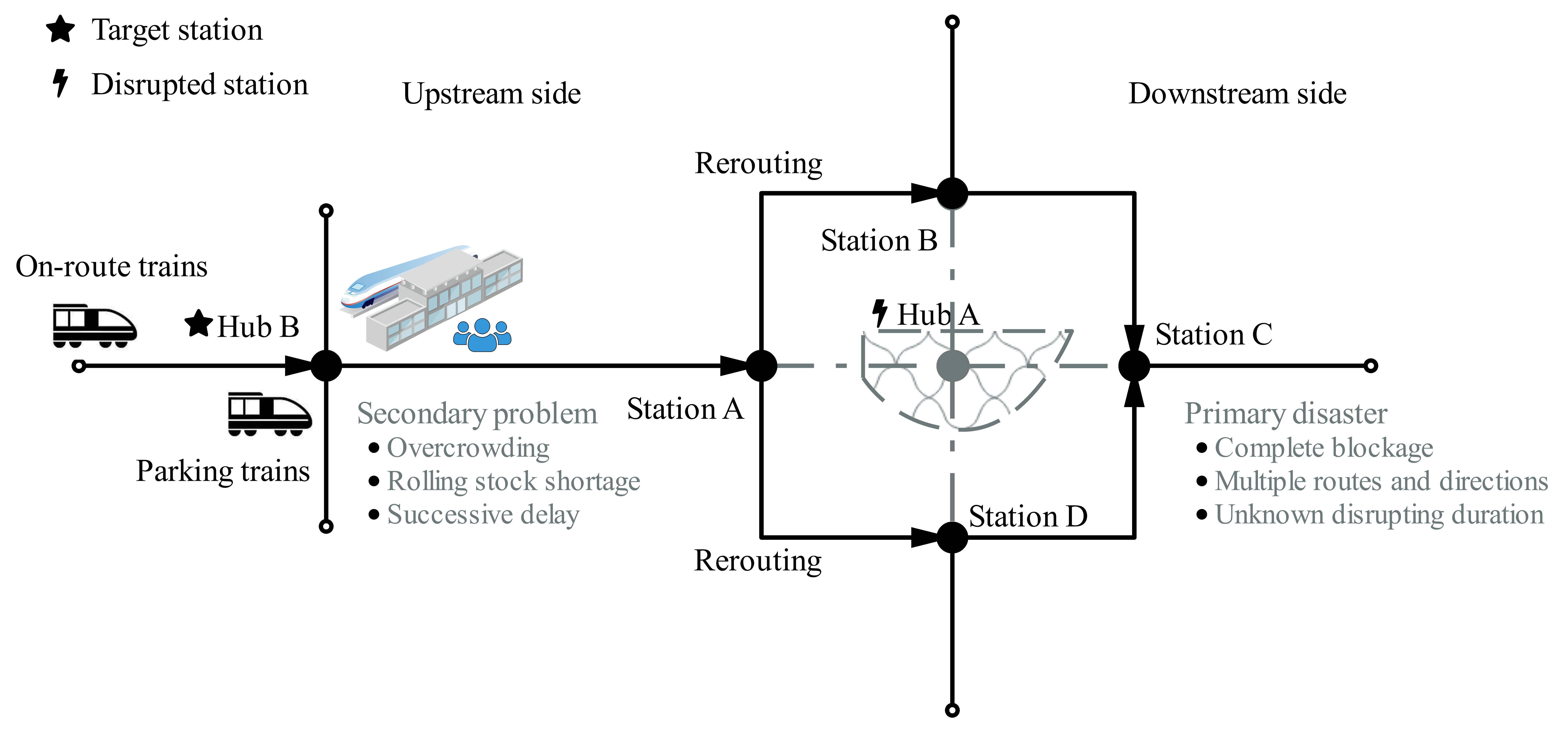}
  \caption{Scenario addressed by this study}
  \label{Fig-scenario}
\end{figure*}

The scenario addressed by our study is demonstrated in Figure \ref{Fig-scenario}. After a disaster occurred in a hub area, multiple railway routes are completely blocked and the recovery time is unknown. Due to the shortage of rolling stock and successive delay of trains, the station hall of upstream hub would suffer the secondary problem related to passenger overcrowding. Fortunately, there are some trains parking in the yard of the upstream hub and some other trains traveling towards the hub unaffected by the disruption. Meanwhile, there are some routes branching before the disrupting area, ensuring the accessibility of downstream routes. Therefore, this paper focuses on rescheduling the routes and timetable of trains from the upstream hub in real time in response to the in-station crowdedness and OD demands of this hub. 

\subsection{Real-time demand-responsive (RTDR) train rescheduling}
\subsubsection{Demand-responsive approach: model-based vs data-driven}

Traditionally, trains on the original timetable with lower demand are cancelled, and passenger groups are recomposed and reassigned to alternative trains to meet the passenger demand as much as possible. This approach neglects the real-world in-station demand, which ultimately limits the flexibility and efficiency of the rescheduling plan. Thus, by using the real-time data of in-station passenger demand, a demand-responsive rescheduling approach can be developed for the post-disrupted situation to meet dynamic demands and improve the capacity utilization.

\begin{figure}[htbp]
\centering
\subfigure[Model-based estimation]{
\includegraphics[width=0.45\linewidth]{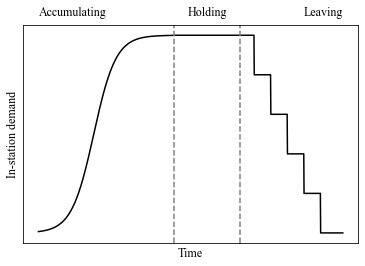}
}
\subfigure[Real-world condition]{
\includegraphics[width=0.45\linewidth]{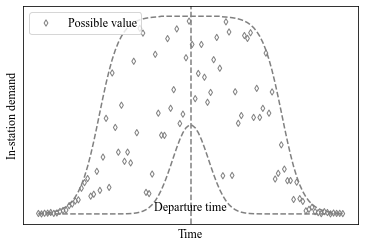}}

\caption{Time-variant demand for a train}
\label{Fig-TVdemand}
\end{figure}

Several model-based methods have been developed to estimate time-variant demand in a station. Typically, a sigmoid function is used to represent passenger arrival and accumulation processes before the original departure time \citep{RN150}, while waiting tolerance is set for different passenger groups to capture the holding and returning processes after the original departure time \citep{VEELENTURF2017133}. Similar demand modelling techniques are also found in relevant areas such as train timetabling with dynamic passenger demand \citep{BARRENA2014134,BARRENA201466}. Based on these assumptions, Figure~\ref{Fig-TVdemand}a presents a visualization of the time-variant demand for a train, using the model-based approach. The dashed lines in the graph divide stages for the demand dynamics, including the accumulation stage, passenger waiting/holding stage, and the stage when passengers start returning. In traditional studies, the value and shape of this figure will vary in terms of total demand, arriving rate and waiting tolerance. 

In reality, passengers' arriving and returning behavior is influenced by multiple factors, including weather conditions, time of the day, and available information on disruptions, etc. As a result, the demand exhibits stochastic and time-variant characteristics within a possible area. Figure~\ref{Fig-TVdemand}b illustrates this variability, where the demand fluctuates over time in a non-deterministic manner. This stochastic nature of passenger behavior highlights the challenges in accurately predicting and managing passenger demand in real-world scenarios. Furthermore, the demand for a train varies every day, which challenges the use of dynamic demand models in real-time rescheduling.

\subsubsection{Real-time rescheduling time horizon: evenly updated vs demand-responsive}

\begin{figure}[htbp]
\centering
\subfigure[Demand propagation of in-station demand]{
\begin{minipage}[t]{0.6\linewidth}
\centering
\includegraphics[width=\linewidth]{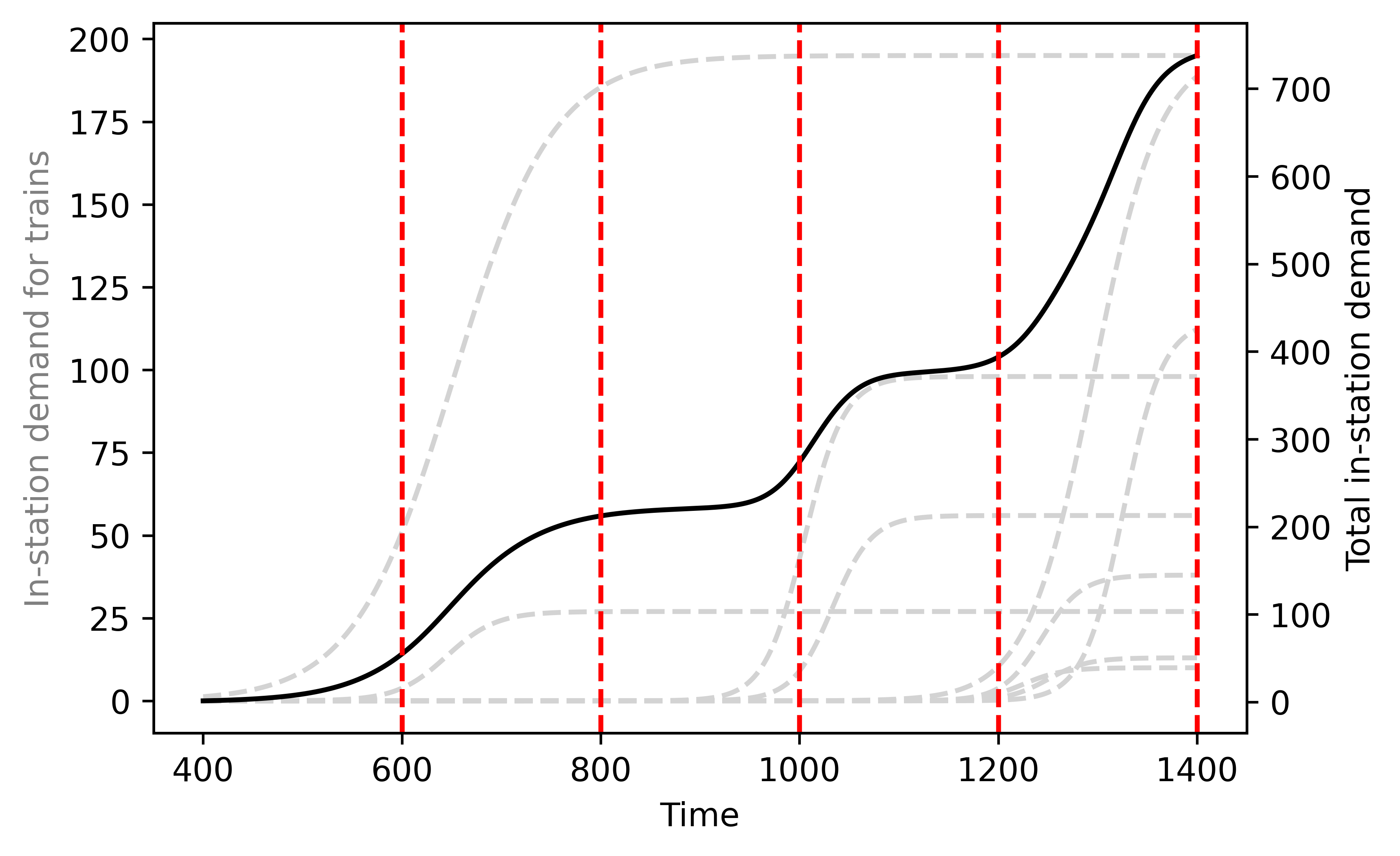}
\end{minipage}
}\\

\subfigure[Demand propagation with uniform interval]{
\begin{minipage}[t]{0.45\linewidth}
\centering
\includegraphics[width=\textwidth]{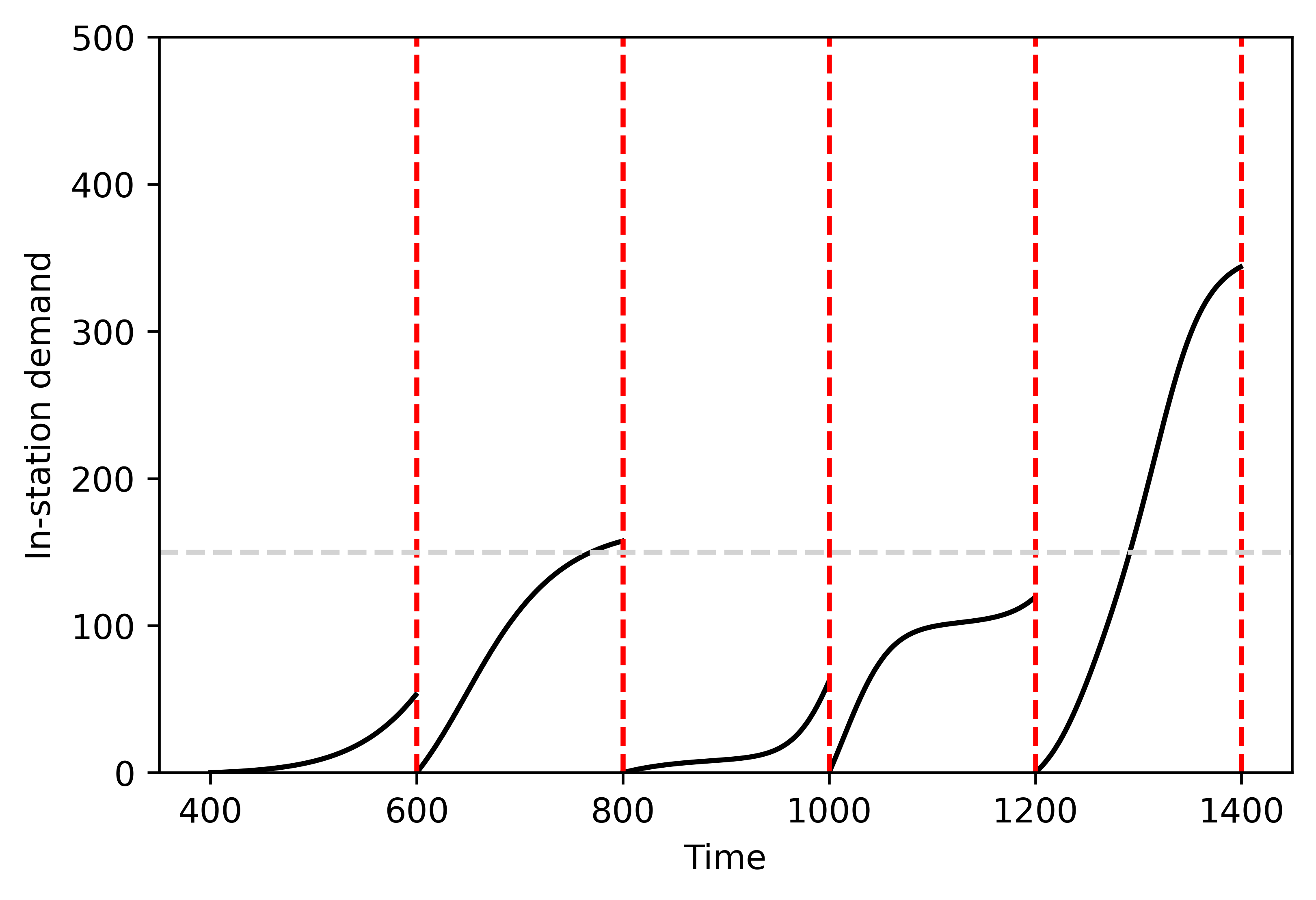}
\end{minipage}
}
\subfigure[Demand propagation with demand response]{
\begin{minipage}[t]{0.45\linewidth}
\centering
\includegraphics[width=\textwidth]{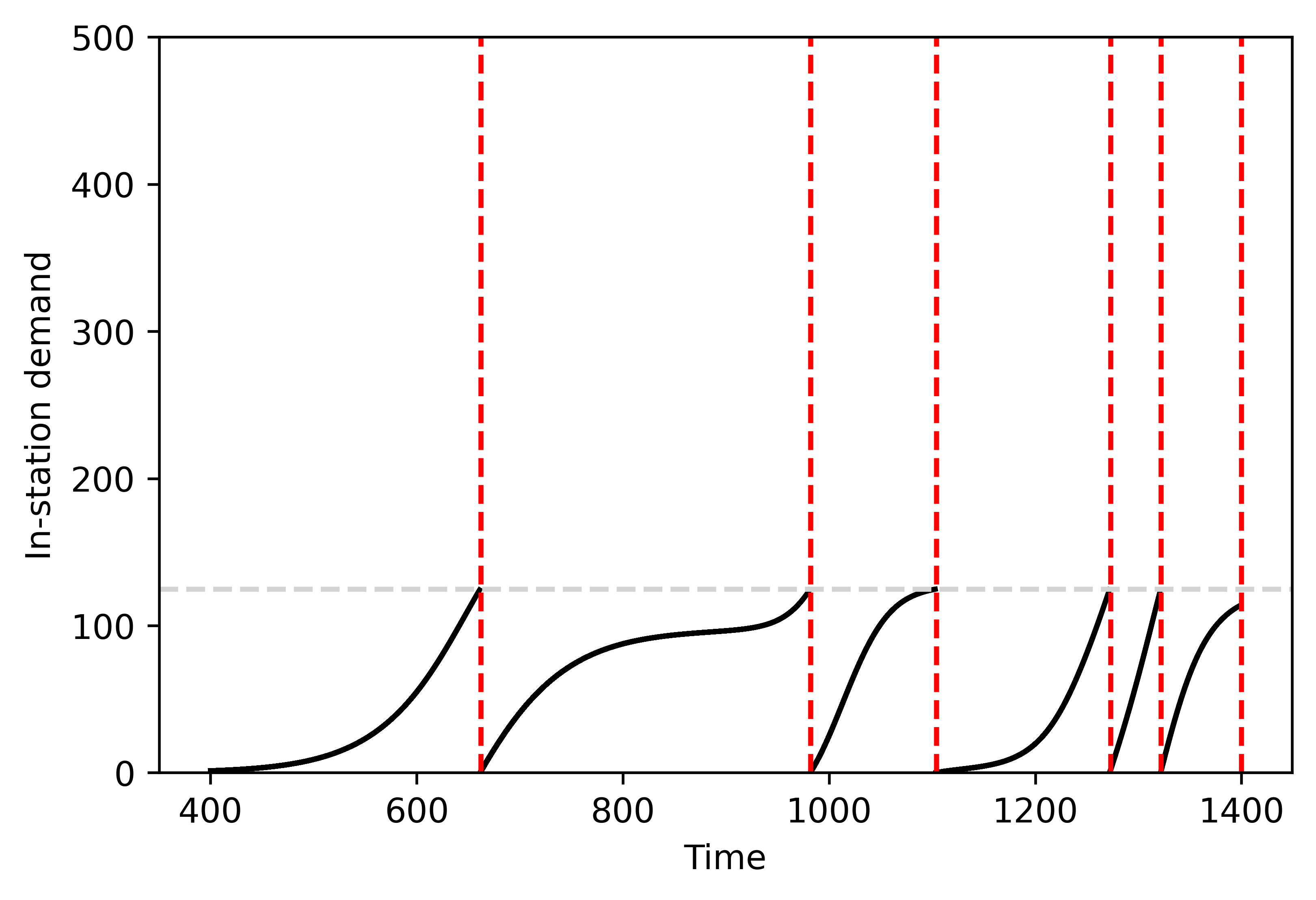}
\end{minipage}
}

\caption{Dynamic of in-station demand under different rescheduling updating principle}
\label{Fig-Demand}
\end{figure}

Schedule updating is critical in real-time rescheduling for long-term disruption. Traditionally, the schedule is often updated by fixed intervals, and the time span is evenly divided into time steps. Then, prediction and optimization approaches are used to provide rescheduling plans for the current or next time step. However, the length of the even intervals is usually a hyperparameter which is determined by testing or experience. In order to illustrate the significance of determining the length of intervals, we present a sample demand propagation pattern within a station, as shown in Figure \ref{Fig-Demand}a. The corresponding timetable is presented in Table \ref{Tab-exa-timetable} where the timetable is on the route between Xi'an and Shanghai during a day. Over a time span from 400 to 1400 minutes of the day (6:40 AM to 11:20 PM), there are 9 trains operating on this route. The train codes, departure time, timestamp of the day (transformed into minutes) and demand data are derived from the MD and timetable dataset dated July 19th, 2021. We used the sigmoid function to simulate the passenger accumulation process for each train, which is depicted by the gray dashed curves. The total in-station demand is represented by the black solid curve. The figure illustrates that the \emph{total} passenger accumulation experiences periods of sharp increase (600-800, 900-1100, and 1200-1400 minutes) and periods of slow increase (400-600, 800-900, and 1100-1200 minutes). This is because the departure times of trains are concentrated between 20:25 and 22:06, resulting in an uneven distribution of demand around this time. 

Figure \ref{Fig-Demand}a is evenly divided by 200 minutes, indicated by the red lines, which is a commonly used approach. In order to simulate the capability of fixed-interval approaches, we make an ideal assumption that the in-station demand can be met in every interval. Consequently, the propagation of in-station passenger quantities is illustrated in Figure~\ref{Fig-Demand}b. At the beginning of each interval, the demand starts from zero, indicating that the demand has been fully met by last interval. However, due to the uneven distribution of demand in each interval, the accumulated passenger demand can be higher, particularly during the sharp-increase hours. The gray dashed horizontal line represents a potential overcrowding threshold. It highlights the risk of overcrowding in the station hall when using the uniform division principle. Additionally, if the capacity of the planning horizon is insufficient during peak hours, some passengers may be denied boarding. In contrast, a demand-responsive updating principle is illustrated in Figure \ref{Fig-Demand}c. This strategy ensures that the in-station demand is cleared whenever the accumulated demand reaches a specific limit, represented by the gray dashed horizontal line in the figure. It adjusts the dispatching plan in a flexible manner, considering the passenger demand and in-station crowdedness, enabling the dispatch of a train before reaching the overcrowding threshold. By removing the restriction of the fixed interval, the in-station demand can be cleared more evenly and strategically.

In addition, the previous example in Figure~\ref{Fig-TVdemand} illustrates the potential risk of overcrowding when the time interval is too long. Shortening the length of interval can address this problem. However, with short intervals, the computational time will increase, inducing the waste of computing resource. Meanwhile, the granularity of the division relies on the real-time algorithm's processing speed. The length of interval could not be less than the solving time of the algorithm. Hence, a novel algorithm is needed to handle the dilemma of the time span division. This algorithm should be able to monitor in-station conditions in real time and update the rescheduling plan as needed. Specifically, it should efficiently conserve computational resources during normal in-station conditions while being capable to generate high-quality rescheduling plans that rival those produced by optimization approaches.

\begin{table}[!ht]
    \centering
    \begin{tabular}{llll}
    \hline
        Train & Departure & Timestamp & Demand \\ \hline
        K362 & 10:45:00 & 645  & 27 \\ 
        Z378 & 10:52:00 & 652  & 196 \\ 
        Z94 & 16:45:00 & 1005  & 98 \\ 
        K559 & 17:13:00 & 1033  & 57 \\ 
        T115 & 20:25:00 & 1225  & 11 \\ 
        Z163 & 20:44:00 & 1244  & 39 \\ 
        Z39 & 20:50:00 & 1250  & 13 \\ 
        Z254 & 21:36:00 & 1296  & 198 \\ 
        K2188 & 22:06:00 & 1326  & 116 \\ \hline
    \end{tabular}
    \caption{Sample of timetable and demand on a route}
    \label{Tab-exa-timetable}
\end{table}

\subsection{Description of mobile data}
This paper infers the in-station demand from an anonymized mobile dataset, which captures the mobile phone users' mobility using signal towers. Each tower oversees a specific area. By analyzing the data from towers that cover the railway station area, it is able to derive information about the mobility of mobile users during the day \citep{RN119}. However, the precision of MD is limited due to the data being captured in a grid network. Additionally, for personal information security, all individual-related mobility and information are protected by the signaling company. Filtering and grouping actions can be performed on the database platform to obtain statistical data without any personal details. 

The data items related to our research include entering time, leaving time, leaving mode, next stop, localization, visiting frequency to this station, and user quantity. A detailed description of these items is provided in Table \ref{Tab-data}. In the table, entering time, leaving time, and next stop are recorded by the signal tower, while leaving mode, localization, visiting frequency, and user quantity are processed and provided by the mobile company, which has sufficient data to identify such features. Methodologically, leaving mode is inferred by mapping users' speed and trajectory to the infrastructure networks to distinguish between the roadway, railway, metro, and airline. Localization is inferred based on the users' residential area and appearance times to determine whether they are residents of the city. Visiting frequency is calculated by summing up the number of times a user has visited the station in a month, and it is categorized as high, medium or low. User quantity is a weighted sum to infer the total number of users from one mobile company's data.

\begin{table}[htbp]
    \centering
    \caption{Data description}
    \begin{tabularx}{\linewidth}{llX}
    \toprule
    Item & Type & Explanation \\
    \midrule
    Entering time & Timestamp & The date and time indicating when a user enters the station area. \\
    Leaving time & Timestamp & The datet and ime indicating when a user exits the station area, including both returning and departure. \\
    Leaving mode & Category & The mode a user takes after leaving the station, including roadway, railway, metro, air.\\
    Next stop & Binary & The next destination after a user leaves the station, whether inside or outside the city. \\
    Localization & Binary & Whether this user is a resident of this city.\\
    Visiting frequency & Category & User's visiting frequency of the station in a month, which is categorized as high, medium, or low.\\
    User quantity & Integer & The total number of users in this category. \\
    \bottomrule
    \end{tabularx}
    \label{Tab-data}
\end{table}

Table \ref{Tab-MD} provides a comparison of MD, OD demand, and ticketing data for the demand estimation in railway system. According to the table, OD demand and ticketing data provide accurate demand quantities for different OD pairs and individual trains. However, they do not include information on in-station passenger quantity and check-in/check-out time. Additionally, the temporal precision of these data is limited to either the day-level or train-level. In contrast, MD contains passengers' check-in/check-out time in minute-level and the real-time in-station passenger quantity. However, passengers' destinations present challenges in demand estimation using MD. Section \ref{Method-Env} addresses these problems by combining the timetable dataset and a model-based method.

\begin{table}[htbp]
  \caption{Comparison of MD, OD demand and ticketing data}
  \label{Tab-MD}
  \begin{tabularx}{\linewidth}{lXXXXXXX}
    \toprule
    Data & TD & IN & OUT & O & D & S & TP \\
    \midrule
    MD & X & \checkmark & \checkmark & \checkmark & X & \checkmark & Minute\\
    OD & X & X & X & \checkmark & \checkmark & X & Day\\
    Ticketing & \checkmark & X & X & \checkmark & \checkmark & X & Train\\
    \bottomrule
    \multicolumn{8}{p{\textwidth}}{Denotation: TD: demand for each train, IN: Check-in time and quantity, OUT: Check-out time and quantity, O: onboard station and quantity,  D: deboard station and quantity, S: in-station passenger quantity, TP: temporal precision}\\
    
  \end{tabularx}
\end{table}

Due to the fact that the MD of every city is controlled by the local mobile company, it is costly to obtain all MD of alongside stations. Therefore, this paper only focuses on one high-demand station to demonstrate the proposed methodology. If the MD data for all alongside cities and stations were available, it could have shed more light on the OD mobility.

\subsection{Challenges and assumptions}

In summary, this paper focuses on a disruption scenario in a hub area where multiple routes in different directions are impacted and the rolling stock circulation is interrupted. As a result, the supply-demand imbalance becomes pronounced, causing consecutive delays and escalating the risk of in-station overcrowding. Among the stations, the high-demand stations located upstream of the disrupted area are particularly susceptible to overcrowding \citep{Liu2022AME}. Thus, the study proposes a real-time demand-responsive (RTDR) approach to reschedule the trains from the target station for mitigating the consequence of the disruption.

The challenges associated with this scenario can be summarized as following aspects, including passenger mobility, in-station crowdedness, rolling stock shortage, open-end disruption duration, and the integrated rescheduling for multiple routes.

\begin{enumerate}[(i)]
    \item Passenger Mobility: Passengers' mobility is considered in real-time, taking into account their arrival and returning times. To accommodate passenger needs, it is essential to manage waiting times effectively. Passengers have a tolerance threshold, and if their waiting time exceeds this limit, they may request a refund.
    \item In-Station Crowdedness: The level of overcrowding within the station is a dynamic factor influenced by the demand for both disrupted and normal trains. Proactive train dispatching is crucial to prevent overcrowding and mitigate the secondary risk due to railway disruption.
    \item Rolling Stock Shortage: The available rolling stock falls short of the original scheduling plan, necessitating the reassignment of passenger demands. Passengers destined for the same location but different trains must be reassigned to rescheduled trains.
    \item Open-end Disruption Duration: The duration of the disruption is uncertain, requiring a contingency plan to ensure that backup rolling stock is sufficient to cover the remaining time in the day. Adequate resources must be allocated to meet the ongoing demand.
    \item Integrated Rescheduling for Multiple Routes: Multiple routes with different directions require rescheduled trains. However, it is necessary to consider the overall rescheduling of rolling stock to accommodate all affected routes effectively.
\end{enumerate}

Our proposed RTDR rescheduling framework and the corresponding HDRL solution method are proposed to specifically address the above five challenges associate with the proposed scenario. In order to maintain the practicality of our approach, we have the following assumptions:

\begin{enumerate}[(i)]
    \item Train Codes Affected by Disruption: We assume that the train codes impacted by the disruption are known or can be identified based on the provided disruption information.
    \item Detour and Rerouting Capability: We assume that the disrupted area can be bypassed by detour and rerouting. This enables trains to continue their scheduled routes in the downstream. Additionally, we assume that the demand for the disrupted hub can be met by diverting passengers to the nearest station in proximity to the hub. In order to simulate their transfer trip, their actual arrival time is extended approximately.
    \item Organization of in-station passengers: We assume that in-station passengers will be announced about the rescheduled plan by the station broadcast service. The in-station passengers are assumed to be willing to leave as soon as possible regardless the specific train.
    \item Availability of rolling stock: Although the rolling stock circulation is disrupted, we assume that some trains have already passed through the disrupted area prior to the onset of the disruption, and there are some trains traveling to the target station from the unaffected area (e.g. from the upstream of the target station). These early trains, along with the trains parking at the target station, are considered to be available for rescheduling.
    \item Sufficient headway in the downstream: Headway safety is considered in our approach. The headway of rescheduled trains dispatched from our target station is imposed to be in a safe range. We also assume that rescheduled trains will not conflict in downstream sections due to significantly sparser track occupancy caused by disruptions compared to the original timetable. This will provide acceptable conditions for our target routes. If needed, additional adjustment can still be made locally in downstream areas without considerably affecting the overall reschedule. 
\end{enumerate}

\section{Hierarchical deep reinforcement learning framework}\label{sec:DRL_framework}
In order to train HDRL in an iterative manner, it is necessary to define four fundamental components: environment simulation, state variables, actions of agent, and reward feedback. The subsequent sections delineate the methodology to overcome the following challenges: 1) the real-time demand for trains should be estimated from an anonyminized MD dataset; 2) the agent should learn the value of dispatching a train during a long-term period according to the in-station conditions; 3) the agent should learn the efficiency of a rescheduling plan under disrupting condition.

\subsection{Environment establishment} \label{Method-Env}

The environment simulates the real-world condition as dynamic factors in the database. The dataset is divided into three subsets for training and testing, which are normal days (most of days from July 1 to 19 2021), days with disturbance due to heavy rain (on July 4, 11, and 18), and the day when disruption happened (on July 20). Although passenger demand varies among days, their arriving process and peak time have a common and regular pattern during the normal days \citep{RN150}, so, this dataset is used for offline training HDRL reacting to the real-time demand in a simulated disrupting scenario. Passengers depart punctually on normal days so that very few data records passengers returning. However, the data from days with disturbances and disruptions contains sufficient passenger returning data, thus, these data is employed to estimate waiting tolerance. Lastly, data on the day when disruption happened is used for online testing the transferability of the trained HDRL in a real-world disrupting condition.

Estimations need to be made for both the offline training dataset and the online testing dataset. As indicated in Table \ref{Tab-MD}, estimating passenger demand using MD poses two challenges: 1) matching passenger demand with trains, and 2) splitting demand in target station to downstream destinations. This section addresses these challenges and provides examples of real-time demand propagation in both stations and trains. Additionally, estimating the passengers' waiting tolerance in the offline training dataset is addressed as well.

\subsubsection{Data preprocessing}

To ensure the quality of the MD, it was necessary to preprocess the data and filter out any incorrect or irrelevant information. Since the dataset contains both passengers and regular workers, it was important to filter out the regular workers and retain only the potential passenger data. The following data are filtered out of the dataset:

\begin{itemize}
    \item \textbf{Missing data}: we defined the data as missing when its weight is 0 or NAN.
    \item \textbf{Error data}: we defined the data is error when the holding time is less than 5 minutes.
    \item \textbf{Duplicate data}: when two data are the same in all parameters, we delete one of them.
    \item \textbf{Regular workers}: we primarily filtered out the users who frequently visited the station and were also identified as local residents.
    \item \textbf{Mode choice}: we only collected data whose mode for the next trip is either by car (returning) or by railway (departing).
    \item \textbf{Time period}: data records around 0 o'clock contain many missings so that we set the study period as 4a.m.-11p.m.
\end{itemize}

To protect the privacy of individual data, potential passengers with the same characteristics are grouped together, and no individual trajectory is output. Passengers are classified into groups based on their entering time, leaving time, and leaving mode. Passenger groups with a leaving mode of railway and a next stop outside the city are classified as successfully departed passengers, while groups with a leaving mode of roadway and a next stop inside the city are classified as potential passengers who abandoned their trip and returned.

\subsubsection{How to match passenger demand with trains?} \label{Method-match}

Under normal circumstances, trains follow a strict timetable, and passengers also leave the station according to the timetable. Therefore, by combining the MD and timetable, passenger demand can be matched to the trains. However, both the offline training dataset, which contains data from normal days, and the online dataset, which contains data from disrupted days, have their limitations.

\begin{enumerate}[(i)]
    \item \emph{Offline training dataset with data from normal days}\\
    Under normal conditions, trains depart punctually according to the predefined timetable. The passenger groups that successfully departed have been filtered in the preprocessing stage. Then, the demand for each train can be estimated by matching the leaving time of each passenger group with the train timetable, and calculating the weighted sum of passenger quantities. In offline training datasets, there is a scarcity of real-world waiting tolerance records for passengers during normal days because most passengers leave the station on time. Conversely, in online testing datasets, most passengers abandon their trip and the duration between their entering and leaving the station can be considered as the real-world waiting tolerance. Therefore, a Random Forest Regression is employed to estimate the waiting tolerance of passengers on normal days using abnormal dataset.
    \item \emph{Online testing dataset with data from abnormal days}\\
    Under disruption and disturbance conditions, trains are either significantly delayed or cancelled, resulting in a majority of passengers' departure times deviating from the timetable. Passenger groups that have successfully departed are those whose trains were not affected by the disruption, whereas returning passenger groups comprise potential passengers who were disrupted by the railway incident. However, matching the leaving time of these groups with the timetable is not feasible for estimating their target train. Therefore, a Random Forest Classifier is employed to predict the target train for a passenger group based on their entering time and the accumulated patterns of the trains using data from normal operating days.
\end{enumerate}

The detailed procedure of estimating waiting tolerance and target trains can be found in Appendix \ref{App-1}. The machine learning process is conducted using the \texttt{sklearn} package \citep{scikit-learn} of Python 3.9. With the estimation results, we can obtain information such as the entering time, leaving time, target train, planned departure time, and passenger quantity for each passenger group in both offline and online dataset. We assumed that the codes of the interrupted trains are provided along with the disruption information. This enables us to filter the potential passenger demand based on their target trains. We consider the passengers who target the interrupted trains as the focus of our strategy, while the passengers who are traveling on the normal operating route can depart based on the actual conditions recorded in the dataset.

The environment of passenger mobility can be summarized in the Table \ref{Tab-Env}. There are 10 factors that capture the mobility of each passenger group $g$ in the station area. These factors are group ID, entering time, planned departure time, returning time, target train code, route number of this train, whether they are currently in the station, the proportion of their demand that has been met, whether they have been denied, and the quantity of their demand. The first three factors are timestamps in minutes, representing the time that the group enters the station $t_A$, their original departure time $t_{Dep}$, and the time which oversteps their waiting tolerance $t_L$. The target train $i$ is indicated by the train code. The route number $r \in R$ indicates the route on which the train travels, and all non-disrupted routes are numbered as 0. The state whether the passenger group is in the station $\xi_{in}$ is indicated by a binary variable. This variable takes a value of 1 when the current time is between the entering and departure time for non-disrupted groups, and between the entering and returning time for disrupted groups. The satisfaction factor $p_{M}$ indicates the proportion of satisfied demand of disrupted group by the previous rescheduled trains. It is needed because the stop pattern is considered in our strategy where a rescheduled train can ether cover all stops or a part or stops on the route. Denying indicates that the remaining in-station passenger gives up the trip, which is defined as denied by the railway system. The denied variable $\xi_{L}$ becomes 1 when the current time is over their returning time. The demand factor $\delta_{g}$ represents the number of passenger in this passenger group with similar feature. Finally, an indexing manner is defined to indicate the information of each passenger group as $g_n()$. For example, $g_1(i) = \mathrm{K000}$ indicates the train code of passenger group 1, $g_1(r) = 1$ indicates the route number of group 1. 

\begin{table}[htbp]
  \caption{Environment of passenger mobility (current time: 800)}
  \label{Tab-Env}
  \begin{tabularx}{\linewidth}{p{0.04\linewidth}p{0.08\linewidth}p{0.08\linewidth}p{0.08\linewidth}p{0.08\linewidth}p{0.08\linewidth}p{0.08\linewidth}p{0.08\linewidth}p{0.08\linewidth}p{0.08\linewidth}}
    \toprule
    ID & Enter  & Depart  & Return  & Train  & Route  & In  & Satisfaction  & Denied  & Demand  \\
    \midrule
    $g$ & $t_A$ & $t_{Dep}$ & $t_L$ & $i$ & $r$ & $\xi_{in}$ & $p_{M}$ & $\xi_{L}$ & $\delta_{g}$ \\
    \midrule
    1 & 560 & 780 & 1000 & K000 & 1 & 1 & 0.5 & 0 & 20\\
    2 & 612 & 890 & 890 & K111 & 0 & 1 & 0 & 0 & 15 \\
    \bottomrule
    
  \end{tabularx}
\end{table}

There are two sampling data in Table \ref{Tab-Env}, and we assume that the current time step is 800 (equals to 13:20). The first line of the data represents a group of passengers who intend to board train K000 on the disrupted route (route 1). The original departure time of K000 is 780, which is equivalent to 13:00. Their waiting tolerance is until 1000, which is equivalent to 16:40. Currently, they are inside the station, and a rescheduled train has covered 50\% of their demand, which included some of the stops of K000. At present, the current time has not yet reached 1000, so they have not been denied. The total demand for this group is 20 passengers, and the remaining demand is 20$\times$(1-0.5)=10. The second line of data pertains to a group of passengers headed towards train K111, which is not disrupted and belongs to route 0. The scheduled departure time for K111 is 890, and currently, this group is waiting in the station. As the current time has not reached the scheduled departure time yet, the in-station demand for this group is 15.

\subsubsection{How to split onboard demand to downstream destinations?}

The method discussed in Section \ref{Method-match} allows us to determine the train demand at our target station. However, this demand includes the demand for all downstream stations. To split the demand from the origin station into OD demand, we employ the gravity model. This model uses the attractiveness of downstream stations for a particular train to divide the total demand from the origin station into OD demand, as shown in Eq. \eqref{Eq. gravity}. In the equation, $\delta^{d'}_g$ represents the demand of group $g$ traveling to a downstream station $d'$. $s_i$ is the set of stops passed by a particular train $i$, and $\alpha^d$ represents the attractiveness factor of station $d$. Previous data-driven research of trip distribution has shown the effectiveness of this model \citep{cordera_trip_2018}. The attractiveness factor can be derived from various statistical data sources, such as historical OD demand \citep{cordera_trip_2018}, Gross Domestic Product (GDP) \citep{Liu2022AME}, etc.

\begin{equation}
    \delta^{d'}_g = \frac{\alpha^{d'}}{\sum_{d \in s_i}\alpha^d}\delta_g
    \label{Eq. gravity}
\end{equation}

Therefore, the total demand for a downstream station $d$ would be the summation of all passenger groups and all trains, as $\delta_d = \sum_g \delta^{d'}_g$. The estimation of real-time OD demand has been accomplished through the demand splitting method. The method has used the MD to derive the demand towards downstream stations. A dynamic dictionary has been employed to store the demand for different routes towards downstream stations. The dictionary uses the route number as the key and a list as the corresponding value to represent the demand for each station on the route, $\delta(r_n) = \{\delta_{d_1},\delta_{d_2},...\}$. As the time window progresses on the environment of passenger mobility, the in-station crowdedness, total demand and OD demand will be automatically changed.

Stop scheduling and passenger reassignment are allowed in this paper, enabling rescheduled trains to cover a portion of the demand from passenger groups that originally targeted specific trains. When a rescheduled train $i'$ covers a subset of stops of train $i$, the in-station demand will be satisfied based on the first-in-first-out principle. If a rescheduled train operates on stops $s_{i'}$ and a group of passengers intend to board the train with stops $s_i$, the demands towards stops that exist on both trains $s_{i'} \cap s_i$ can be accommodated, subjecting to the available capacity of this train $\kappa_i$, which is indicated by $\sum_d \delta_g^{d' \in s_{i'} \cap s_i} \leq \kappa_i$. The proportion of demand that can be met for a passenger group is calculated using Equation \ref{Eq. metdemand}.

\begin{equation}
    p_M = \frac{\sum_d \delta_g^{d' \in s_{i'} \cap s_i}}{\delta_g}.
    \label{Eq. metdemand}
\end{equation}

\subsection{Hierarchical DQN agent}

In order to train the agent understanding the current condition in the time-varying environment, state variables should be dynamically updated from the environment dataset. The real-time passenger environment has been established using the MD dataset. These state variables from the passenger side include in-station crowdedness, total in-station demand on interrupted routes and dynamic OD demands. The states from railway operating side include dynamic quantity of available rolling stock and headway. 

The dynamics of in-station passenger has been explained, which relates to their mobility. Additionally, the quantity of available rolling stock is treated as time-variant in our study as well. The rolling stock comes from two ways that the trains parking in the yard and the trains stopped and waiting at the station due to the disruption ahead. The first kind of rolling stock train (with an amount of $N_{RS}$) provides a static number of rolling stock and these trains are flexible to be scheduled onto any route. The second kind of trains $RS^*$, which comes from other non-disrupted routes, provides a dynamic number of rolling stock and these trains are restricted to be scheduled on their original routes. Their available time is up to the arrival time at the target station. They are restricted on the routing choice because their on-board passenger demands heading to downstream stations should be met.

States (state variables) are explained below:

\begin{itemize}
    \item \textbf{In-station crowdedness $\delta^\text{in}_t$}: The total number of passengers present in the station at the current time, including those on normal and interrupted routes. This state represents the risk of overcrowding in the station, $\delta^\text{in}_t = \sum \delta_g (1-p_M)  \xi_{in} $.
    \item \textbf{In-station demand $\delta^{\text{in},r}_t$}: The total passenger demand for only interrupted routes. This state represents the urgency of dispatching a train to alleviate the disruption, $\delta^{\text{in},r}_t = \sum \delta_g  (1-p_M)  \xi_{in} $ where $g(i,r) \neq 0$.
    \item \textbf{OD demand $\delta^d_t$}: A dictionary of demand for each downstream station. This state represents the urgency of dispatching a train on a specific route, $\delta^d_t = \delta_d (1-p_M)  \xi_{in}$.
    \item \textbf{Available number of rolling stock $\bar{N}_{RS}$}: The number of rolling stock waiting in the station for rescheduling instructions. This state represents the capacity available for rescheduling.
    \item \textbf{Potential number of rolling stock $\bar{N}_{RS}^*$}: The number of rolling stock has not arrived at the station for rescheduling instructions. This state represents the potential capacity available for rescheduling.
    \item \textbf{Headway $H^r_t$}: A list of the up-to-date headway on each route. The on-route headway is continually updated at each time step. After each round, the headway for routes without a dispatched train is increased by a time interval, while for routes with a newly dispatched train, the headway is reset to zero. This state ensures that trains are dispatched with a safe headway between them.
\end{itemize}



\begin{figure*}[htbp]
  \centering
  \includegraphics[width=0.5\linewidth]{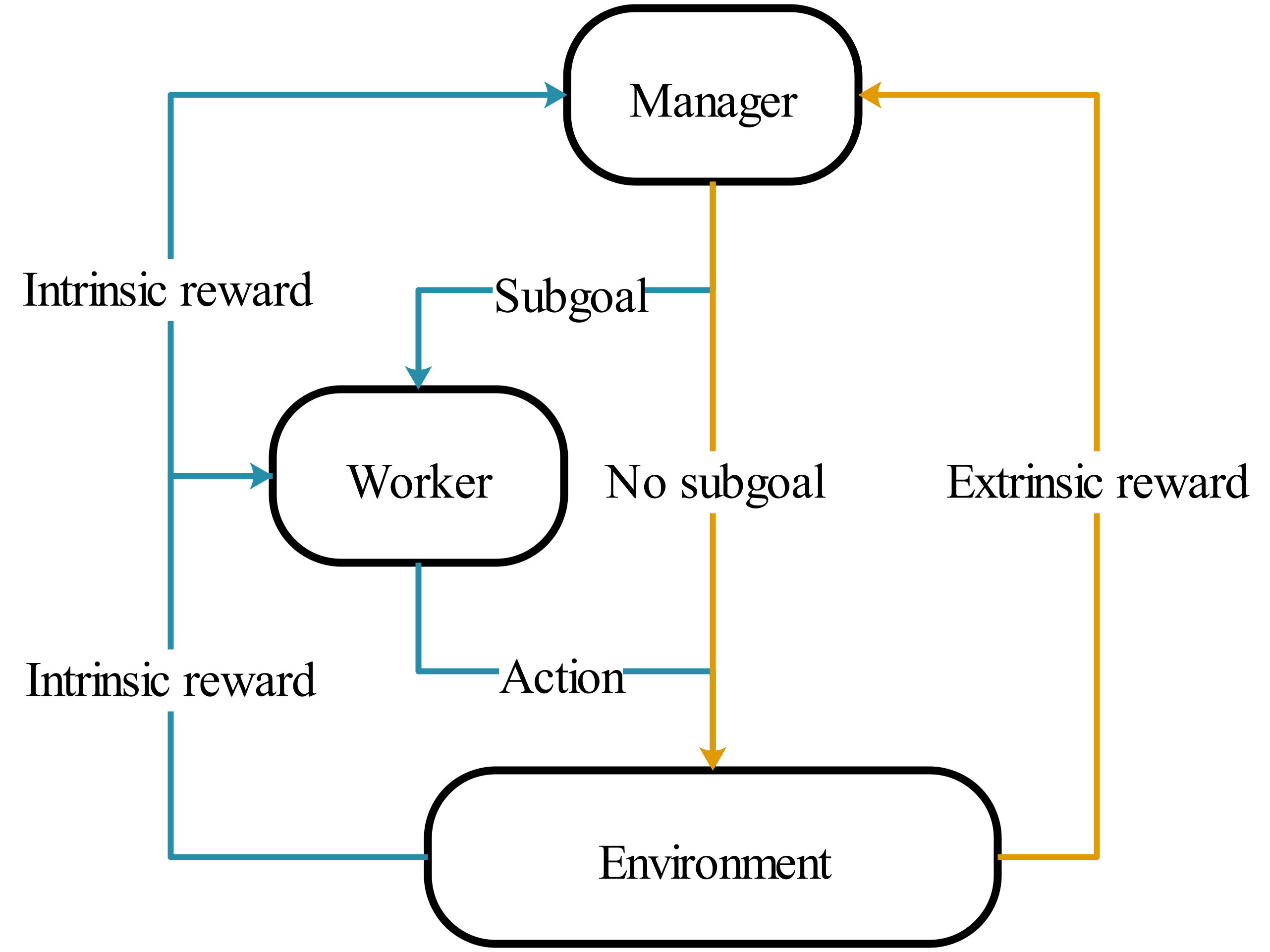}
  \caption{Feudal hierarchy structure}
  \label{Fig-FH}
\end{figure*}

Due the complexity of state variables and sparse reward issue, the agent is implemented as a feudal hierarchy structure with two sub-agents. In the context of a feudal hierarchy, each sub-agent is assigned a distinct level of abstract information. The upper-level agent plays the role of a ``manager'', responsible for deriving subgoals based on the state variables it monitors. The lower-level agent operates as a ``worker'', tasked with achieving the subgoal provided by the upper level. The feudal hierarchy structure is illustrated in Figure \ref{Fig-FH}.  If the subgoal is set, the worker will make an action and get the reward from environment. This reward is defined as the \emph{intrinsic reward} in our study. The manager will get a reward from environment no matter a subgoal is set. This reward is defined as the \emph{extrinsic reward}. In our scenario, the manager monitors in-station conditions to determine whether a train is required to clear the in-station demand effectively, whose decision is the subgoal for the lower level. Meanwhile, the worker decides which route necessitates a train with urgency and efficiency. The subgoal in our problem is not predefined but it is learned by the agent itself. This learning is based on the fundamental principle of accumulating the maximum reward throughout the long-term decision-making process (similar settings can be found in \cite{Zhong2023AHF, Casanueva2018FeudalRL}).

The states of the upper-level agent $s_1$ include the current in-station crowdedness, demand, availability of rolling stock, potential number of rolling stock, and the current time step. The actions are two instructions $a_1 \in \{0,1\}$ where 1 indicates to dispatch a train and 0 indicates to keep holding. In the lower level, the states $s_2$ include the OD demand and headway on each route. The agent selects a predefined \emph{rerouting plan} $r$ and \emph{stop pattern} $p_n, n=1,2,3$ from the sets of routes and stops, i.e., $a_2 = (r,p_n), r \in R, p_n \in P(r)$. We assumed that rerouting plans are given to detour the disrupted area. These rerouting plans replace the original routes of trains during disruption. Even on the same route, different trains have distinct stop patterns. Some significant stations are stopped by all trains, while some less significant stations are only served by one or two trains. So, not all stations on the route need to be stopped by every rescheduled train. The stop patterns need to be rescheduled on the rerouting plan as well. 

The  stop pattern  $p_n \in P(r)$ on each route $r$ is determined by the significance of the station according to the following Algorithm \ref{Alg-stop}. There are multiple trains operate on each route with different stop pattern. In our study, the significance of a station is reflected by the number of trains passing this station every day where the more train passes the station, the more important the station is. The stop patterns on each route are categorized into three types: Type $p_1$ corresponds to fast trains that only stop at significant stations, which are present in every train on the route. Type $p_2$ represents normal trains that make stops at some less important stations, which are visited by more than two trains. Type $p_3$ refers to slow trains that stop at all stations along the route.

\begin{algorithm}
\caption{Stop pattern determination}\label{Alg-stop}
\begin{algorithmic}[1]
\Require Timetable, Route set $R$, Train set $I$, Station/Stop set $\Sigma$.
\ForAll{route $r \in R$}
\State Collect the trains on route $r$ from the timetable, $I_r := \{i \in I : Route(i) = r\}$, where $Route(i)$ gives the route of train $i$;
\State Collect the set of train stops on this route $\Sigma_r:=  \bigcup_{i \in I_r} Stops(i) $, where $Stops(i)=\{\sigma \in \Sigma : \sigma \in Route(i) \} $;
\ForAll{stop $\sigma \in \Sigma$}
\State Calculate $NT(\sigma) = |\{i \in I_r: \sigma \in r \} | $, i.e., the sum of the number of trains passing station $\sigma$;
\EndFor
\ForAll{stop $\sigma \in \Sigma$}
\State Determine the significance of $\sigma$ according to $NT(\sigma)$:
\State \quad -- Significant stations: $\Sigma_1 = \{ \sigma :  NT(\sigma) = |I_r| \}$, i.e., all trains pass this station;
\State \quad -- Less significant stations: $\Sigma_2 = \{\sigma : 2 \leq NT(\sigma)<|I_r|$\}, i.e., at least 2 trains pass this station;
\State \quad -- Insignificant stations: $\Sigma_3 = \{\sigma :  NT(\sigma)=1 \}$, i.e., only 1 train passes this station;
\EndFor
\State Output 3 stop patterns:
\State \quad $p_1$: only stop at $\Sigma_1$;
\State \quad $p_2$: stop at $\Sigma_1$ and $\Sigma_2$;
\State \quad $p_3$: stop at $\Sigma_1$, $\Sigma_2$ and $\Sigma_3$;
\State A set of stop pattern on $r$ is produced as $P(r) = \{p_1,p_2,p_3\}$;
\EndFor
\State Output: $P(r), \forall r \in R$: A dictionary of stop patterns on all disrupted routes.
\end{algorithmic}
\end{algorithm}

Though the action space is discrete in a low dimensionality, all state variables of upper and lower levels are high dimensionality. Their complexity increases with the range of value in exponential pattern making the table-based Q-learning approach result in a 'curse of dimensionality'. Therefore, a DQN approach is used to formulate the both agents.

\subsection{Reward and penalty principle}

Based on the nature of long-term disruption, the disrupting time is divided into time steps, and the agent interacts with the environment in every time step. Real-time rescheduling during a long-term disruption induces the sparse reward issue in DRL training. In most time, the agent just monitors the states but does nothing. Only it dispatching a train or in some special conditions, will it get a reward from the environment. Therefore, the technique of ``reward shaping'' is applied using a \emph{step reward}, given after every time step, and an \emph{episode reward}, given at the terminal of episode \citep{Ng1999PolicyIU,reward_shaping_thesis2004}. 

\subsubsection{Step reward}

At each time step, the agent takes an action and the impact of this action in the environment is evaluated to determine the rewards and penalties that the agent receives. The extrinsic rewards include the in-station crowdedness and the quantity of denied passengers. These rewards are received by the upper-level agent regardless of whether a train is dispatched. Only in instances when a train is dispatched, indicated by $a_1=1$, does the upper-level agent gain intrinsic rewards. These intrinsic rewards are contingent on the quality of the rescheduling plan, guided by the actions of the lower-level agent. The lower-level agent is trained only with the intrinsic rewards. The step reward is a weighted sum of these factors, calculated in Eq. \eqref{Eq.reward-step}. 

\begin{equation}\label{Eq.reward-step}
    R_S = \omega_1\delta^-_{t+1}+\omega_2\delta^\text{in}_t+a_1(\omega_3\delta_i+\omega_4u_i+\omega_5\Delta_i+\omega_6H^r_t),
\end{equation}

where $\delta^-_{t}$ is the sum of the denied passenger groups. $\delta^\text{in}_t$ measures the current in-station crowdedness. $\delta_i$ is the satisfied demand by train $i$. $u_i$ is the capacity utilization, $u_i=\delta_i/\kappa_{RS}$, where $\kappa_{RS}$ is the capacity of the rolling stock used in this scenario. When the utilization is lower than a threshold, it will get a penalty with $1-u_i$. $\Delta_i$ is the total delay induced by rescheduled train $i$. The original arrival time of a passenger group $g$ at downstream station $d$ is indicated by $\widehat{arr}_d^{g}$, and the actual arrival time is indicated by $arr_d^{g}$. The total delay is indicated by $\Delta_i=\sum_g\sum_d \max(0,(arr^{g}_d - \widehat{arr}_d^{g}))$. $H^r_t$ measures the reward related to the headway on route $r$. If the departure time of a train induce an unsafe headway, $H^r_t$ will be a penalty for this plan.

\subsubsection{Episode reward}

To assess the performance of the agent for an episode, an evaluation is conducted to provide feedback, based on the principle of satisfying enough demand, making efficient use of rolling stocks and avoiding in-station overcrowding. The episode reward is set as an exponential type, given in Eq. \eqref{Eq.reward-epi}.

\begin{equation}\label{Eq.reward-epi}
    R_E = \beta_1e^{\beta_2\delta+\beta_3 N_{RS}} - \omega_{CR},
\end{equation}

where $\beta_1, \beta_2$ and $\beta_3$ are coefficients, $\delta$ represents the total satisfied demand by the episode, and $N_{RS}$ is the total number of rolling stock used in the episode. The value of $\omega_{CR}$ is the penalty of overcrowding, given in Eq.~\eqref{Eq.crowdedness}, which will only be valid when the maximum quantity of in-station passenger is greater than the threshold of overcrowding $\Theta_{CR}$. Otherwise, no penalty is given.

\begin{equation}\label{Eq.crowdedness}
    \omega_{CR} = 
    \begin{cases}
                \max_t(\delta^\text{in}_t), \quad \max_t(\delta^\text{in}_t) > \Theta_{CR}, \\
        0, \quad  \mathrm{otherwise}.\\
    \end{cases}
\end{equation}

\subsection{HDRL training method}

\begin{figure*}[htbp]
  \centering
  \includegraphics[width=\linewidth]{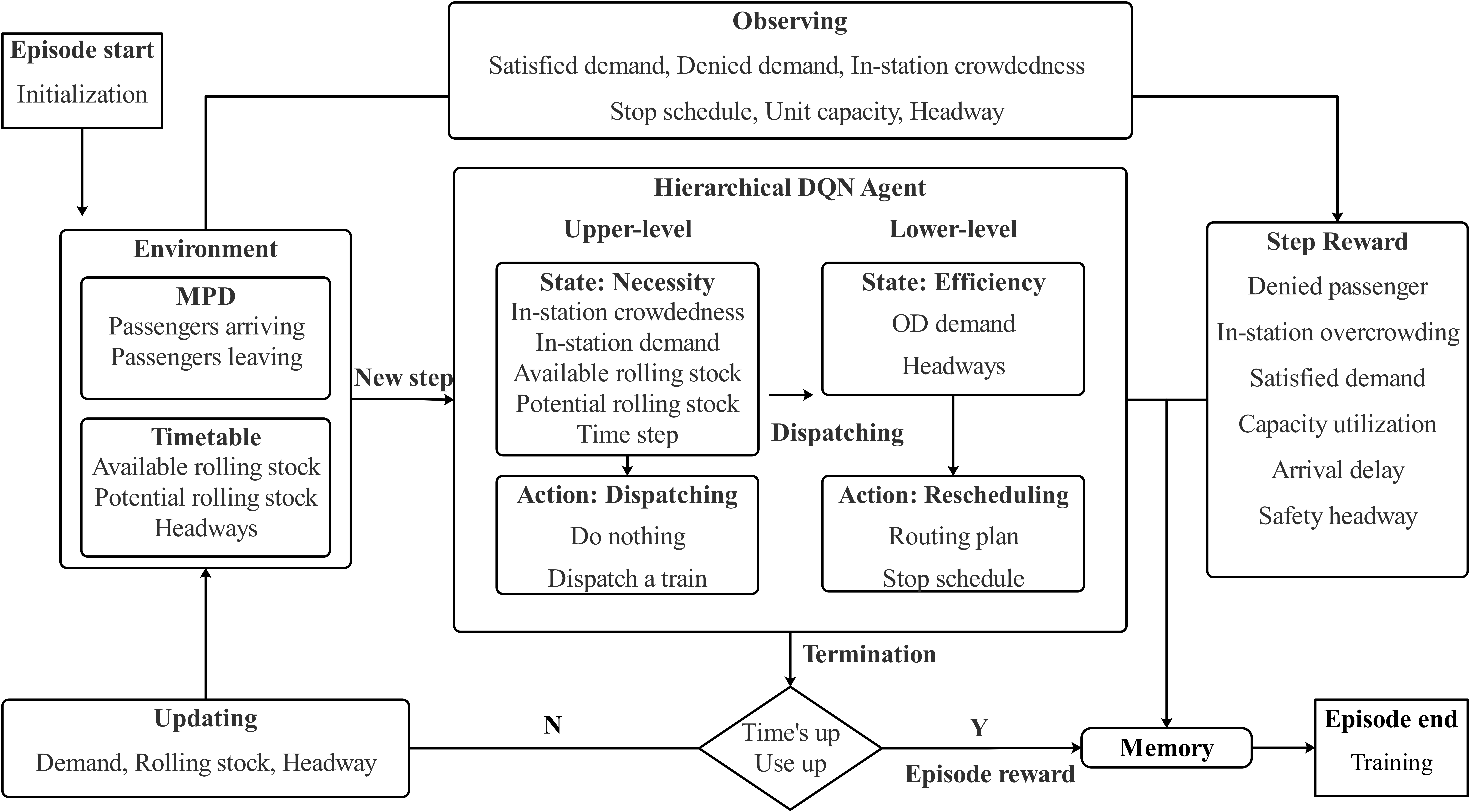}
  \caption{Learning framework of hierarchical DQN}
  \label{Fig-QL}
\end{figure*}

The foundation components of HDRL framework enable the agent dynamically interacting with the environment, and learning the value of dispatching and rescheduling in an iterative manner. A DQN approach is used in both levels to facilitate the high-dimensional states and discrete actions. Figure~\ref{Fig-QL} illustrates the closed-loop training framework. The parameters are initialized at the start of each episode, and the DQN is trained using the memory replay method at the end of each episode. The continuous time dimension is discretized into time steps, and the agent acquires new information from the environment at every time step, providing actions to the environment and getting feedback. Two termination conditions are established. First, the episode ends when the time limit is reached (``time's up''). Second, the episode ends when no  rolling stock resource is available (``used up''). After each episode, the state, action, reward, and next action are saved in the memory set. At the end of an episode, the memory set, along with the episode reward, is used to train the hierarchical DQN. To encourage the agent to learn the short-term and long-term benefits, the transition set of an entire episode with sequential step rewards and final episode reward is used for training.


The HDRL training method follows the psuedocode in Algorithm \ref{Alg-DQN}, which is a hierarchy structure modified from DQN training. The weights of ANNs are initialized at the beginning, including the weights of agent networks $Q_1(\theta_1),Q_2(\theta_2)$ and of target networks $\hat{Q_1}(\theta_1^{-}),\hat{Q_2}(\theta_2^{-})$ for upper-level and lower-level agent, respectively, and the iteration times is set as $E$. The iteration starts from the initial state $s_1^1$. An epsilon-greedy method is used to balance exploration and exploitation. The upper agent acts based on random or experienced choices. If its action at the current state is 1, indicated by $a_1^t = 1$, the lower-level agent will act following epsilon-greedy and get a reward feedback from the environment; otherwise, the upper-level agent will not get this intrinsic reward. The total step reward $R_S$ will be summed up accordingly, and the transition of this step will be recorded into the memory $M$ as a tuple $(s_1^t,a_1^t,s_2^t,a_2^t,R_{S},s_1^{t+1},s_2^{t+1})$. At the end of the episode, the terminal states $s_1^T, s_2^T$ will be used to calculate the episode reward $R_E$, and this tuple will also be stored into the memory $M$. Finally, the memory of transitions will be used to train the upper-level and lower-level agent respectively. Iteratively, the weights of target networks are updated to be a soft copy of the weights of the agent network. 

The DQN training method with the gradient decent policy is used in both agents. The target Q-value $y^t$ are obtained by taking the reward from the experience $R_S$ or $R_E$ and adding the discount factor $\gamma$ and maximum Q-value of the next state $\hat{Q}(s^{t+1},a^{t+1};\theta^{-})$ by target network, which is derived from the Bellman function in Eq. \eqref{Eq.DNQ-1}. The loss function calculate the mean square error for the target Q-value $y^t$ and predicted target Q-values $Q(s^t,a^t;\theta)$ in Eq. \eqref{Eq.DNQ-2}. At last the weight of the agent ANNs is updated by a gradient descent method $\theta \leftarrow \theta - \alpha\nabla\mathcal{L}$ where $\alpha$ is the learning rate.

\begin{equation}\label{Eq.DNQ-1}
    y^t=r^t+\gamma\max_{a^{t+1}}\hat{Q}(s^{t+1},a^{t+1};\theta^{-})
\end{equation}

\begin{equation}\label{Eq.DNQ-2}
    \mathcal{L}(\theta)=\frac{1}{|M|}\sum_{t}(y^t-Q(s^t,a^t;\theta))^{2}
\end{equation}

\begin{algorithm}[H]
\caption{Hierarchical DQN training with an episode experience}
\label{Alg-DQN}
\begin{algorithmic}[1]
\State Initialize replay memory $M$
\State Initialize upper-level and lower-level Q-network $Q_1,Q_2$ with random weights $\theta_1, \theta_2$
\State Initialize target networks $\hat{Q_1},\hat{Q_2}$ with weights $\theta^{-}_1=\theta_1,\theta^{-}_2=\theta_2$
\For{episode $= 1$ to $E$}
\State Initialize state $s_1^1$
\For{timestep $= 1$ to $T$}
\State With probability $\varepsilon_1$ select a random action $a_1^t$, otherwise $a_1^t=\arg\max_{a}Q_1(s_1^t,a_1;\theta_1)$
\If {$a_1^t=1$}
\State With probability $\varepsilon_2$ select a random action $a_2^t$, otherwise $a_2^t=\arg\max_{a}Q_2(s_2^t,a_2;\theta_2)$
\State Execute lower action $a_2^t$ and observe intrinsic reward and next state $s_2^{t+1}$
\EndIf
\State Calculate the total step reward $R_S$
\State Store transition $(s_1^t,a_1^t,s_2^t,a_2^t,R_{S},s_1^{t+1},s_2^{t+1})$ in $M$
\State Evaluate the episode reward $R_E$, and store the terminal transition $(s_1^T,a_1^T,s_2^T,a_2^T,R_E)$
\EndFor
\State Sequentially pick transitions from a replay memory $M$
\State Update the weights of $Q_1(\theta_1),Q_2(\theta_2)$
\State Periodically update the weights of $\hat{Q_1}(\theta_1^{-}),\hat{Q_2}(\theta_2^{-})$
\EndFor
\end{algorithmic}
\end{algorithm}

The epsilon-greedy method is adopted to balance exploration and exploitation during the learning process. The main idea behind the epsilon-greedy method is to make a trade-off between taking actions that are estimated to be the best and taking random actions to explore the environment. The parameter epsilon ($\varepsilon$) controls the exploration rate. Initially, when the agent has limited knowledge, a higher epsilon value is set to encourage exploration. As the learning progresses, the epsilon value is gradually decreased, shifting the agent's focus towards exploitation of the learned knowledge.

To further enhance the decay of epsilon in the epsilon-greedy method, this paper introduces a sigmoid function to modulate the exploration rate over time in Eq.~\eqref{Eq. epsilon}. The sigmoid function offers a smooth and gradual transition in epsilon values, enabling more fine-grained control over the exploration-exploitation trade-off.

\begin{equation}\label{Eq. epsilon}
    \varepsilon = \varepsilon_0 - \frac{\varepsilon_0 - \varepsilon_{\min}}{1+e^{-\gamma_{\varepsilon}((t-\gamma_t) - 0.5)}}
\end{equation}

where $\varepsilon$ represents the exploration rate at a given time step $t$. $\varepsilon_0$ is the initial exploration rate, $\varepsilon_{\min}$ is the minimum exploration rate, and $\gamma_{\varepsilon}$ controls the rate of decay and $\gamma_t$ controls the iteration times near the $\varepsilon_0$.

Given the nature of HDRL, the issue of non-stationarity happens due to the accuracy of extrinsic rewards depending on the stationary of lower-level sub-agent \citep{Levy2017LearningMH}. The lower-level DQN should be trained before the upper-level agent. The upper level's $\varepsilon$ is initially set to 1, implying entirely random dispatching decisions. Simultaneously, the weights of the upper-level ANN remain unchanged. Only the lower-level $\varepsilon$ decays using the given sigmoid function, and its weights are updated accordingly. After the performance of lower-level agent becoming stationary, the upper level's $\varepsilon$ starts to decay, meanwhile, the lower level's agent is set in a lower value, which ensures the agent making decisions based on the experience. This stepwise training manner ensures the step reward to be reliable for the upper-level agent.

\section{Computational experiments}\label{sec:experiments}

\subsection{Case description}
Zhengzhou, located in the central region of China, serves as a vital railway hub. However, from July 20th to July 23rd, 2021, the city was severely affected by devastating floods. The local railway stations faced significant challenges during this period, leading to disruptions in the railway services. The Longhai railway, which connects the east and west parts of the country, as well as the Jingguang railway, connecting the north and south, had to be shut down for several weeks starting from July 20th. Consequently, a critical railway route connecting the northwest inland and the southeast coast was completely disrupted, leading to 137 trains stopping at nearby stations, cancelling, or delay.

\begin{figure*}[htbp]
  \centering
  \includegraphics[width=0.8\linewidth]{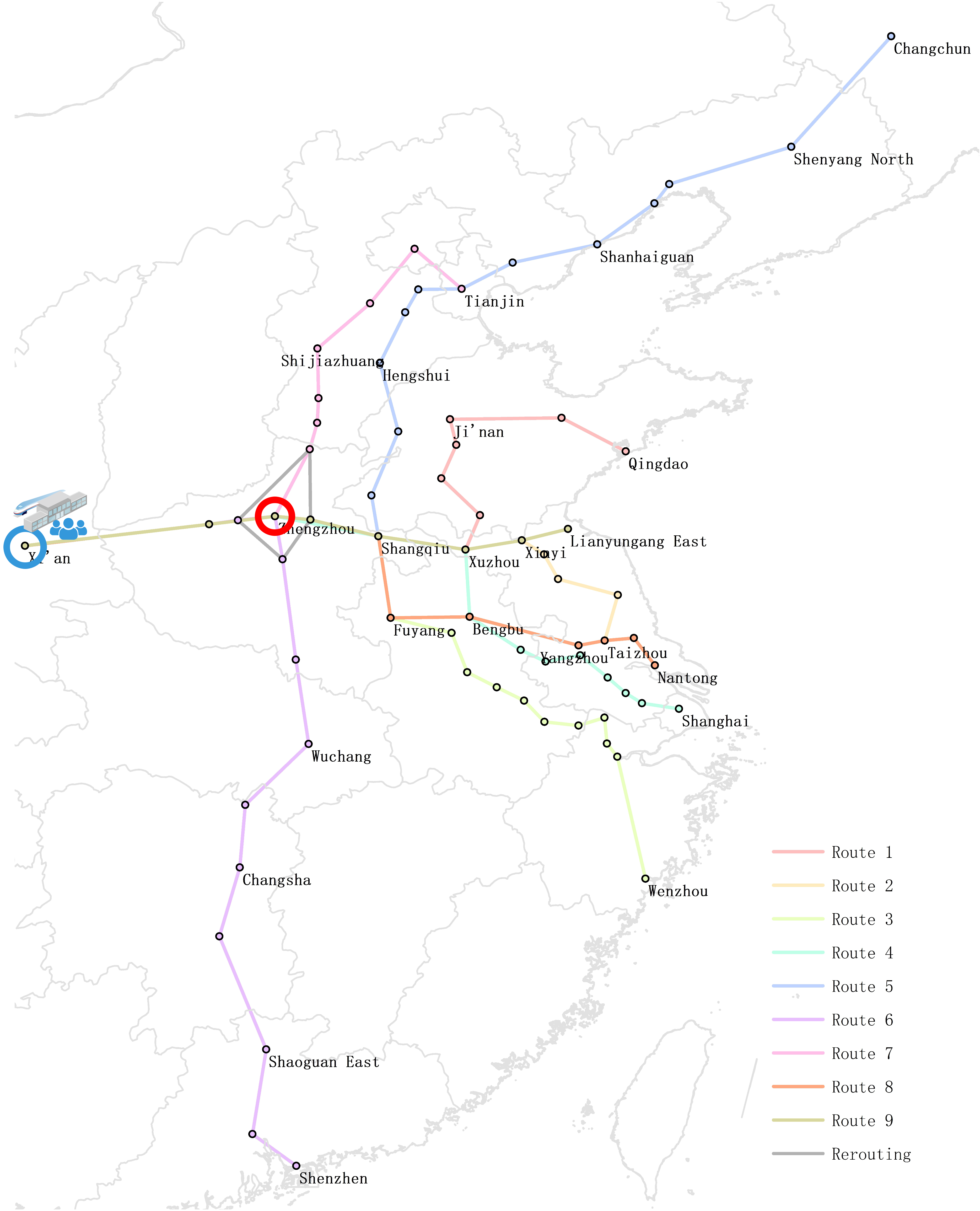}
  \caption{Railway network disrupted by the Zhengzhou flood}
  \label{Fig-map}
\end{figure*}

Xi'an, an important railway hub situated to the west of Zhengzhou, relies on the Longhai railway for connectivity. As a consequence of the flood-induced disruptions, all train services heading towards the northeast, east coast, and southeast regions had to be either cancelled or urgently halted at nearby stations. Nine railway routes passing through Xi'an and Zhengzhou were directly affected by the floods, as depicted in Figure~\ref{Fig-map}. Among these routes, there are a total of 33 trains interrupted by the flood.

\begin{figure*}[htbp]
  \centering
  \includegraphics[width=\linewidth]{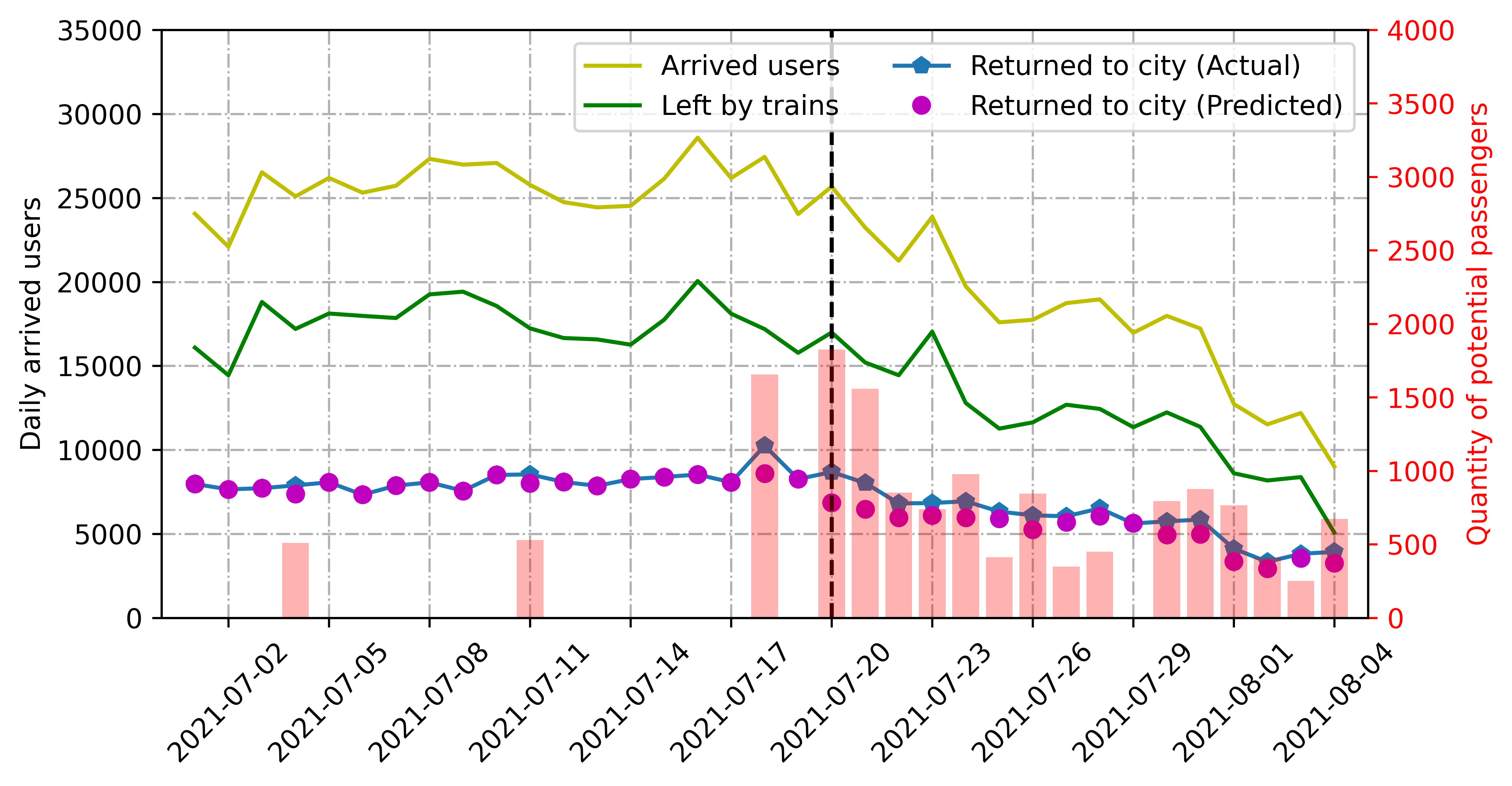}
  \caption{Daily movement of users who arrived at Xi'an station.  The right y-axis (red) is a scale for the bar chart representing the quantity of potential passengers who abandoned their plans and returned}
  \label{Fig-return}
\end{figure*}

Following the disruption, a large number of passengers gathered in the station hall of Xi'an and remained there for an extended period. Eventually, most of them abandoned their trips and left the station. According to the analysis of MD, the quantity of passengers who stayed within the station premises and ultimately chose to return was noticeably higher than the average on regular days. Figure \ref{Fig-return} showcases the movement of station users from July 1st to August 4th before and during the flood, where the methods and models employed are introduced in Appendix \ref{App-2.1}. The left y-axis of the graph displays the total number of individuals who arrived at the station (yellow), left the city by train (green), returned to the city area (blue), and the regular users identified through the model (purple). The discrepancy between the actual number of returned users and those identified by the model reflects the quantity of interrupted passengers who initially intended to travel but ultimately gave up due to the disruption. The right y-axis (red) depicts a bar chart representing the quantity of potential passengers who abandoned their plans and returned.

Figure \ref{Fig-return} is split from July 20th where the left part reflects the normal condition and the right part reflects the disruption period. As for the railway travelers, their daily quantity was between 15,000 to 20,000 before the disruption, which decreased during the disruption by about 10,000. There was only a small proportion of users who gave up and returned without boarding a train (referred to as ``returned'' hereafter) on normal day except July 4th, 11th and 18th. On those days, trains suffered long-last delays due to the heavy rainfall \citep{Weather_Xian} but the downstream Zhengzhou Station was still functioning. However, after the downstream Zhengzhou Station failed, the upstream Xi'an Station began to suffer from over-saturated passengers, who might not been timely informed about the emergency and stranded in the railway station. They finally returned to the city without boarding. The detailed consequences resulting from the unexpected disruption on July 20th have been discussed in Appendix \ref{App-2.2}, which addresses the issues of prolonged waiting times in Figure \ref{Fig-waiting} and overcrowding within the train stations in Figure \ref{Fig-instation}.

\subsection{Offline training}

To train the agent to understand the value of dispatching and rescheduling, a simulated environment was created using data from a normal day, specifically July 19th, which was one day before the disruption. The MD data from July 19th and the real-world railway timetable in July were used to establish the environment. The network setup is illustrated in Figure \ref{Fig-map}, consisting of 9 disrupted routes, which are physical railway routes with specific terminations. Throughout the day, 33 trains traveled along these disrupted routes, all of which were disrupted. The disrupted train timetable is summarized in Table \ref{Tab-33train} in Appendix \ref{App-3.1}.

A set of rescheduling plans with specific stop pattern and timetable, from which the lower-level action selects, is predefined by Algorithm \ref{Alg-stop}. The stop pattern for each route has three options, corresponding to $s_1, s_2$ and $s_3$ in the algorithm. For example, on route 4, the stops and travel times are outlined in Table \ref{Tab-RnSn} in Appendix \ref{App-3.2}. The travel times are calculated based on the average travel time from the original timetable. In the disrupted area near Zhengzhou, the travel time is prolonged by a factor of 1.5 to simulate the slow-down caused by heavy rainfall. 

According to the timetable in Table \ref{Tab-33train}, there are a total of 10 trains departing from Xi'an and 23 trains passing through. For the purpose of our case, the timetable is divided by a time step at 17:00. In this scenario, we assume that trains arriving before 17:00 can be potentially rescheduled, while those arriving later are considered unavailable due to the disruption. It simulates the impact of the disruption on rolling stock circulation. Therefore, the number of flexible rolling stock is 10, which depart from Xi'an, and the number of restricted rolling stock is also 10.

In the simulated environment, the parameters are set as follows. The time step is set to 5 minutes to simulate the real-time supervision. The agent monitors the in-station conditions from 4:00 to 24:00, resulting in 240 possible actions. In each action, the agent has 21 choices of route and stop patterns to make decisions. Though the total number of choices is 9 $\times$ 3 = 27, some routes have only one or two trains passing through, eliminating the need for three choices. There are 121 downstream stations on the 9 disrupted routes. The total demand throughout the day for the 33 disrupted trains is 1867 passengers. In the gravity model, the attractiveness of each station is determined based on the average GDP of the corresponding city from 2010 to 2019 \citep{GDP_national}. The flexible rolling stock is assigned a capacity of 250 passengers, while the restricted units have a capacity of up to 200. These capacity are discounted based on the gravity model, considering the previous stops of the trains. For the purpose of normalization, the weights of step reward function are set as $\omega_1 = -1, \omega_2 = -1, \omega_3 = 1, \omega_4 = 100, \omega_5 = -0.01, \omega_6 = 1$, and the weights of episode reward function are set as $\beta_1 = 4, \beta_2 = 3, \beta_3 = 5$. The threshold of overcrowding is set as $\Theta_{CR} = 1600$, which is 75\% level of the maximum in-station quantity according to Figure \ref{Fig-instation}. The headway limit is set as $\Bar{H}^r_t = 10$, and the penalty for actions that unmet the headway limit is set as $H^r_t = -500$. The parameters in the epsilon decay function are set as $\varepsilon_0 = 1, \varepsilon_{\min} = 0.1$, and the iteration times is set as 5000. The epsilon-decay rates, $\gamma_{\varepsilon}, \gamma_t$, are different in upper and lower level.

Our HDRL is implemented in Python on ARC4, part of the High Performance Computing facilities at the University of Leeds, UK with Intel Xeon Gold 6138 CPUs (2.0GHz) \citep{HPC_ARC4}. The neural network used in our hierarchical DQN framework is implemented using TensorFlow. In the upper level, the input consists of 5 state variables: disrupted train in-station demand, in-station crowdedness, available rolling stock, potential remaining rolling stock, and the current time. The lower level has 2 action variables. Specifically, the ANN for the upper level comprises an input channel with 5 units in the input layer, followed by 2 fully-connected hidden layers with 200 and 100 units. The output layer is a fully-connected linear layer with 2 outputs. In the lower level, the input consists of 130 state variables, including information on 121 stations and 9 on-route headways. The ANN for the lower level consists of two fully-connected hidden layers with 400 and 200 units. The output layer is a fully-connected linear layer with 21 outputs. For the training process, a learning rate of the both ANNs is set as $10^{-4}$ initially, and decays during iteration by the \texttt{ExponentialDecay} function in Tensorflow where the decay step is 500 episode and the decay rate is 0.1. The discount factor in Bellman equation is $\gamma = 0.9$.

\begin{figure}[htbp]
\centering
\subfigure[Epsilon decay]{
\includegraphics[height=3.5cm]{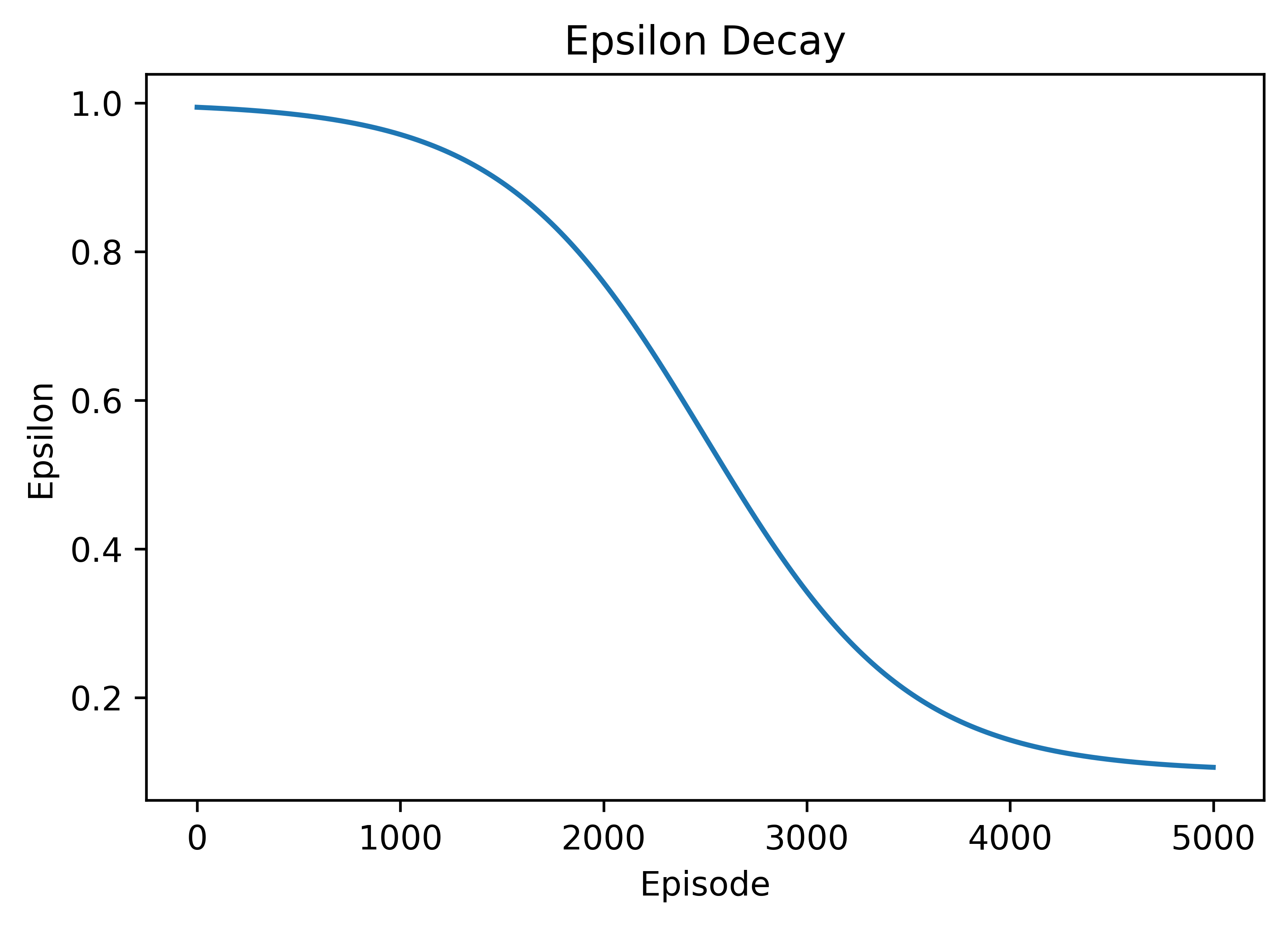}
}
\subfigure[Average reward]{
\includegraphics[height=3.5cm]{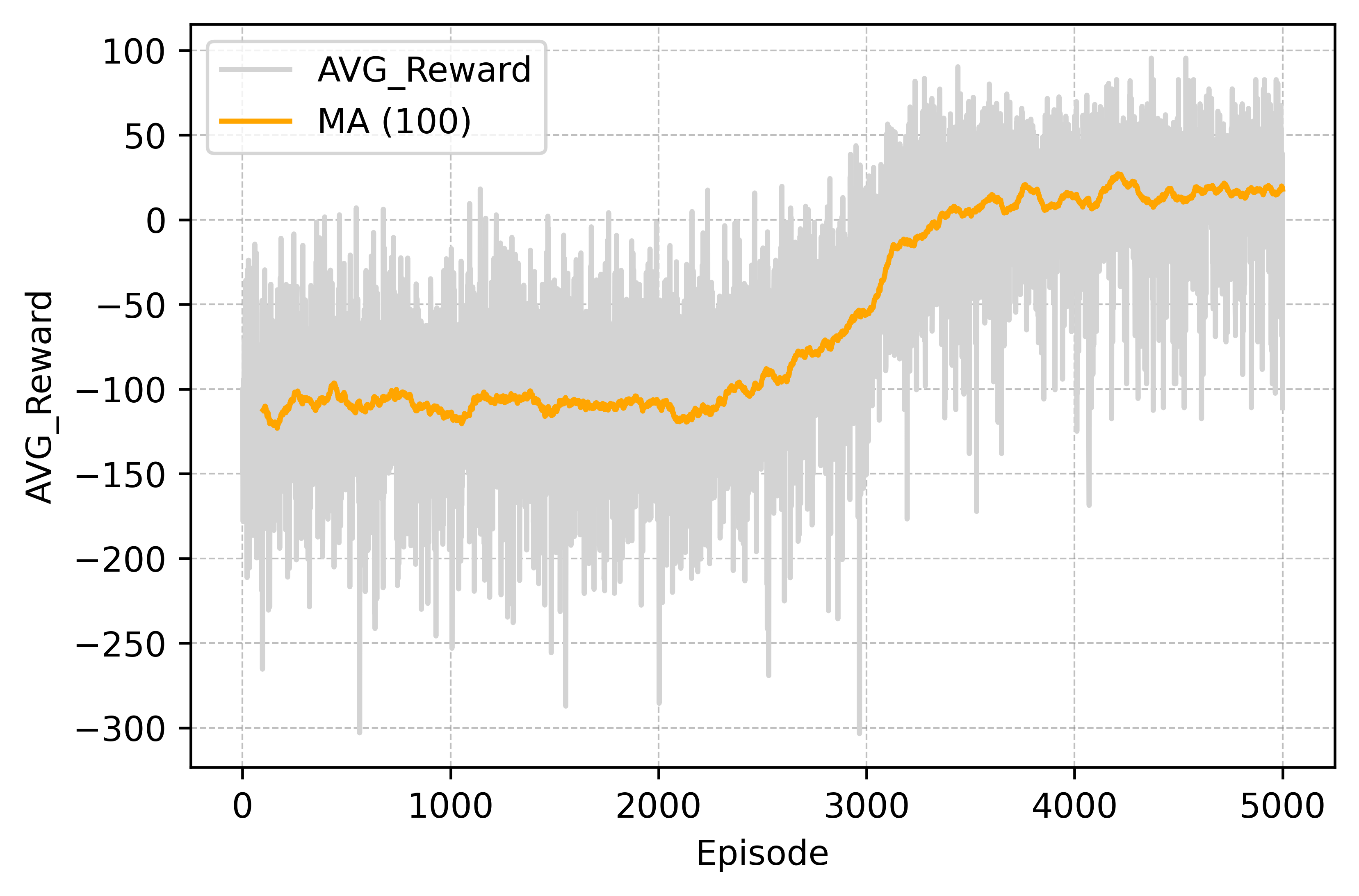}
}
\subfigure[Average demand satisfaction]{
\includegraphics[height=3.5cm]{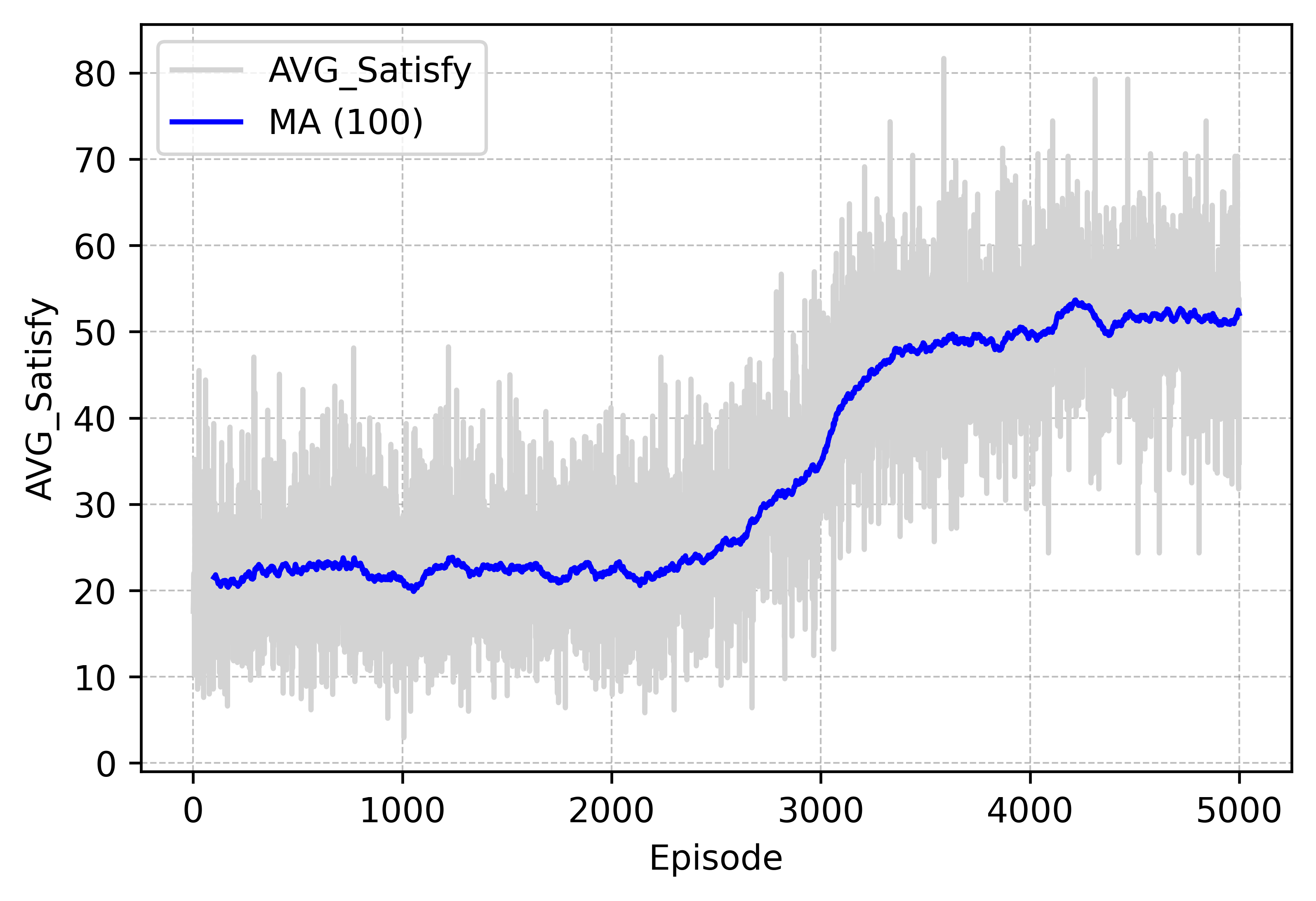}
}
\caption{Convergence of lower-level agent}
\label{Fig-low-con}
\end{figure}

Given the nature of HDRL, the issue of non-stationarity happens due to the accuracy of extrinsic rewards depending on the stationary of lower-level sub-agent \citep{Levy2017LearningMH}. A crucial pretraining phase is necessary for the lower level. The upper level's $\varepsilon$ is initially set to 1, implying entirely random dispatching decisions. Simultaneously, the weights of the upper-level ANN remain unchanged. Only the lower-level $\varepsilon$ decays using the given sigmoid function, and its weights are updated accordingly. To prevent the agent from getting stuck at the initial stages, we extended the time step to 30 minutes. This adjustment ensures that the agent can access information spanning the entire time duration. Because of the number of actions and states is relatively large, the agent needs longer exploring time to get sufficient attempts. So, the epsilon-decay rates are given as $\gamma_{\varepsilon} = 0.005, \gamma_t = 3000$, and the process is given in Figure \ref{Fig-low-con}a. Due to the upper level's randomness, the total value of indicators cannot adequately reflect the lower level's capabilities. Instead, we consider the average reward and the satisfaction for each dispatched train as indicators of the lower level's proficiency. The training and convergence progression of the lower level's ANN are illustrated in Figure \ref{Fig-low-con}b,c. Grey lines are the value of the indicators at every episode, and colored lines are the moving average of 100 points (MA(100)). To analyze the convergence process, the iteration is divided into exploring stage ($Episodes \in [0,2000]$), transition stage ($Episodes \in [2000,4000]$) and exploiting stage ($Episodes \in [4000,5000]$).

At the initial exploration stage, $\varepsilon$ is maintained at a high level, permitting only random actions. Consequently, the reward remains at a low level. During this phase, the agent assimilates the outcomes of various rescheduling plans. However, due to the penalty of low utilization, most attempts get negative rewards. Subsequently, $\varepsilon$ commences descent, marking the transition stage. As $\varepsilon$ diminishes, the agent gradually bases its decisions on learned experiences. Concurrently, the reward begins to rise, signifying the effectiveness of training. At the exploiting stage, the moving average of reward becomes positive, indicating that the lower-level agent can stably make efficient decision at every step.

\begin{figure}[htbp]
\centering
\subfigure[Reward]{
\includegraphics[height=4.5cm]{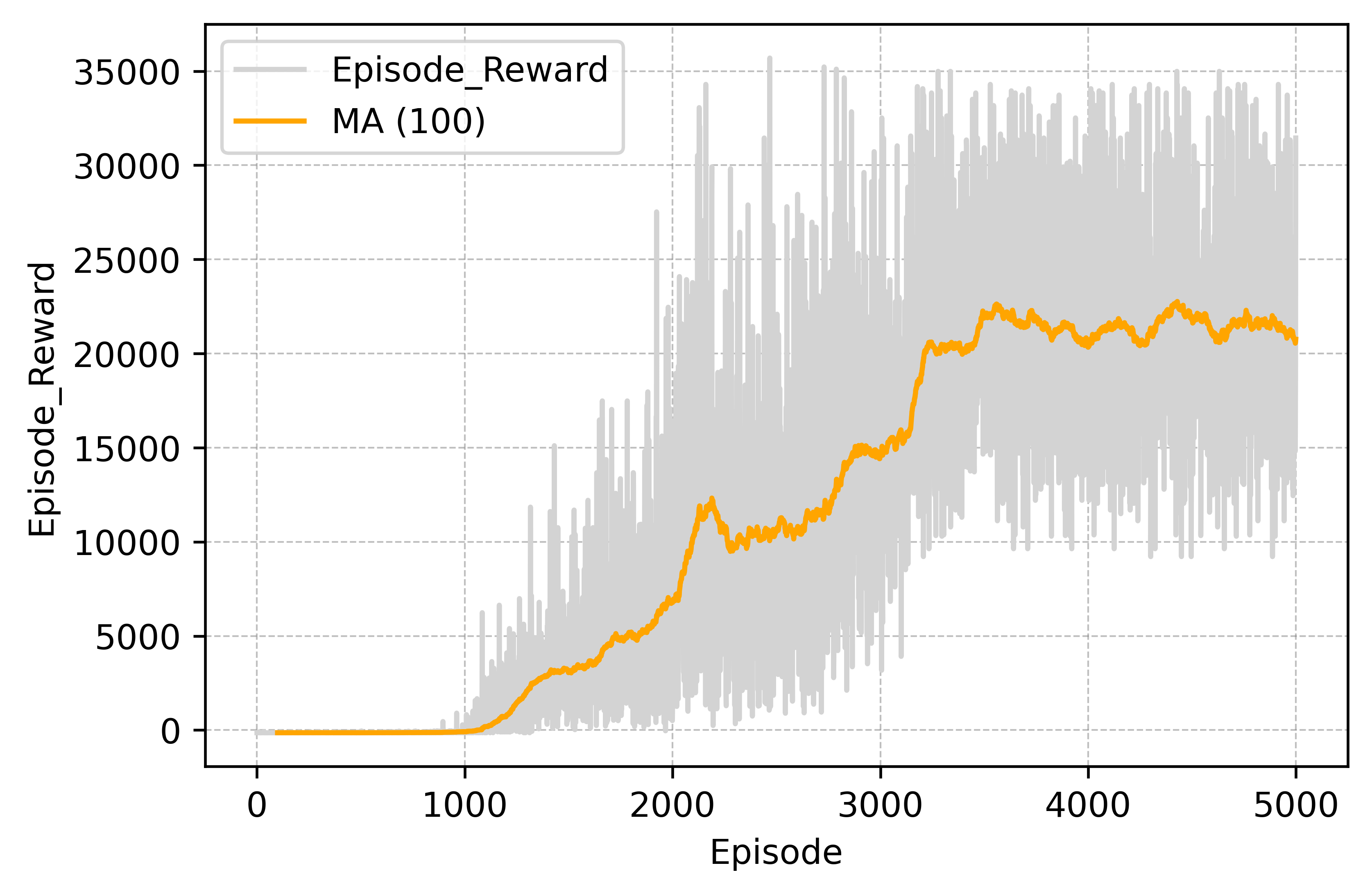}
}
\subfigure[Demand satisfaction]{
\includegraphics[height=4.5cm]{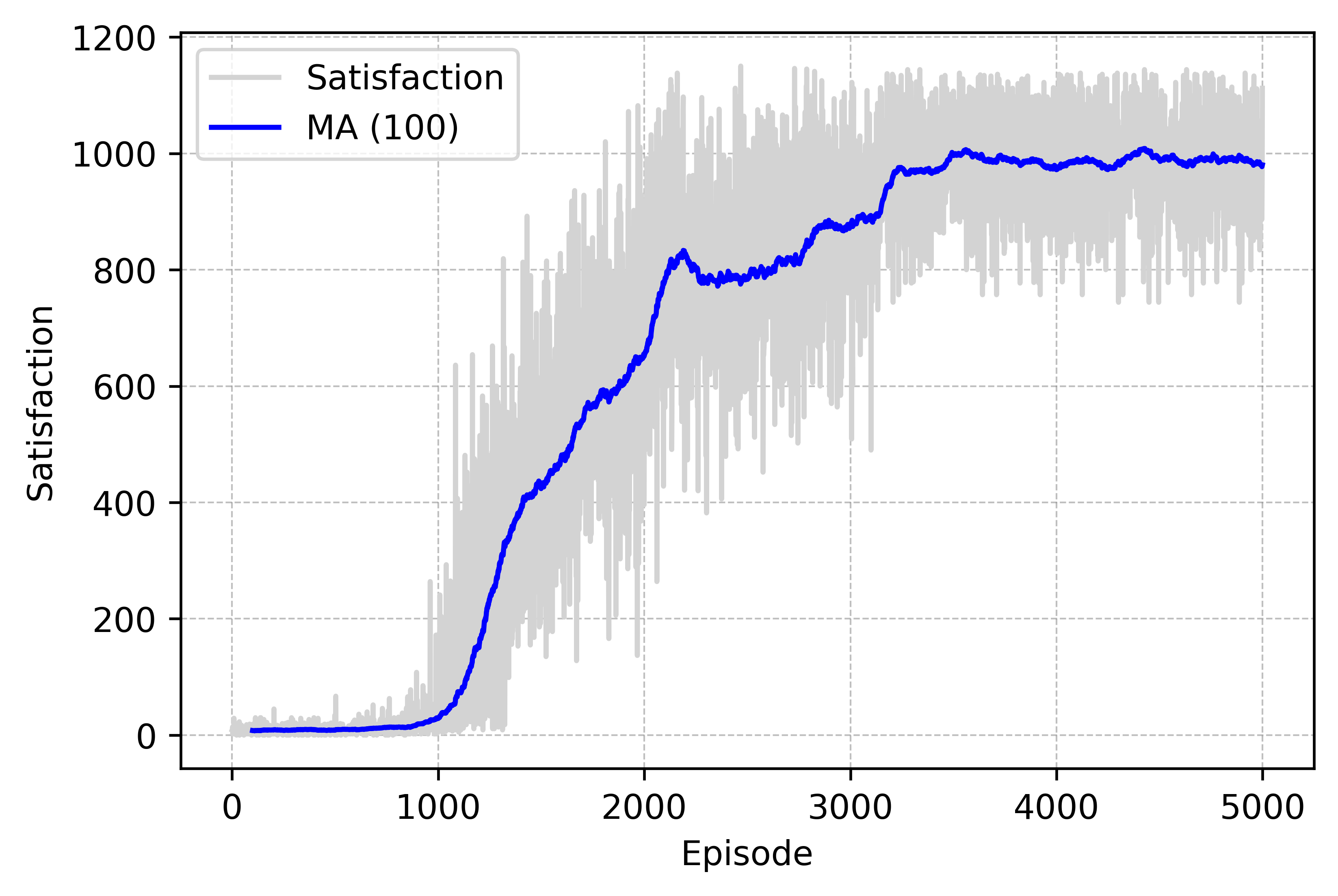}
}
\subfigure[Applied rolling stock]{
\includegraphics[height=4.5cm]{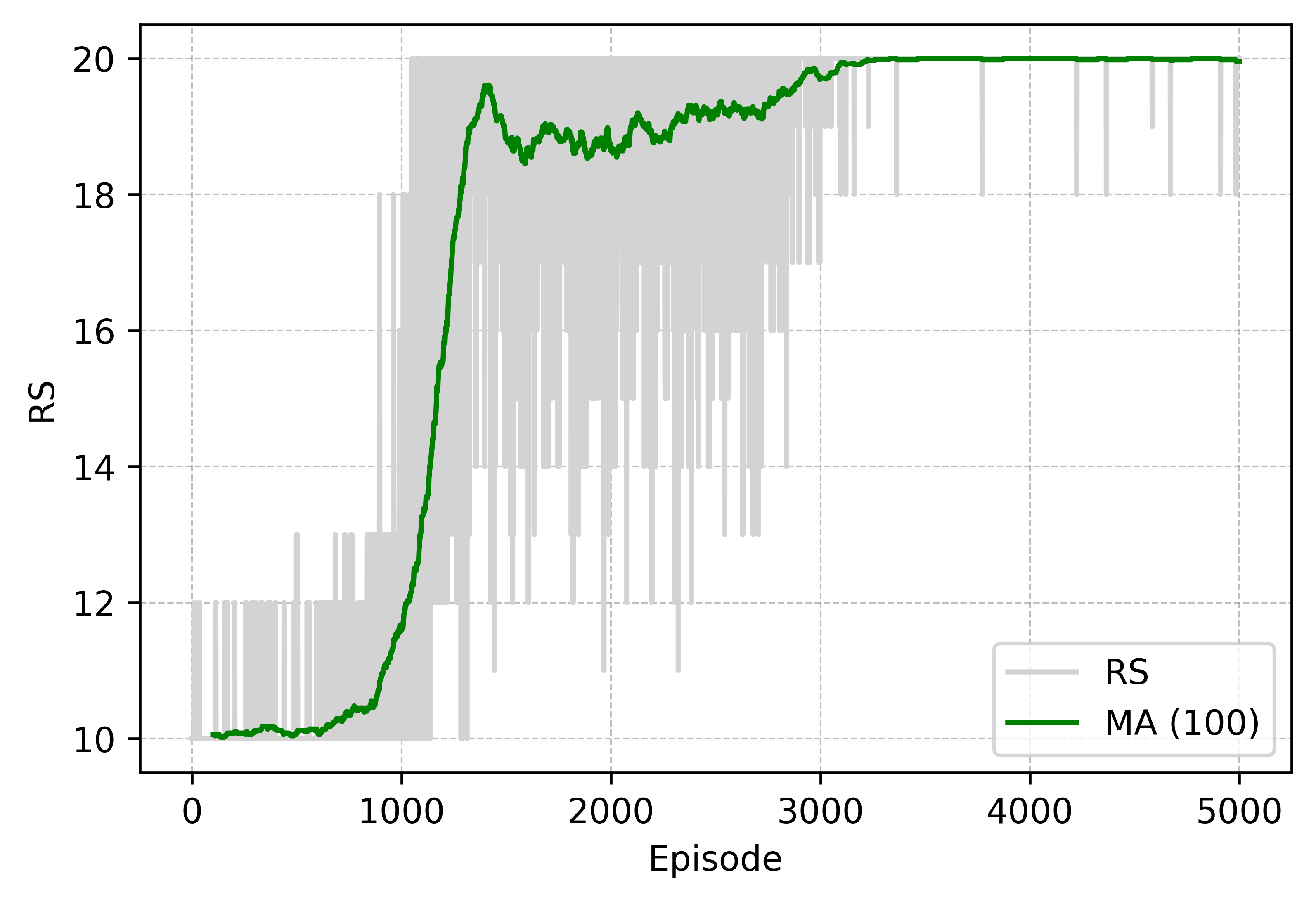}
}
\subfigure[Played time steps]{
\includegraphics[height=4.5cm]{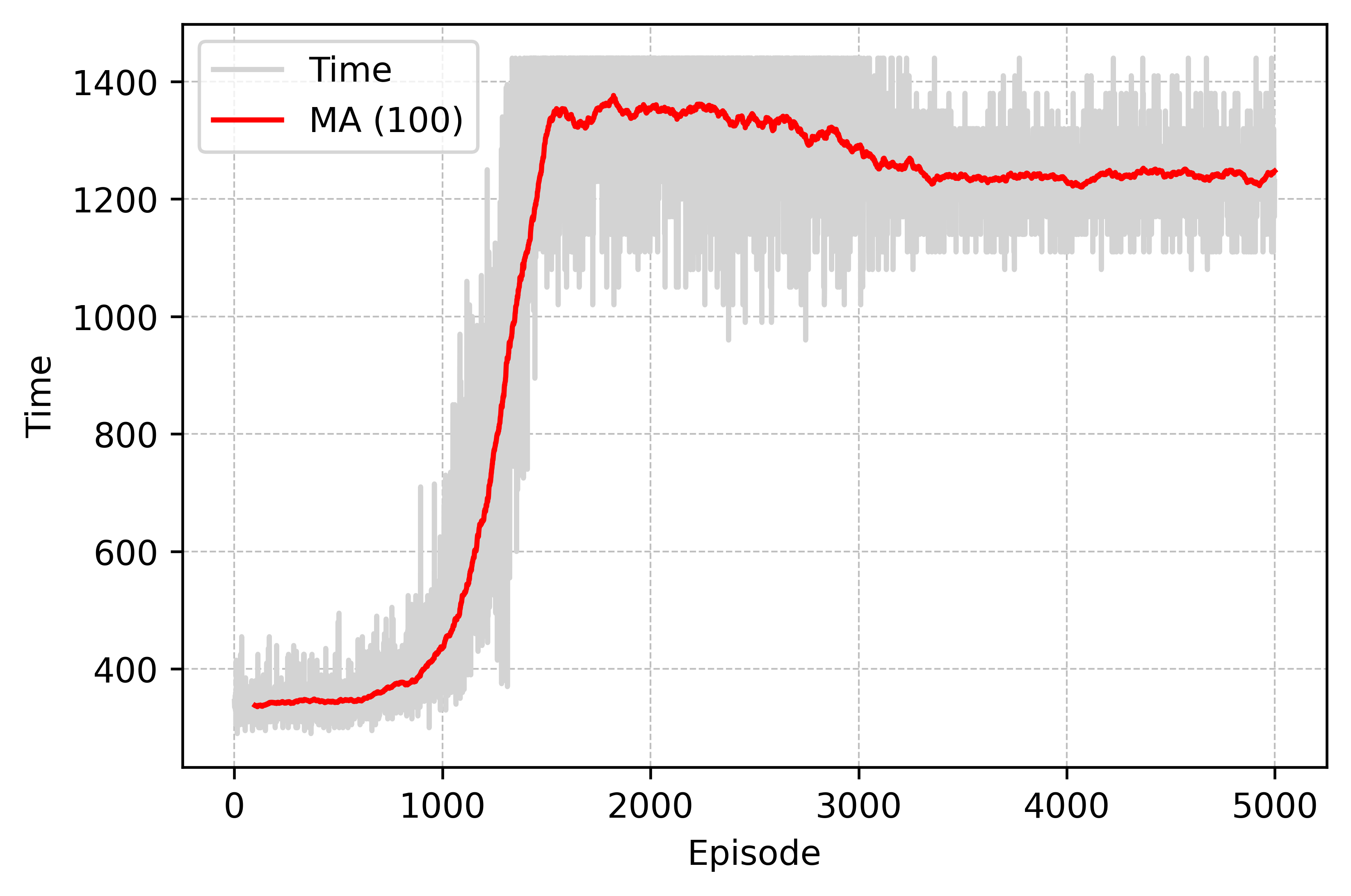}
}
\subfigure[Maximum in-station crowdedness]{
\includegraphics[height=4.5cm]{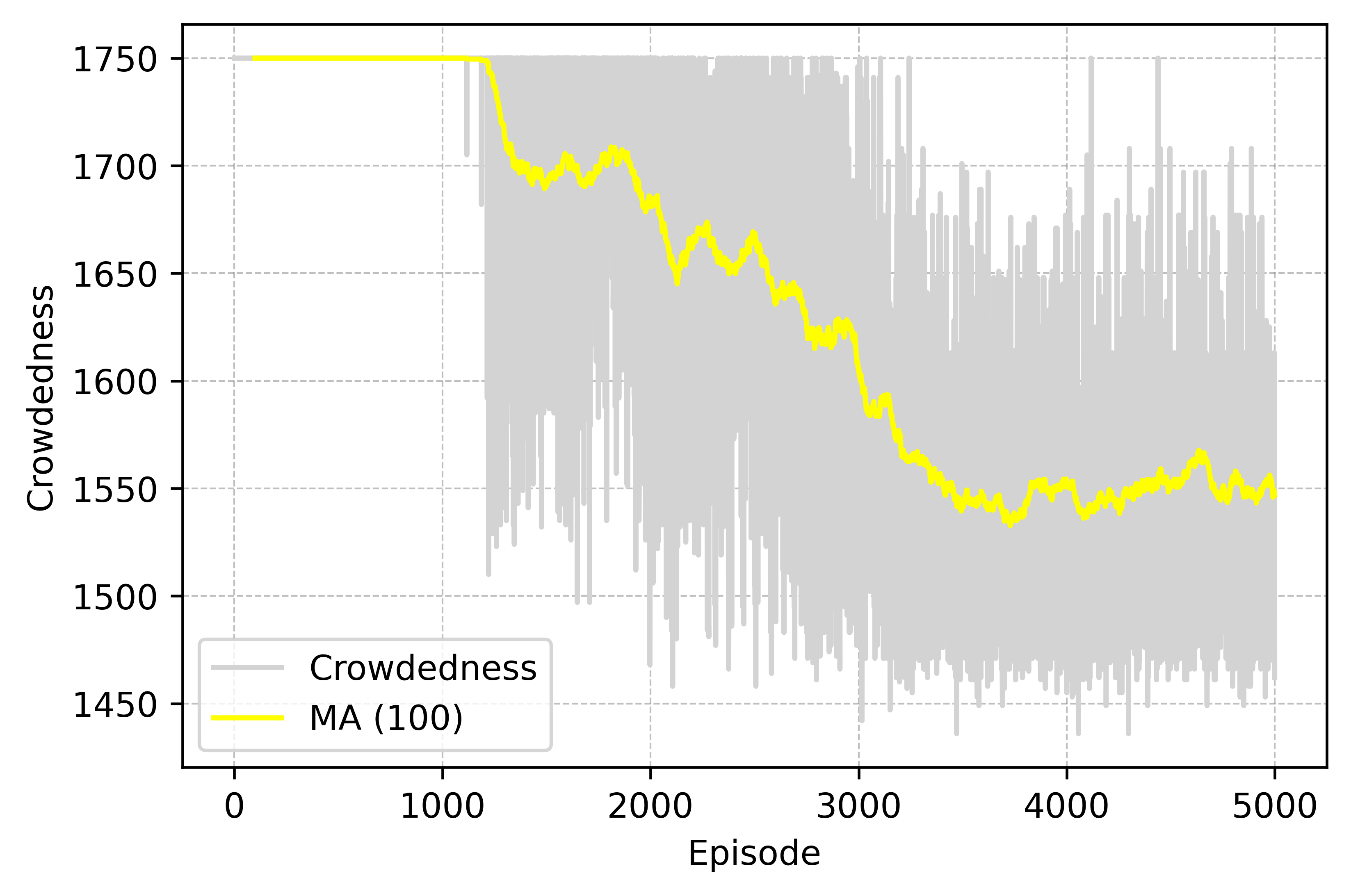}
}
\subfigure[Epsilon decay]{
\includegraphics[height=4.5cm]{Fig-E_decay.png}
}
\caption{Convergence of factors during offline training}
\label{Fig-off-con}
\end{figure}

The upper-level agent is trained after the lower-level agent being stationary. $\varepsilon_2$ is fixed as 0.1 and the learning rate for the lower-level agent is fixed as $10^{-6}$. The epsilon-decay rate for the upper level is set as $\gamma_{\varepsilon} = 0.005, \gamma_t = 2500$. The epsilon decay process is drawn in Figure \ref{Fig-off-con}f. In order to avoid the agent stuck at the beginning stage for a long time,  $\varepsilon_1$ decays slower and smoother. Initially, during the first thousand steps, the agent explored the environment to gather information. Subsequently, in the following thousand steps, the agent gradually reduced its exploration rate and began making decisions based on exploitation. From the 3000th step on, the exploration rate stabilized at around 0.2, indicating that the agent primarily relied on its accumulated experience to make decisions. The training process resulted in the convergence of  the episode reward, total satisfied demand, applied rolling stock, played time steps, and maximum in-station crowdedness,  as illustrated in \ref{Fig-off-crowd}a-e respectively. In Figure \ref{Fig-off-con},  grey lines represent the values of each episode, while the colored lines represent the moving average with a window size of 100. 

During the initial exploring stage, both the reward and satisfied demand remained at lower levels. As seen in Figure \ref{Fig-off-con}c and d, the agent frequently exhausted the available rolling stock in the early steps of the episode. In most cases, the agent only used the flexible rolling stock to operate within the first 350 time steps, resulting in failure to handle subsequent arriving trains. This inadequate action plan led to potential overcrowding during the evening peak hours. Then, as epsilon decay in the transition stage, the agent interacted with the environment and received feedback. Over time, the agent learned the importance of preserving trains and made better use of the rolling stock. Consequently, the agent gradually became capable of effectively managing the entire episode, strategically conserving trains until the end of the day ($Time = 1440$). This improvement was reflected in the increased episode reward and satisfied demand, as well as the decreased maximum in-station crowdedness. In summary, during the exploring phase, the agent lacked an understanding of the long-term benefits and urgency associated with dispatching trains. As a result, it frequently depleted the available rolling stock before the episode ended, or even before the evening peak hours, demonstrating a short-sighted strategy. This behavior had the potential to cause overcrowding during later hours and was rendered helpless due to the wasted resources. We consider the end of the \textit{first half} of the exploring stage to be reached when the agent learned to save rolling stock until the episode concluded  ($Episodes \in [0,1300]$) . During the first half, the agent was exploring an unfamiliar and dynamic environment until it gained a comprehensive understanding of the entire time-variant conditions. The agent's strategy is transitioned from aggressive to conservative in the first half. In certain aspects, we can acknowledge that the agent demonstrates capability in finding feasible solutions during this stage. 

The conservative strategy was advantageous as it preserved resources to address unexpected future conditions. However, it came at the expense of satisfying less demand and leaving some resource at the end of episode. Subsequently, the training entered the \textit{second half} ($Episodes \in [1300,2000]$). As observed in Figure~\ref{Fig-off-con}c and d,  the agent adopts a conservative strategy at the beginning of the second half. It frequently leaves a portion of the available rolling stock unused, indicating that there are trains remaining until the end of the episode. The episode often ends in the final time steps, indicating the agent tends to preserve trains for long-term benefits. Throughout transition stage ($Episodes \in [2000,3000]$), the agent progressively recognizes the advantages of utilizing rolling stock to meet less obvious demands under appropriate conditions, such as overcrowding and higher in-station demand. Additionally, the episode reward function  Eq.~\eqref{Eq.reward-epi} stimulates the agent to strive for meeting more demand and optimizing the usage of rolling stock. At the exploitation stage ($Episodes \in [3000,5000]$), the agent becomes more courageous and astute in dispatching rolling stock. The number of applied rolling stock increases and stabilizes around the total backup number. Moreover, the episodes often conclude before the end of the day, as there are typically few passengers arriving at midnight. In these cases, the agent dispatches the last train after the evening peak hour and concludes the episode.

Throughout the exploration and exploitation, the in-station crowdedness consistently decreases, and the moving average approaches the safety limit and remains below the overcrowding threshold. Regardless of the aggressiveness of the strategy, the agent learns to recognize peak hours and dispatches trains to alleviate overcrowding to some extent. Furthermore, after the exploration stage, the total satisfied demand stabilizes, as indicated by the moving average, with an average satisfaction of approximately 1000 (54\%). The optimal solution we have found in this process is capable of meeting 1150 passenger demand (61.6\%). It is worth highlighting that the available rolling stock is dynamic and, at most, comprises 60.6\% of the original planning.

\subsection{Comparison with GA}
In order to showcase the efficiency of the optimal solution, DRL approaches are often compared with heuristic algorithms \citep{LI2022230,ying_actor-critic_2020}. In this study, a genetic algorithm (GA) is designed for the same environment with a discrete time step. The planning horizon is divided into time steps where each time step requires a decision. The decisions consist of an action set containing the number of rescheduling plans and a 0 representing doing nothing. The length of the chromosome is 240, with 20 rescheduling plans that can be chosen; the rest are all set to 0. The sum of step rewards and episode reward is defined as the fitness in GA. The psuedocode of our GA is given in Appendix \ref{App-4}.

It should be noted that the GA employed for comparison purposes is not suitable for real-time applications, as it can only take the complete information on demand, rolling stock and crowdedness once for all as its input. Therefore, the comparison is intended solely to provide a benchmark, showcasing the quality of solutions generated by the HDRL. Unlike the GA, our HDRL approach offers real-time applicability based on dynamically updated information.

The simulation of the environment operates in the same way as in DRL. The fitness is calculated for the given rescheduling plan within every chromosome. Subsequently, crossover and mutation are processed to generate a new population iteratively. The parameters are set as follows: the population size is 50, the number of iterations is 100 to match the training times of the HDRL. The crossover rate is 0.2, and the mutation rate is 0.05. A check is performed for every new population that all the chromosomes with more than 20 non-zero decisions are randomly chosen and reduced to 20, to make sure the number of rescheduling plans does not exceed 20, and the rescheduling plan should be feasible for the rolling stock coming from the upstream.

\begin{figure}[htbp]
\centering
\subfigure[Demand satisfaction]{
\includegraphics[height=3.5cm]{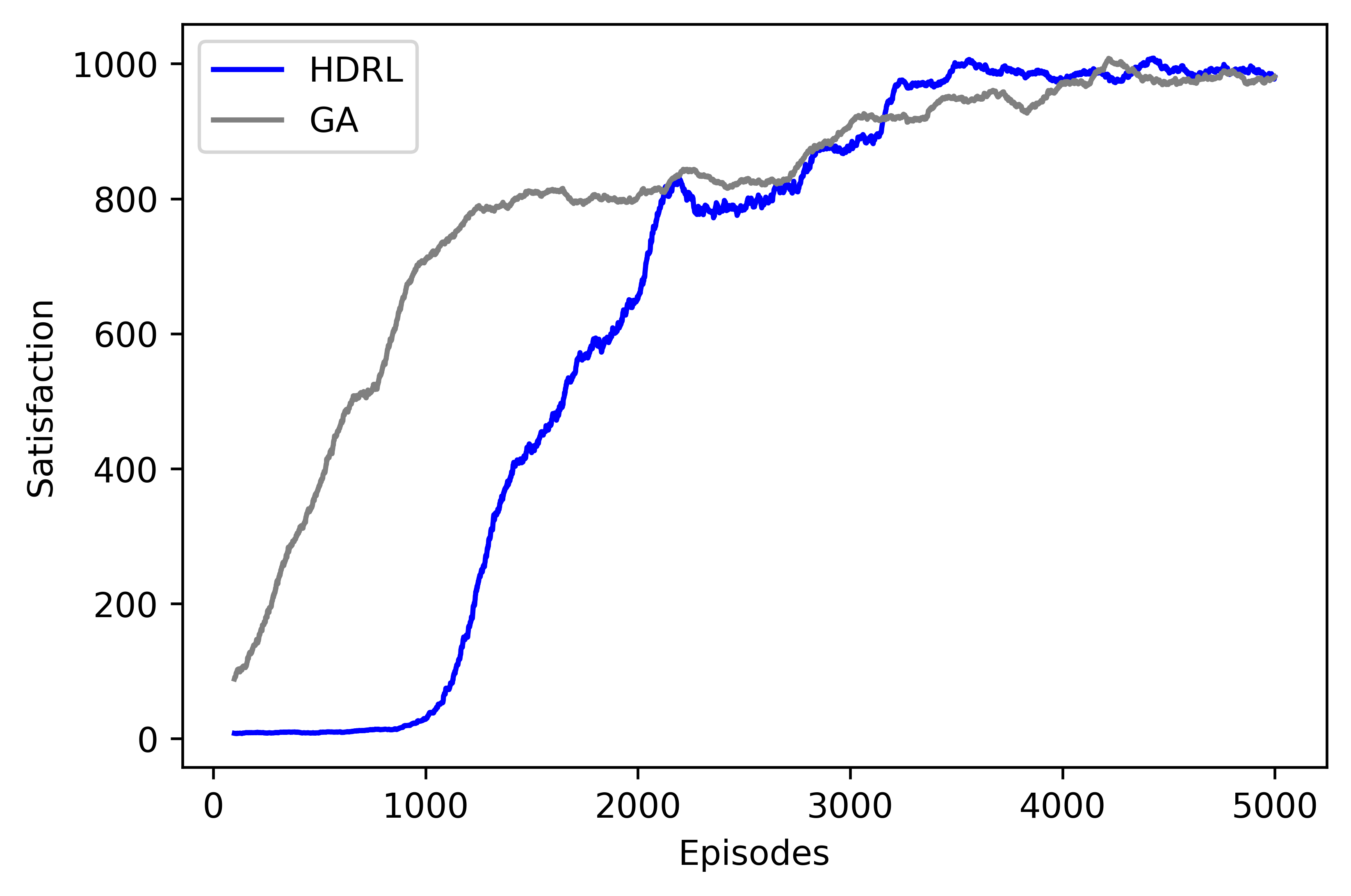}
}
\subfigure[Applied rolling stock]{
\includegraphics[height=3.5cm]{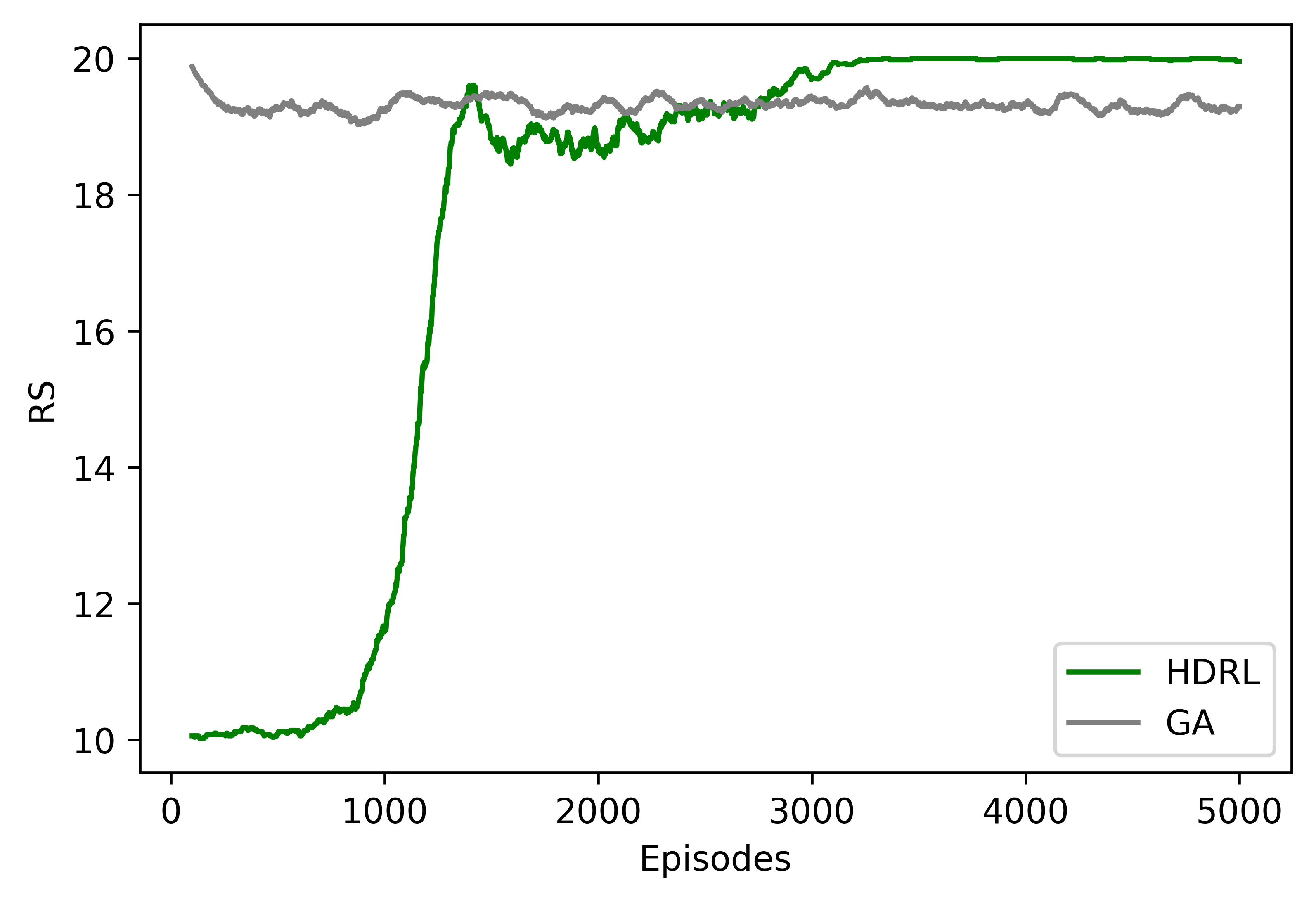}
}
\subfigure[Episode reward]{
\includegraphics[height=3.5cm]{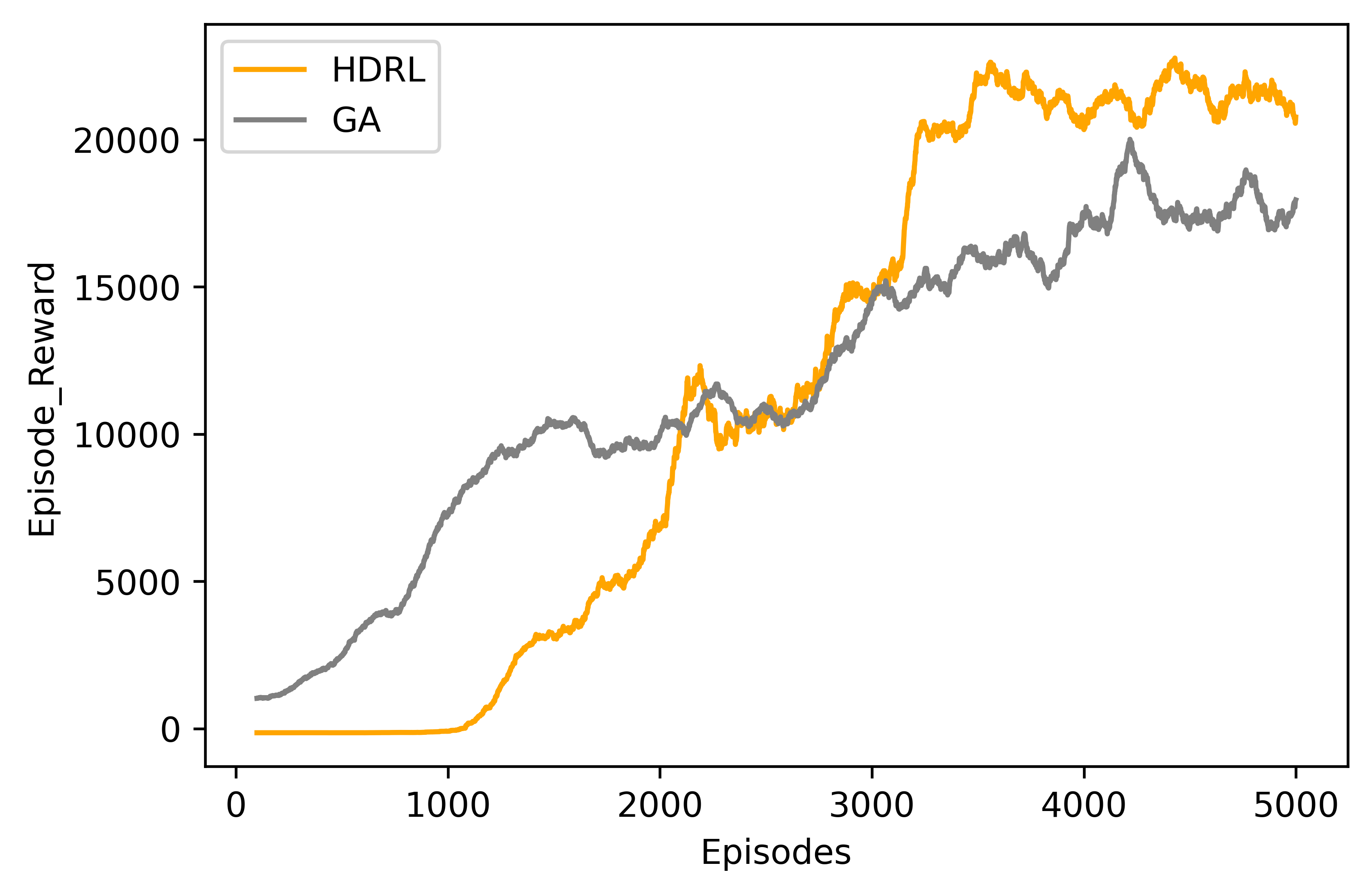}
}
\caption{Comparison of the convergence between HDRL and GA}
\label{Fig-bench-con}
\end{figure}

The iteration process of GA is presented in Figure \ref{Fig-bench-con}. Three important indicators are shown:  the demand satisfaction, the number of used rolling stock and the episode reward. The moving average is depicted in the figure, and GA results are all in grey. The ending time is not included because GA needs to traverse the time steps for all chromosomes. Crowdedness is not included either because overcrowding only occurs a few times, likely due to enough trains being uniformly assigned throughout the time span. Based on demand satisfaction, GA learns faster than HDRL. This could be because GA is assumed to have knowledge of the complete information for the entire time span, which is not applicable in real-time conditions where future information is unknown. The number of satisfied demands is similar for GA and HDRL. Regarding the number of used rolling stock, GA initially uses 20, but with crossover and mutation, this number is reduced. As a result, GA cannot guarantee optimal resource utilization every time.

Consequently, based on the evaluation criteria in this paper, HDRL can achieve a higher episode reward than GA after the same number of training iterations. This indicates the satisfactory quality of our proposed HDRL and that it has better comprehensive abilities in preventing overcrowding, satisfying more demand, and optimizing the use of rolling stock. 

\subsection{Adaptive rescheduling strategies learned by the agent}

The optimal solution yielded a detailed rescheduling plan that highlights the actions and corresponding outcomes achieved by the agent. 
The rescheduled timetable and demand satisfaction during the disrupted period, as well as the in-station crowdedness, are worthy to discuss. In this period, the scarcity of rolling stock presents a dilemma in timetabling and capacity allocation, specifically, determining the dispatch frequency and route selection. The agent faces decisions such as which trains to cancel, how to allocate rolling stock across multiple routes, how many spare trains to assign to each route, and when to dispatch a train on a specific route. These decisions are made by the agent considering the OD demand and headway of each route, ultimately aiming to maximize the accumulated rewards. These aspects warrant a detailed analysis and examination in understanding the quality of rescheduling during the disrupted period.

\begin{figure*}[htbp]
  \centering
  \includegraphics[width=\linewidth]{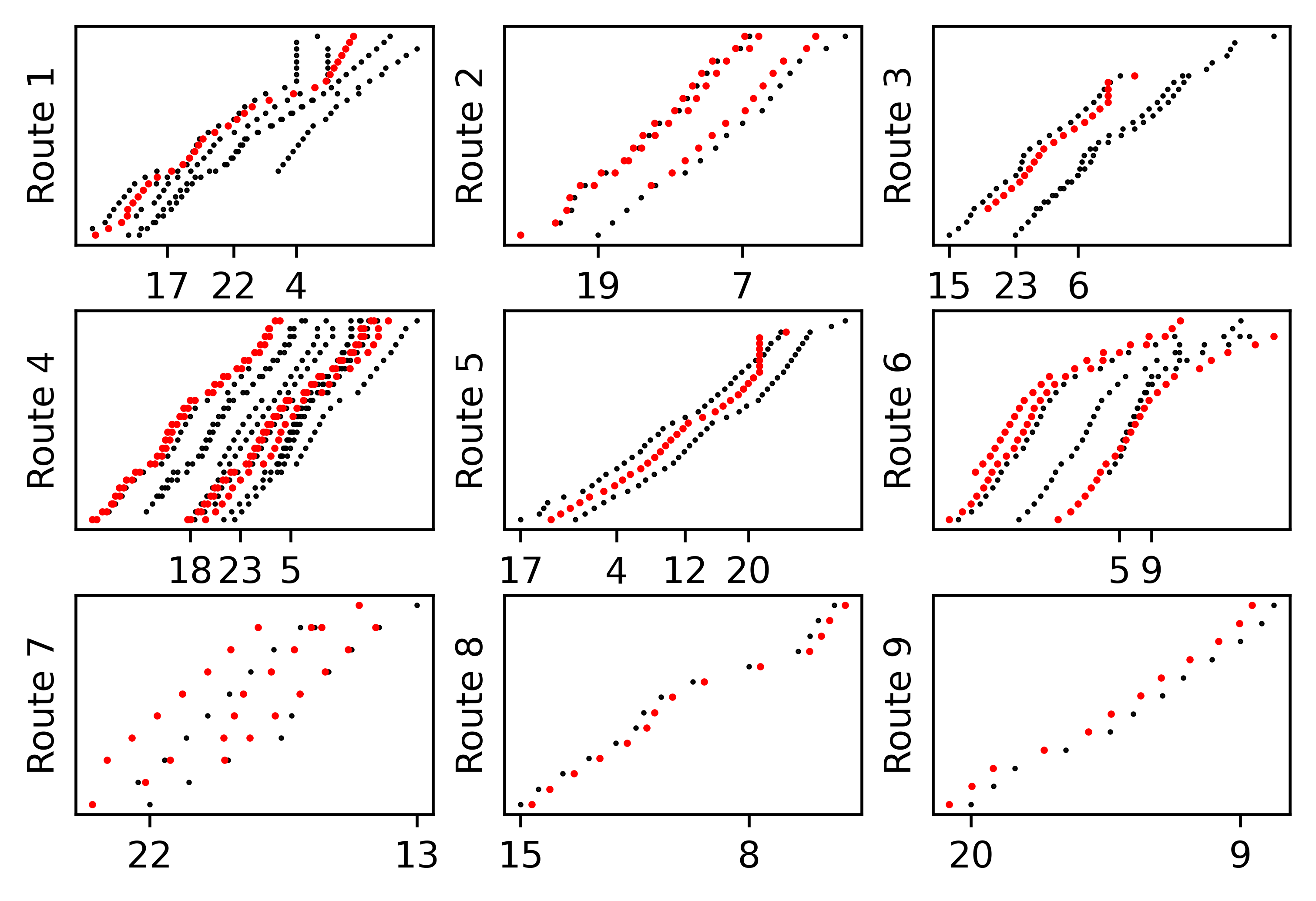}
  \caption{Comparison between original and rescheduled timetable (Horizontal axis represents arrival time;  vertical axis represents downstream stations.)}
  \label{Fig-off-delay}
\end{figure*}

Figure \ref{Fig-off-delay} provides a comprehensive comparison between the rescheduled timetable and the original timetable for the 9 routes to illustrate the agent's timetabling strategy. Each subfigure depicts an operating diagram on a specific route. The horizontal axis represents the arrival time, while the vertical axis represents the downstream stations. The original arrival times of trains are indicated by black dots, while the red dots represent the arrival times of trains dispatched by the agent. 

On high-frequency routes like route 1, 4, and 6, which experience high demand, it is crucial to allocate more attention and resources. Interestingly, there are slight differences in the timetable distribution between route 1 and routes 4 and 6. Route 1 exhibits a dense timetable during the noon period, while routes 4 and 6 have a more evenly distributed timetable throughout the day. The agent has learned distinct strategies to address these variations.

For route 1, the agent strategically dispatches one train in the middle of the original timetable. This rescheduled train serves as a consolidation point for passengers from different trains with varying departure times. By redistributing the demand across this rescheduled train, the agent effectively recomposes the passenger distribution.
On routes 4 and 6, the agent dispatches two batches of trains during the day, specifically during the morning and evening peaks. This rescheduling strategy enables passengers who were originally targeting different trains during peak hours to be accommodated on the rescheduled trains. This approach facilitates an enhanced demand-responsive timetabling by employing passenger reassignment and demand recomposition techniques.
A similar rescheduling strategy is also observed on lower-traffic routes such as route 3 and 5. These routes have fewer trains during the day, and their departure times are in close proximity. To recompose the demand on different trains, the agent dispatches one train in the middle of the departure times of these trains, providing a consolidated option for passengers traveling during similar time frames.

In contrast, on lower-traffic but high-demand routes like route 2 and 7, the agent focuses on maintaining adherence to the original timetable. Meanwhile, it also considers the potential en-route delays resulting from rerouting and rainfall. To mitigate any potential disruptions, the agent prioritizes dispatching the trains to depart slightly earlier than its original scheduled time. Finally, on routes with only one train scheduled during the day, such as route 8 and 9, the agent accurately identifies the departure time of these trains and dispatches a train to ensure their timely fulfillment.

\begin{figure*}[htbp]
  \centering
  \includegraphics[width=0.95\textwidth]{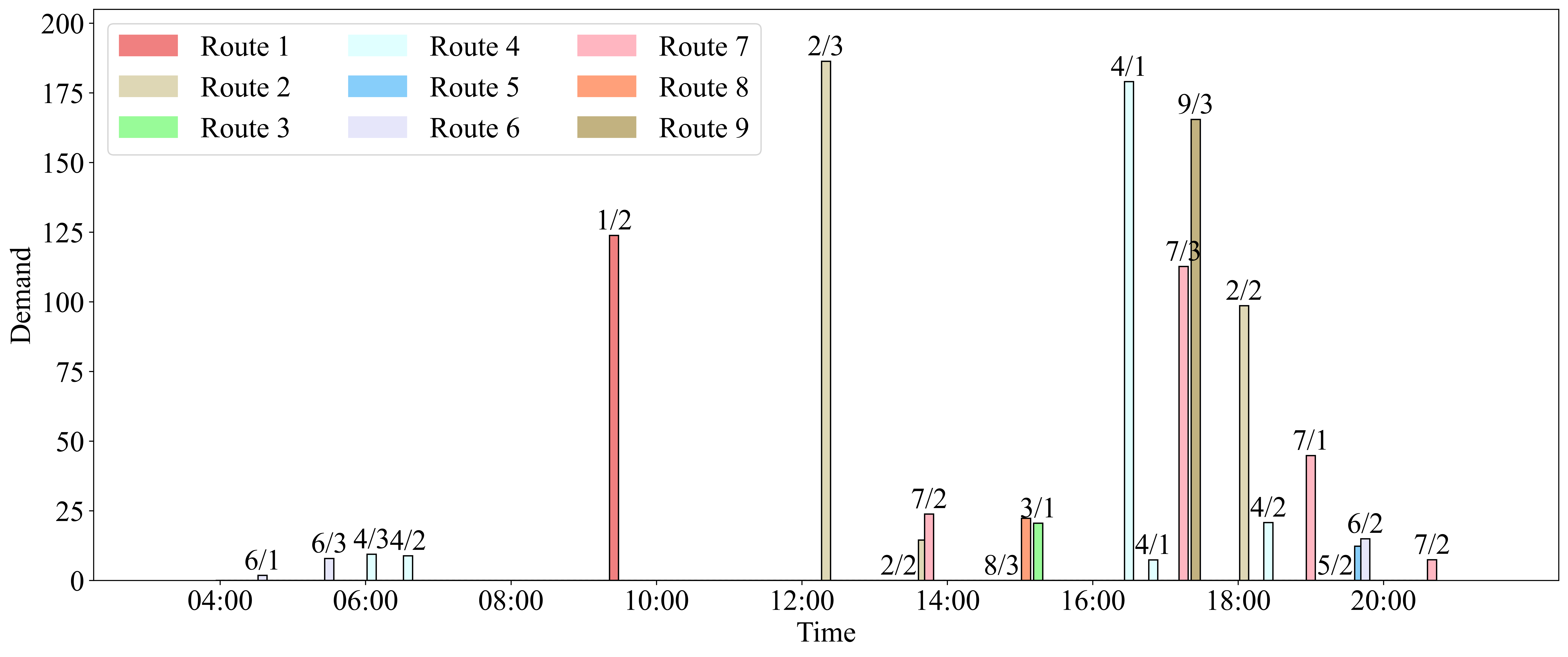}
  \caption{Satisfied demand by the rescheduled trains (Bars  are labeled with ``route ($r$) / stop pattern ($p_1,p_2,p_3$)''  )}
  \label{Fig-off-satisfy}
\end{figure*}

The timetabling diagram presented in Figure \ref{Fig-off-delay} provides a visual representation of the agent's effective utilization of demand recomposition and delay management strategies during the rescheduling process. The demand satisfaction diagram depicted in Figure \ref{Fig-off-satisfy} offers a comprehensive depiction of the agent's rescheduling strategy, specifically focusing on its effectiveness in meeting passenger demand on individual trains throughout the day. This diagram provides valuable insights into the agent's demand-responsive approach and its impact on rescheduling outcomes. The color coding represents the routes, corresponding to Figure \ref{Fig-map}. The bars in the diagram are labeled with the corresponding ``route ($r$) / stop pattern ($p_1,p_2,p_3$)'' information, where, for example, ``2/3'' indicates that a train operates on route 2 and follows the stop pattern denoted by $p_3$.

For routes with concentrated timetables, such as route 1, 3, and 5, the agent employs a demand recomposition and passenger reassignment strategy. On route 1, with higher demand, a train is dispatched to accommodate the substantial demand, strategically operating in stop pattern 2 to balance station coverage and travel speed and abandon stations with fewer demand. On the other hand, routes 3 and 5, with relatively lower demand, exhibit passenger reassignment, highlighting the efficacy of demand-responsive strategies in optimizing capacity allocation on low-demand routes.

Furthermore, the figure shows the strategy that the agent's handling of high-demand routes, specifically route 2 and 7. Their original departure time and frequencies are retained to accommodate the passenger demand. Notably, despite the increased departure frequency, the dispatched trains still onload a higher level of passenger demand. This finding suggests that the demand on these routes is substantial, making it advantageous to avoid train cancellations. Additionally, the agent demonstrates its ability to recognize the importance of different stop patterns during rescheduling. It strategically dispatches a train operating with the more comprehensive stop pattern, such as $p_3$, during peak hours to accommodate the high demand across all stops. Subsequently, during hours with lower demand, the agent dispatches faster trains operating with less comprehensive stop patterns, such as $p_1$ or $p_2$, to effectively meet the demand for specific OD pairs. This adaptive approach highlights the agent's understanding of the varying demand patterns throughout the day and its ability to deploy trains with different stop patterns based on the demand intensity.

Moreover, for routes 4 and 6 where trains originally operate during both morning and evening peak hours, the agent uses a combination rescheduling strategy. During the morning peak, the agent dispatches a combination of fast trains and full-stop trains to efficiently cater to the demand during this period, avoiding passenger dissatisfaction and prolonged waiting times. In the evening, the agent modifies its strategy to strike a balance between meeting passenger demand, considering their waiting tolerance, and delay reduction. This adaptable approach ensures an optimal trade-off between passenger satisfaction and operational efficiency during the evening hours.

During the daytime, the agent adopts a reactive approach by dispatching trains primarily on busy routes in the morning. This strategy ensures that an adequate number of rolling stock is available for rescheduling during peak hours and unexpected increases in demand. In contrast, during the evening, the agent switches to a proactive approach by increasing the frequency of dispatches and even allocating trains to lower-demand routes. This proactive strategy aims to optimize the utilization of backup rolling stock, meet passenger demand effectively, and minimize capacity wastage.

\begin{figure*}[htbp]
  \centering
  \includegraphics[width=0.6\linewidth]{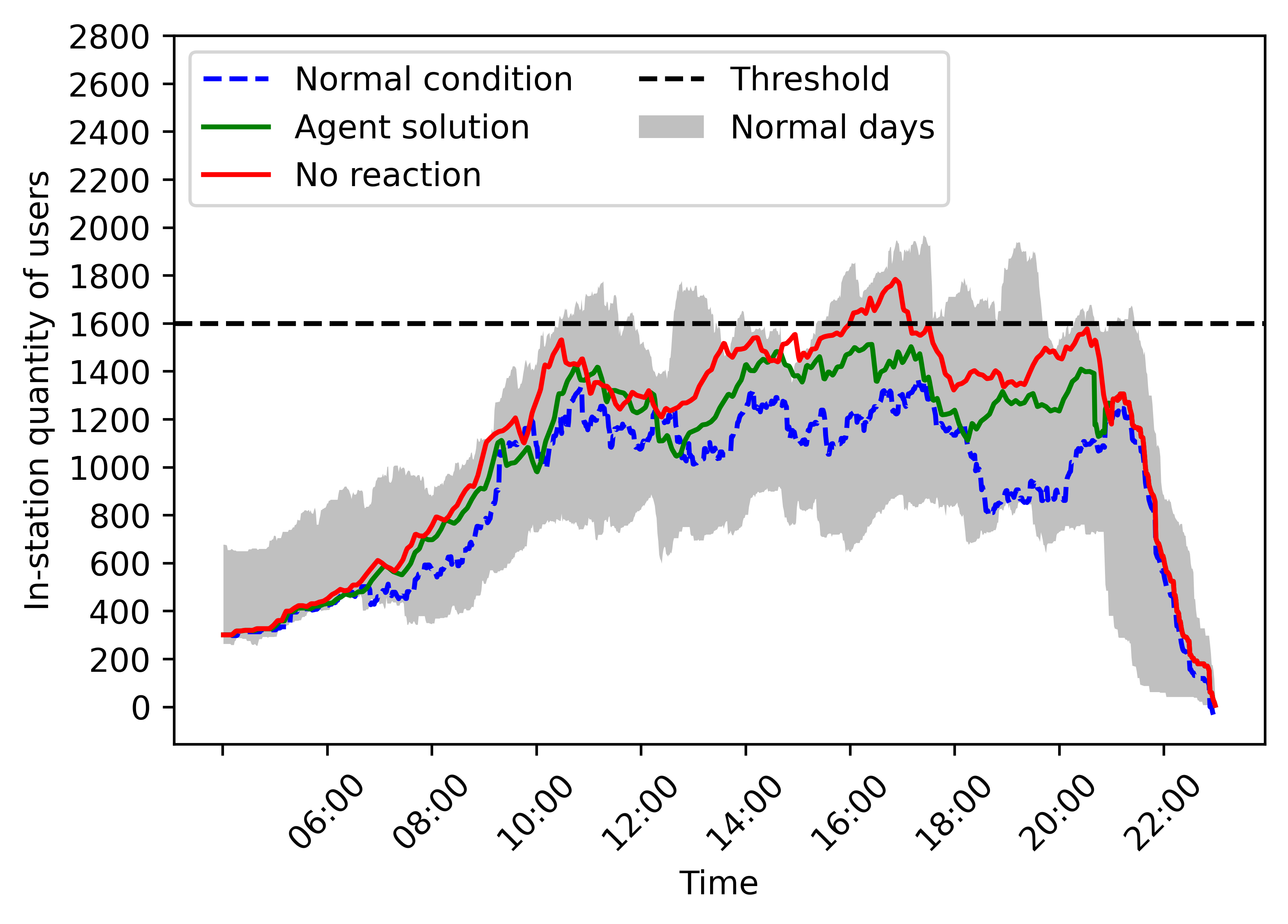}
  \caption{In-station crowdedness during offline training case}
  \label{Fig-off-crowd}
\end{figure*}

Figure \ref{Fig-off-crowd} presents the in-station crowdedness during the disruption period, with the aid of the rescheduling. The grey shaded area represents the real-time in-station crowdedness during normal days (July 1-19th), where the quantity value corresponds to the number of passengers who ultimately depart by trains. This quantity reflects the passenger quantity within the station hall. The threshold for overcrowding is represented by the black dashed line. Additionally, the figure includes real-world conditions observed on July 19th, denoted by the blue dashed line. The railway operation is in normal condition, so, most passengers depart on time. The in-station passenger quantity on this day falls within an intermediate range compared to the normal days.
The disruption condition is simulated on July 19th, passengers are deemed to return when the waiting time surpasses their tolerance. The red line illustrates the in-station passenger number if there is not any reactions taken. In another word, trains on the interrupted routes are cancelled. Passengers can only stay inside the station and return. The passenger quantity keeps in a high level and breaks the overcrowding threshold at the evening peak hour. The green line illustrates the optimal solution generated by the agent. Notably, despite the lack of rolling stock, the agent successfully avoids overcrowding in the station's waiting hall throughout the entire disruption period. However, the overall crowdedness remains higher than the normal condition. This demonstrates the agent's ability to accurately identify the distribution of passenger demand and anticipate peak hours, allowing for proactive rescheduling of trains to prevent overcrowding.

\subsection{Transferability}

The training process has reached convergence, when the epsilon value stays at a lower level, indicating its capability to make real-time decisions through the ANN. To assess the agent's online performance, a new environment was created using the data from the first day of the real-world flood event, namely July 20. The new environment features an unnoticed demand distribution and passengers' waiting tolerance based on real-world records. The agent's objective is to mitigate the consequences of the Zhengzhou flood at the Xi'an station by using at most 60\% capacity. The agent was applied repeatedly in this new environment for 1000 simulations, without any learning or updating of the ANN weights, to simulate an online testing scenario. The goal was to statistically analyze the agent's performance in this novel environment, which presented a total demand of 2787, 1.5 times higher than that of the training environment. All other settings remain the same as during offline training, including a fixed epsilon value of 0.1 for both upper- and lower-level agent.

\begin{figure}[htbp]
\centering
\subfigure[Reward]{
\includegraphics[width=0.18\textwidth]{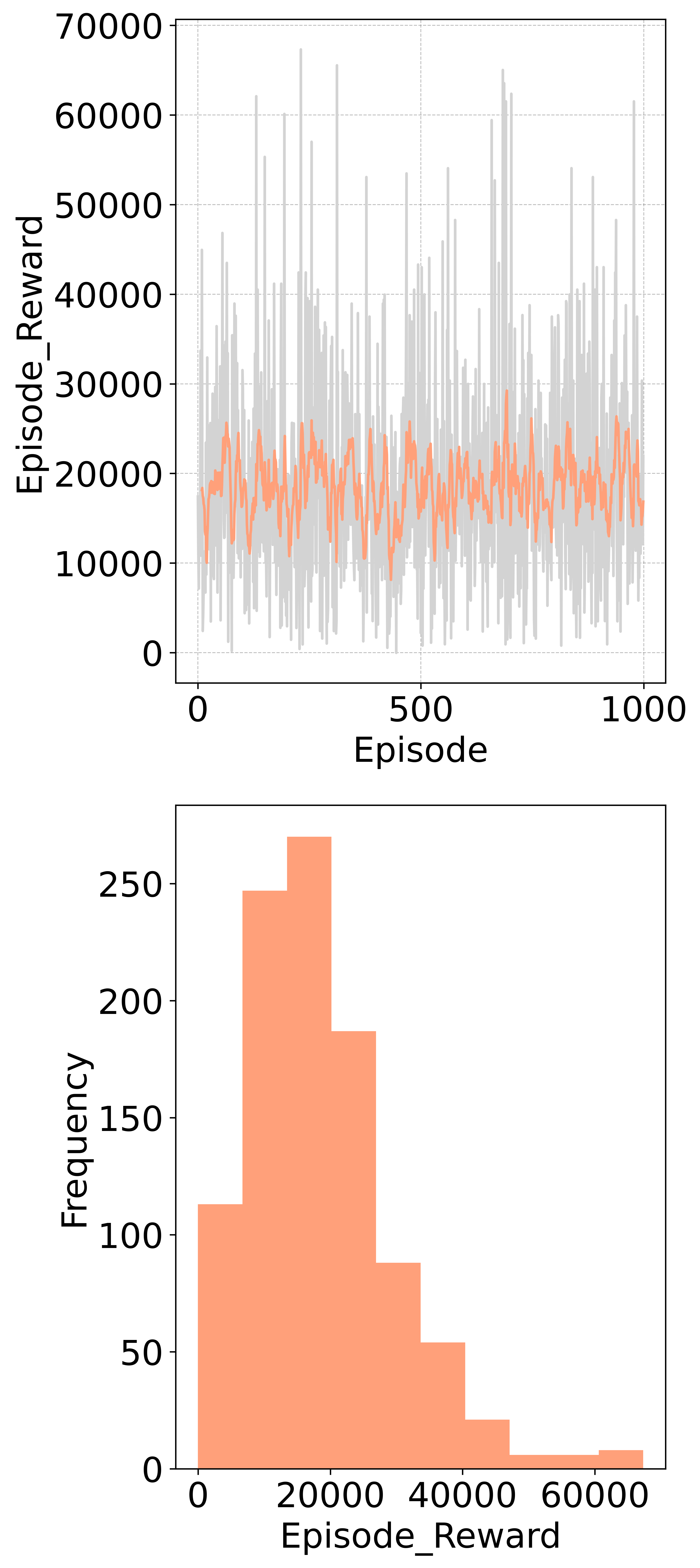}
}
\subfigure[Demand]{
\includegraphics[width=0.18\textwidth]{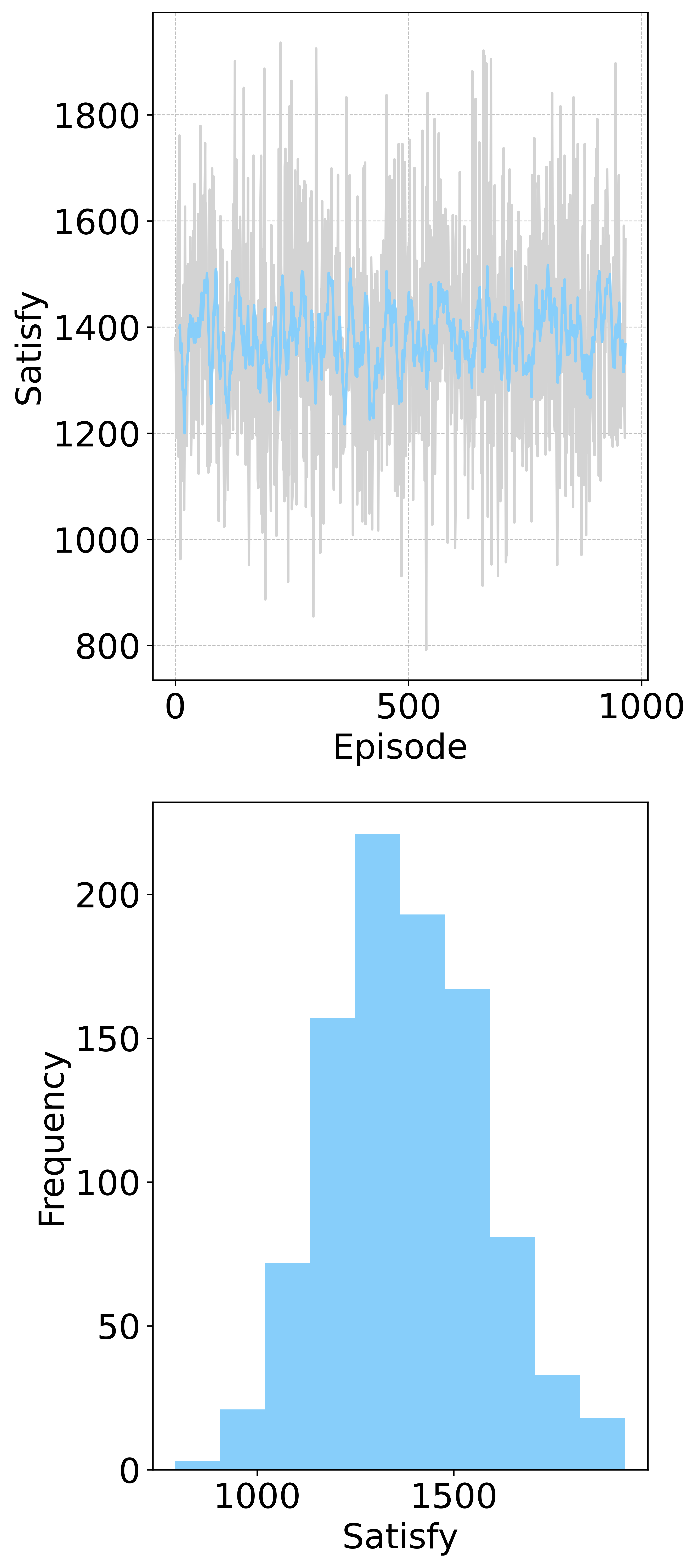}
}
\subfigure[Rolling stock]{
\includegraphics[width=0.18\textwidth]{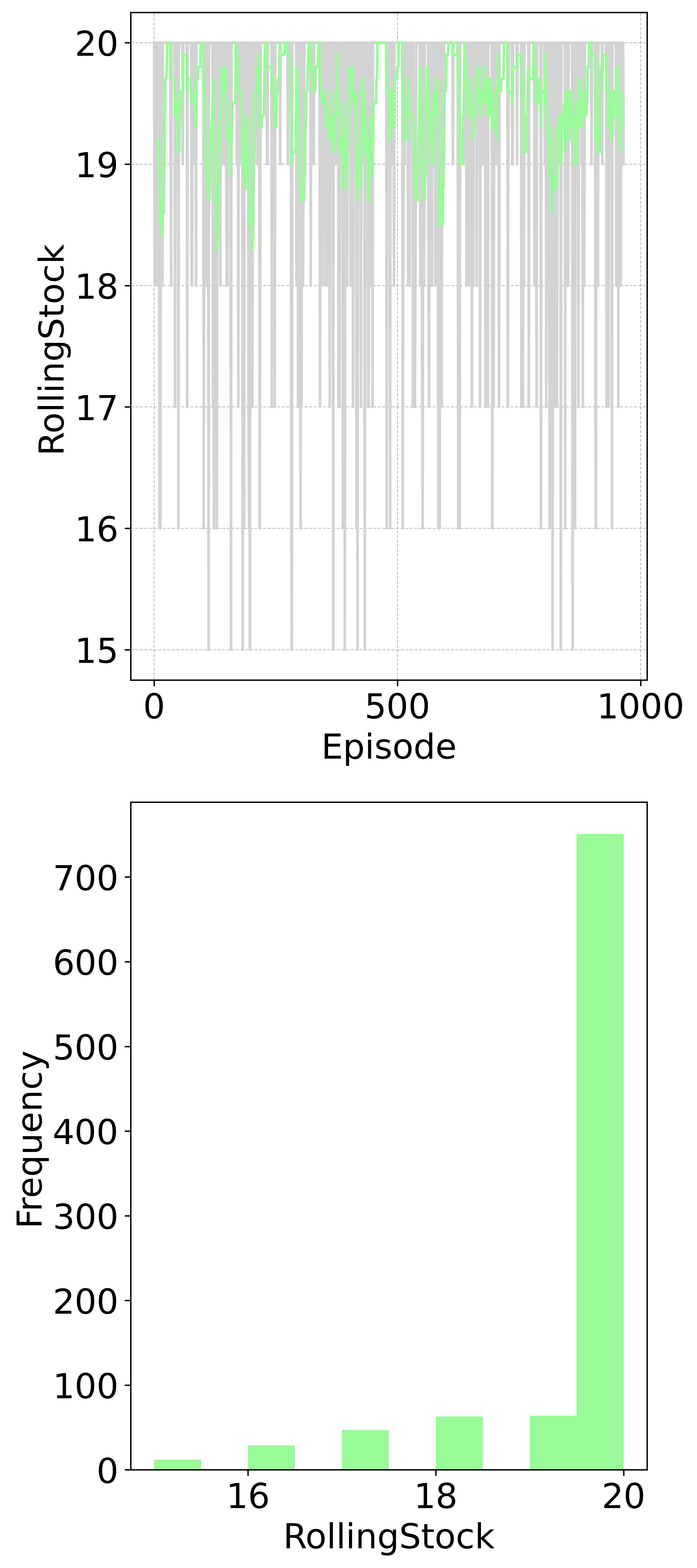}
}
\subfigure[Time steps]{
\includegraphics[width=0.18\textwidth]{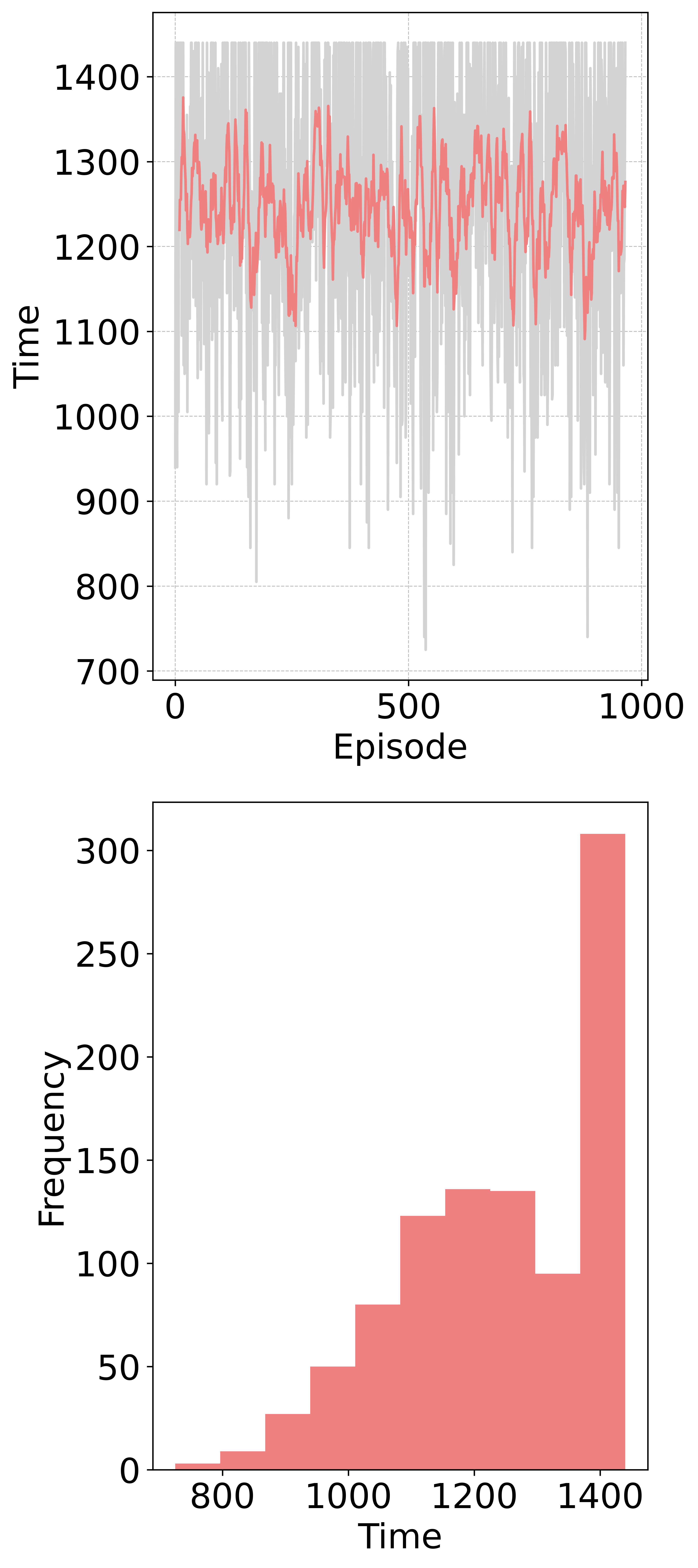}
}
\subfigure[Crowdedness]{
\includegraphics[width=0.18\textwidth]{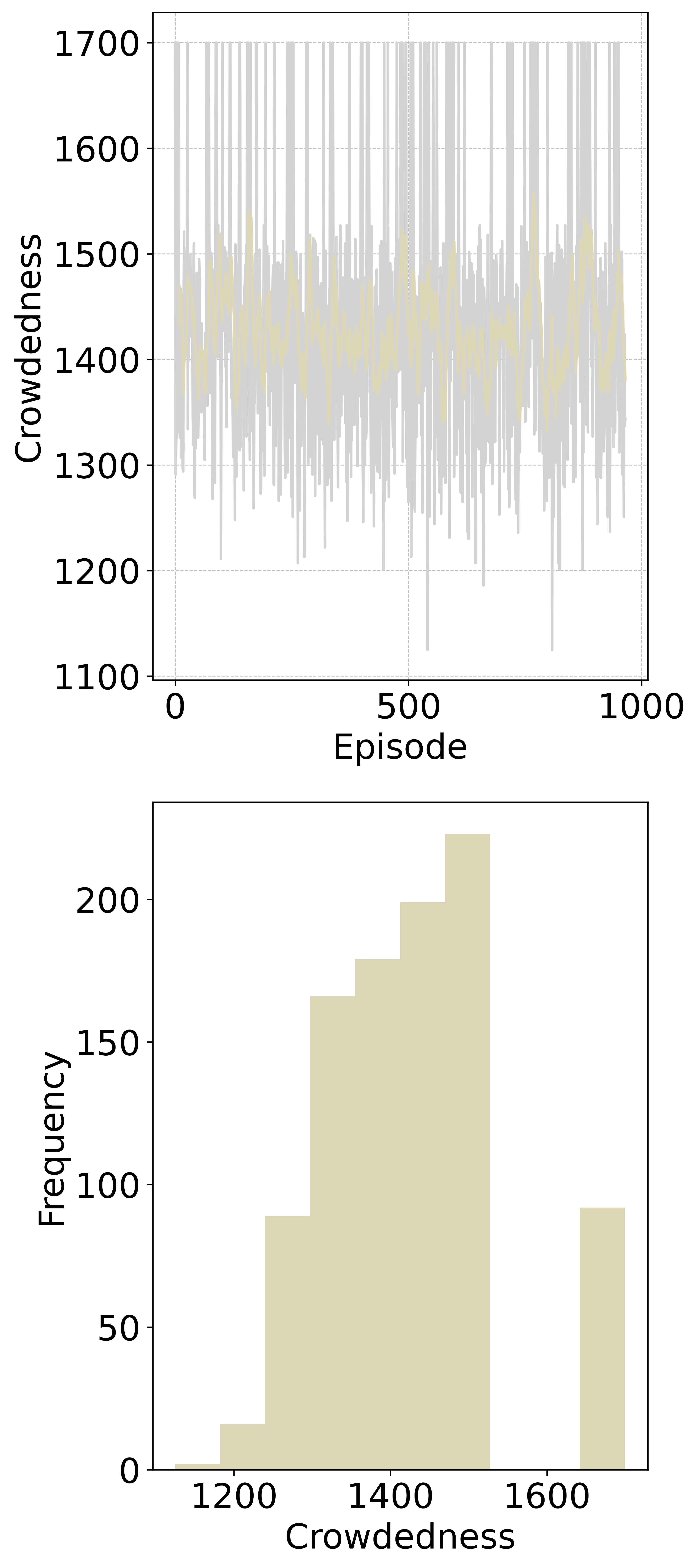}
}

\caption{Statistic of agent's performance in the transferred online environment}
\label{Fig-on-con}
\end{figure}

Figure \ref{Fig-on-con} provides a statistic of the agent's performance during online testing in the transferred environment. It highlights the values and histogram diagrams of key factors, including the reward, satisfied demand, applied rolling stock, played time steps, and maximum in-station crowdedness, observed across 1000 attempts. 90.4\% attempts found the feasible solutions which ends the episode after the evening peak at 18:00. The analysis reveals several important findings. Firstly, the agent consistently achieves positive rewards throughout the episodes, indicating a favorable overall performance. Secondly, the satisfied demand is concentrated around 1400, representing approximately 50\% of the total demand. This level of performance aligns closely with the stable satisfaction rate of around 55\% observed during the offline training process. Notably, the agent achieves a maximum satisfied demand of 1936, accounting for 69.4\% of the total demand. These results demonstrate the agent's ability to effectively handle higher demand scenarios. Furthermore, the agent demonstrates efficient utilization of rolling stock resources, with minimal instances of unused rolling stock. Moreover, the episodes consistently conclude near the end of the day, indicating the agent's strategic approach to long-term rolling stock allocation for maximum benefit. Lastly, despite the transferred environment experiencing a total demand 1.5 times higher than the training environment, the feasible solutions successfully manage in-station crowdedness, keeping it consistently below the predefined threshold. This highlights the agent's proficiency in mitigating overcrowding issues and ensuring a smooth passenger flow.

Overall, the agent's performance in the new online environment showcases its ability to adapt and effectively address challenges associated with heavy demand and potential overcrowding. By intelligently managing rolling stock and continuously monitoring conditions, the agent contributes to a smoother and more efficient operation during periods of disruption.

\section{Conclusions}\label{sec:conclusions}
This paper bridges the gap between \emph{real-time}  and \emph{demand-responsiveness} by defining a real-time demand responsive (RTDR) train rescheduling problem considering station crowdedness for the first time. To solve this challenging RTDR problem, a novel hierarchical deep reinforcement learning (HDRL) approach is proposed and tested based on real-world cases.  Two deep Q-network (DQN) for upper and lower level agents are trained to determine the dispatching and rescheduling plan during a long-term disruption. 
The mobile data (MD) of the focused  station area is used to estimate the real-time demand, allowing for capturing passengers' behaviour and in-station crowdedness through a set of machine learning methods. 

The effectiveness and advantages of the HDRL for are demonstrated via a real-world case in which a railway hub city experiences a complete disruption caused by a flood. The focus is on mitigating the impact of the disruption at an upstream heavy-demand station. The disruption affects a total of 9 routes and 33 trains in one direction that connect the target station to the failed hub. The agent demonstrates its efficacy in addressing challenges of passenger independent returning, in-station overcrowding, limited rolling stock, open-ended disruption duration, integrated rescheduling on multiple routes, and delay reduction due to detour. Notably, the agent is able to satisfy up to 62\% demand on the 9 routes with only 61\% of the original rolling stock, and avoids in-station overcrowding during the whole period of the disruption. When the agent is transferred to a new environment with higher demand, it stably maintains its solving capability similar to the offline training results, which shows its adaptability in handling unforeseen disruptions in real-time settings.

During the offline training phase, the agent's behavior evolves as it progresses from the exploration stage to the exploitation stage. Initially, the agent demonstrates a proactive approach by actively dispatching trains to meet early demand, resulting in a rapid consumption of resources. As it gains knowledge and experience, the agent transitions to a reactive approach, preserving the rolling stock for the evening rush hours, but potentially resulting in underutilizing resources. Eventually, the agent develops a high level of proficiency in accurately identifying the propagation of passenger demand and effectively utilizing the available backup rolling stock. The convergence of the agent is achieved within approximately 5000 iterations, with each iteration of finishing the total episode taking approximately 5 minutes on a High Performance Computing facility. At the end of the training, the agent is capable of consistently finding feasible solutions which avoid in-station overcrowding, monitor the duration of the disruption until the end of the day, and meet approximately 55\% of the total demand with stability.

By analyzing the optimal solution found by the agent, we observe that it is capable of performing simultaneous timetabling on 9 routes using strategies of passenger reassignment, retiming, rerouting, and stop scheduling. The agent employs the combination of strategies to optimize train dispatch and improve overall performance. During the morning and evening peak hours, it increases the number of dispatched trains on busy routes. The agent also prioritizes maintaining train services on high-demand routes to ensure more passengers reach their destinations. For less busy routes, it combines the passenger demand on different trains by dispatching one train in the middle of the original departure time. Additionally, the agent ensures the dispatch of trains on routes that have only one train passing through, guaranteeing connectivity. By rescheduling trains with earlier departure times, the agent effectively reduces en-route delays caused by detours. The agent also learns to strike a balance between demand satisfaction and delay reduction through stop scheduling. It strategically uses stops that encompass all stations, ensuring comprehensive service coverage, and selectively includes trains that only stop at significant stations based on the observed in-station demand patterns. During the daytime, the agent adopts a reactive approach by prioritizing train dispatches on busy routes in the morning, while in the evening, it switches to a proactive approach by increasing dispatch frequency.

The agent is trained using data from one day preceding the real-world disaster, where the total demand is recorded as 1867. Following the training phase, the agent is deployed in the actual disruption environment, characterized by a higher demand of 2787, constituting a 49\% increase compared to the training environment. Remarkably, even without any additional training or feedback specific to the new environment, the agent consistently performs in a manner akin to the training results. It consistently offers feasible solutions and demonstrates the ability to meet a maximum of 69\% of the demand without online training.

This paper represents an initial exploration into deriving railway demand from MD and introduces a stable version of the DRL framework for RTDR rescheduling. While the approach shows promise, there are several areas that warrant further improvement to enhance estimating accuracy, solution quality, and online applicability. For example, incorporating seat reservation data along with MD could improve demand estimation. Additionally, expanding the analysis to encompass multiple cities along different railway routes would enable network-level rescheduling. Furthermore, online fine-tuning during long-term disruptions are avenues for future research.

\section*{Acknowledgement}
The support provided by the China Scholarship Council (CSC) during the first author's visit to the University of Leeds (No. 202106560015) is gratefully acknowledged. We would like to also appreciate the provision of mobile data by Smart Step Digital Technology Co., Ltd., China Union. This work was undertaken on ARC4, part of the High Performance Computing facilities at the University of Leeds.

\bibliography{paper}

\begin{thebibliography}{}

\bibitem[Aghabayk et~al., 2021]{Aghabayk2021EffectsOC}
Aghabayk, K., Esmailpour, J., and Shiwakoti, N. (2021).
\newblock Effects of covid-19 on rail passengers’ crowding perceptions.
\newblock {\em Transportation Research. Part A, Policy and Practice}, 154:186
  -- 202.

\bibitem[ARC4, 2023]{HPC_ARC4}
ARC4 (2023).
\newblock Initial hardware configuration.
\newblock \url{https://arcdocs.leeds.ac.uk/systems/arc4.html}.

\bibitem[Barrena et~al., 2014a]{BARRENA201466}
Barrena, E., Canca, D., Coelho, L.~C., and Laporte, G. (2014a).
\newblock Exact formulations and algorithm for the train timetabling problem
  with dynamic demand.
\newblock {\em Computers \& Operations Research}, 44:66--74.

\bibitem[Barrena et~al., 2014b]{BARRENA2014134}
Barrena, E., Canca, D., Coelho, L.~C., and Laporte, G. (2014b).
\newblock Single-line rail rapid transit timetabling under dynamic passenger
  demand.
\newblock {\em Transportation Research Part B: Methodological}, 70:134--150.

\bibitem[Besinovic et~al., 2022]{besinovic_matheuristic_2022}
Besinovic, N., Wang, Y., Zhu, S., Quaglietta, E., Tang, T., and Goverde, R.
  M.~P. (2022).
\newblock A {Matheuristic} for the {Integrated} {Disruption} {Management} of
  {Traffic}, {Passengers} and {Stations} in {Urban} {Railway} {Lines}.
\newblock {\em IEEE Transactions on Intelligent Transportation Systems},
  23(8):10380--10394.

\bibitem[Binder et~al., 2021]{BINDER2021103368}
Binder, S., Maknoon, M., {Sharif Azadeh}, S., and Bierlaire, M. (2021).
\newblock Passenger-centric timetable rescheduling: A user equilibrium
  approach.
\newblock {\em Transportation Research Part C: Emerging Technologies},
  132:103368.

\bibitem[Binder et~al., 2017]{binder_2017}
Binder, S., Maknoon, Y., and Bierlaire, M. (2017).
\newblock The multi-objective railway timetable rescheduling problem.
\newblock {\em Transportation Research Part C: Emerging Technologies},
  78:78--94.

\bibitem[Cacchiani et~al., 2014]{CACCHIANI201415}
Cacchiani, V., Huisman, D., Kidd, M., Kroon, L., Toth, P., Veelenturf, L., and
  Wagenaar, J. (2014).
\newblock An overview of recovery models and algorithms for real-time railway
  rescheduling.
\newblock {\em Transportation Research Part B: Methodological}, 63:15--37.

\bibitem[Canca et~al., 2014]{RN150}
Canca, D., Barrena, E., Algaba, E., and Zarzo, A. (2014).
\newblock Design and analysis of demand-adapted railway timetables.
\newblock {\em Journal of Advanced Transportation}, 48(2):119--137.

\bibitem[Casanueva et~al., 2018]{Casanueva2018FeudalRL}
Casanueva, I., Budzianowski, P., Su, P.~H., Ultes, S., Rojas-Barahona, L.~M.,
  Tseng, B.~H., and Gasic, M. (2018).
\newblock Feudal reinforcement learning for dialogue management in large
  domains.
\newblock {\em ArXiv}, abs/1803.03232.

\bibitem[Cordera et~al., 2018]{cordera_trip_2018}
Cordera, R., Sanudo, R., dell'Olio, L., and Ibeas, A. (2018).
\newblock Trip distribution model for regional railway services considering
  spatial effects between stations.
\newblock {\em Transport Policy}, 67:77--84.

\bibitem[Corman et~al., 2017]{corman_integrating_2017}
Corman, F., D’Ariano, A., Marra, A.~D., Pacciarelli, D., and Samà, M.
  (2017).
\newblock Integrating train scheduling and delay management in real-time
  railway traffic control.
\newblock {\em Transportation Research Part E: Logistics and Transportation
  Review}, 105:213--239.

\bibitem[Corman and Quaglietta, 2015]{CORMAN201515}
Corman, F. and Quaglietta, E. (2015).
\newblock Closing the loop in real-time railway control: Framework design and
  impacts on operations.
\newblock {\em Transportation Research Part C: Emerging Technologies},
  54:15--39.

\bibitem[Dayan and Hinton, 1992]{Dayan1992FeudalRL}
Dayan, P. and Hinton, G.~E. (1992).
\newblock Feudal reinforcement learning.
\newblock In {\em Neural Information Processing Systems}.

\bibitem[Demissie et~al., 2016]{RN129}
Demissie, M.~G., Phithakkitnukoon, S., Sukhvibul, T., Antunes, F., Gomes, R.,
  and Bento, C. (2016).
\newblock Inferring passenger travel demand to improve urban mobility in
  developing countries using cell phone data: A case study of senegal.
\newblock {\em IEEE Transactions on Intelligent Transportation Systems},
  17(9):2466--2478.

\bibitem[Dollevoet et~al., 2017]{DOLLEVOET2017203}
Dollevoet, T., Huisman, D., Kroon, L.~G., Veelenturf, L.~P., and Wagenaar,
  J.~C. (2017).
\newblock Application of an iterative framework for real-time railway
  rescheduling.
\newblock {\em Computers \& Operations Research}, 78:203--217.

\bibitem[Ghaemi et~al., 2018]{ghaemi_macroscopic_2018}
Ghaemi, N., Cats, O., and Goverde, R.~M. (2018).
\newblock Macroscopic multiple-station short-turning model in case of complete
  railway blockages.
\newblock {\em Transportation Research Part C: Emerging Technologies},
  89:113--132.

\bibitem[Hong et~al., 2021]{HONG2021103025}
Hong, X., Meng, L., D'Ariano, A., Veelenturf, L.~P., Long, S., and Corman, F.
  (2021).
\newblock Integrated optimization of capacitated train rescheduling and
  passenger reassignment under disruptions.
\newblock {\em Transportation Research Part C: Emerging Technologies},
  125:103025.

\bibitem[Kroon et~al., 2015]{Kroon2015}
Kroon, L., Mar\'{o}ti, G., and Nielsen, L. (2015).
\newblock Rescheduling of railway rolling stock with dynamic passenger flows.
\newblock {\em Transportation Science}, 49(2):165--184.

\bibitem[Laud, 2004]{reward_shaping_thesis2004}
Laud, A.~D. (2004).
\newblock {\em Theory and Application of Reward Shaping in Reinforcement
  Learning}.
\newblock PhD thesis, USA.

\bibitem[Levy et~al., 2017]{Levy2017LearningMH}
Levy, A., Konidaris, G.~D., Platt, R.~W., and Saenko, K. (2017).
\newblock Learning multi-level hierarchies with hindsight.
\newblock In {\em International Conference on Learning Representations}.

\bibitem[Li and Ni, 2022]{LI2022230}
Li, W. and Ni, S. (2022).
\newblock Train timetabling with the general learning environment and
  multi-agent deep reinforcement learning.
\newblock {\em Transportation Research Part B: Methodological}, 157:230--251.

\bibitem[Li et~al., 2023]{LI202330}
Li, X., Shi, L., Tang, J., Yang, C., Zhao, T., Wang, Y., and Wang, W. (2023).
\newblock Determinants of passengers' ticketing channel choice in rail transit
  systems: New evidence of e-payment behaviors from xi'an, china.
\newblock {\em Transport Policy}, 140:30--41.

\bibitem[Liao et~al., 2021]{Liao_deep_2021}
Liao, J., Yang, G., Zhang, S., Zhang, F., and Gong, C. (2021).
\newblock A deep reinforcement learning approach for the energy-aimed train
  timetable rescheduling problem under disturbances.
\newblock {\em IEEE Transaction on Transportation Electrification},
  7(4):3096--3109.

\bibitem[Liu et~al., 2022]{Liu2022AME}
Liu, E., Barker, K., and Chen, H. (2022).
\newblock A multi-modal evacuation-based response strategy for mitigating
  disruption in an intercity railway system.
\newblock {\em Reliability Engineering System \& Safety}, 223:108515.

\bibitem[Meng and Zhou, 2019]{meng_integrated_2019}
Meng, L. and Zhou, X. (2019).
\newblock An integrated train service plan optimization model with variable
  demand: {A} team-based scheduling approach with dual cost information in a
  layered network.
\newblock {\em Transportation Research Part B: Methodological}, 125:1--28.

\bibitem[Mnih et~al., 2013]{mnih_playing_2013}
Mnih, V., Kavukcuoglu, K., Silver, D., Graves, A., Antonoglou, I., Wierstra,
  D., and Riedmiller, M. (2013).
\newblock Playing atari with deep reinforcement learning.

\bibitem[Mo et~al., 2022]{MO2022102628}
Mo, B., Koutsopoulos, H.~N., and Zhao, J. (2022).
\newblock Inferring passenger responses to urban rail disruptions using smart
  card data: A probabilistic framework.
\newblock {\em Transportation Research Part E: Logistics and Transportation
  Review}, 159:102628.

\bibitem[{National~Bureau~of~Statistics~of~China}, 2023]{GDP_national}
{National~Bureau~of~Statistics~of~China} (2023).
\newblock National data.
\newblock \url{https://data.stats.gov.cn/index.htm}.

\bibitem[Ng et~al., 1999]{Ng1999PolicyIU}
Ng, A., Harada, D., and Russell, S.~J. (1999).
\newblock Policy invariance under reward transformations: Theory and
  application to reward shaping.
\newblock In {\em International Conference on Machine Learning}.

\bibitem[Nielsen et~al., 2012]{RN76}
Nielsen, L.~K., Kroon, L., and Maróti, G. (2012).
\newblock A rolling horizon approach for disruption management of railway
  rolling stock.
\newblock {\em European Journal of Operational Research}, 220(2):496--509.

\bibitem[Ning et~al., 2019]{ning_deep_2019}
Ning, L., Li, Y., Zhou, M., Song, H., and Dong, H. (2019).
\newblock A deep reinforcement learning approach to high-speed train timetable
  rescheduling under disturbances.
\newblock In {\em 2019 {IEEE} Intelligent Transportation Systems Conference
  ({ITSC})}, pages 3469--3474. {IEEE}.

\bibitem[Niu and Zhou, 2013]{niu_optimizing_2013}
Niu, H. and Zhou, X. (2013).
\newblock Optimizing urban rail timetable under time-dependent demand and
  oversaturated conditions.
\newblock {\em Transportation Research Part C: Emerging Technologies},
  36:212--230.

\bibitem[Pateria et~al., 2023]{pateria_hierarchical_2022}
Pateria, S., Subagdja, B., Tan, A.-h., and Quek, C. (2023).
\newblock Hierarchical reinforcement learning: A comprehensive survey.
\newblock {\em {ACM} Computing Surveys}, 54(5):1--35.

\bibitem[Pedregosa et~al., 2011]{scikit-learn}
Pedregosa, F., Varoquaux, G., Gramfort, A., Michel, V., Thirion, B., Grisel,
  O., Blondel, M., Prettenhofer, P., Weiss, R., Dubourg, V., Vanderplas, J.,
  Passos, A., Cournapeau, D., Brucher, M., Perrot, M., and Duchesnay, E.
  (2011).
\newblock Scikit-learn: Machine learning in {P}ython.
\newblock {\em Journal of Machine Learning Research}, 12:2825--2830.

\bibitem[Sun et~al., 2014]{sun_demand-driven_2014}
Sun, L., Jin, J.~G., Lee, D.-H., Axhausen, K.~W., and Erath, A. (2014).
\newblock Demand-driven timetable design for metro services.
\newblock {\em Transportation Research Part C: Emerging Technologies},
  46:284--299.

\bibitem[Sutton and Barto, 2018]{RN157}
Sutton, R.~S. and Barto, A.~G. (2018).
\newblock {\em Reinforcement Learning: An Introduction}.
\newblock MIT Press, Cambridge, MA, USA, 2nd edition.

\bibitem[{Time and date}, 2021]{Weather_Xian}
{Time and date} (2021).
\newblock {Past Weather in Xi'an, Shaanxi, China — July 2021}.
\newblock
  \url{https://www.timeanddate.com/weather/china/sian/historic?month=7&year=2021}.

\bibitem[Veelenturf et~al., 2017]{VEELENTURF2017133}
Veelenturf, L.~P., Kroon, L.~G., and Maróti, G. (2017).
\newblock Passenger oriented railway disruption management by adapting
  timetables and rolling stock schedules.
\newblock {\em Transportation Research Part C: Emerging Technologies},
  80:133--147.

\bibitem[Wang et~al., 2015]{Wang2015ModelingAS}
Wang, J., Zhang, L., Shi, Q., Yang, P., and Hu, X. (2015).
\newblock Modeling and simulating for congestion pedestrian evacuation with
  panic.
\newblock {\em Physica A-statistical Mechanics and Its Applications},
  428:396--409.

\bibitem[Wang et~al., 2021a]{wang_policy-based_2021}
Wang, Y., Lv, Y., Zhou, J., Yuan, Z., Zhang, Q., and Zhou, M. (2021a).
\newblock A policy-based reinforcement learning approach for high-speed railway
  timetable rescheduling.
\newblock In {\em 2021 {IEEE} International Intelligent Transportation Systems
  Conference ({ITSC})}, pages 2362--2367. {IEEE}.

\bibitem[Wang et~al., 2021b]{RN116}
Wang, Z., Wang, S., and Lian, H. (2021b).
\newblock A route-planning method for long-distance commuter express bus
  service based on od estimation from mobile phone location data: the case of
  the changping corridor in beijing.
\newblock {\em Public Transport}, 13(1):101--125.

\bibitem[Wu et~al., 2021]{wu_deep_2021}
Wu, W., Yin, J., Pu, F., Su, S., and Tang, T. (2021).
\newblock A deep reinforcement learning approach for the traffic management of
  high-speed railways.
\newblock In {\em 2021 {IEEE} International Intelligent Transportation Systems
  Conference ({ITSC})}, pages 2368--2373. {IEEE}.

\bibitem[Yin et~al., 2019a]{yin_hybrid_2019}
Yin, Y., Li, D., Bešinović, N., and Cao, Z. (2019a).
\newblock Hybrid {Demand}-{Driven} and {Cyclic} {Timetabling} {Considering}
  {Rolling} {Stock} {Circulation} for a {Bidirectional} {Railway} {Line}:
  {Hybrid} demand-driven and cyclic timetabling.
\newblock {\em Computer-Aided Civil and Infrastructure Engineering},
  34(2):164--187.

\bibitem[Yin et~al., 2019b]{Yin2019HybridDA}
Yin, Y., Li, D.~W., Besinovic, N., and Cao, Z. (2019b).
\newblock Hybrid demand‐driven and cyclic timetabling considering rolling
  stock circulation for a bidirectional railway line.
\newblock {\em Computer‐Aided Civil and Infrastructure Engineering}, 34.

\bibitem[Ying et~al., 2020]{ying_actor-critic_2020}
Ying, C.-s., Chow, A. H.~F., and Chin, K.-S. (2020).
\newblock An actor-critic deep reinforcement learning approach for metro train
  scheduling with rolling stock circulation under stochastic demand.
\newblock {\em Transportation Research Part B: Methodological}, 140:210--235.

\bibitem[Zhan et~al., 2016]{zhan_2016}
Zhan, S., Kroon, L.~G., Zhao, J., and Peng, Q. (2016).
\newblock A rolling horizon approach to the high speed train rescheduling
  problem in case of a partial segment blockage.
\newblock {\em Transportation Research Part E: Logistics and Transportation
  Review}, 95:32--61.

\bibitem[Zhan et~al., 2015]{RN105}
Zhan, S.~G., Kroon, L.~G., Veelenturf, L.~P., and Wagenaar, J.~C. (2015).
\newblock Real-time high-speed train rescheduling in case of a complete
  blockage.
\newblock {\em Transportation Research Part B: Methodological}, 78:182--201.

\bibitem[Zhan et~al., 2022]{RN93}
Zhan, S.~G., Wong, S.~C., Shang, P., and Lo, S.~M. (2022).
\newblock Train rescheduling in a major disruption on a high-speed railway
  network with seat reservation.
\newblock {\em Transportmetrica A: Transport Science}, 18(3):532--567.

\bibitem[Zhang et~al., 2023]{zhang_2023}
Zhang, P., Zhao, P., Qiao, K., Wen, P., and Li, P. (2023).
\newblock A multistage decision optimization approach for train timetable
  rescheduling under uncertain disruptions in a high-speed railway network.
\newblock {\em {IEEE} Transactions on Intelligent Transportation Systems},
  pages 1--15.

\bibitem[Zhang et~al., 2020]{Zhang2020RailwaySR}
Zhang, Q., Zhuang, Y., Wei, Y., Jiang, H., and Yang, H. (2020).
\newblock Railway safety risk assessment and control optimization method based
  on fta-fpn: A case study of chinese high-speed railway station.
\newblock {\em Journal of Advanced Transportation}, 2020:1--11.

\bibitem[Zhao et~al., 2021]{zhao_integrated_2021}
Zhao, S., Yang, H., and Wu, Y. (2021).
\newblock An integrated approach of train scheduling and rolling stock
  circulation with skip-stopping pattern for urban rail transit lines.
\newblock {\em Transportation Research Part C: Emerging Technologies},
  128:103170.

\bibitem[Zhong et~al., 2019]{RN119}
Zhong, G., Yin, T., Zhang, J., He, S., and Ran, B. (2019).
\newblock Characteristics analysis for travel behavior of transportation hub
  passengers using mobile phone data.
\newblock {\em Transportation}, 46(5):1713--1736.

\bibitem[Zhong et~al., 2023]{Zhong2023AHF}
Zhong, J., He, Z., Wang, J., and Xie, J. (2023).
\newblock A hierarchical framework for passenger inflow control in metro system
  with reinforcement learning.
\newblock {\em IEEE Transactions on Intelligent Transportation Systems},
  24:10895--10911.

\bibitem[Zhu and Goverde, 2020]{zhu_2020}
Zhu, Y. and Goverde, R.~M. (2020).
\newblock Dynamic and robust timetable rescheduling for uncertain railway
  disruptions.
\newblock {\em Journal of Rail Transport Planning \& Management}, 15:100196.

\bibitem[Zhu and Goverde, 2019]{RN132}
Zhu, Y. and Goverde, R. M.~P. (2019).
\newblock Railway timetable rescheduling with flexible stopping and flexible
  short-turning during disruptions.
\newblock {\em Transportation Research Part B: Methodological}, 123:149--181.

\bibitem[Zhu and Goverde, 2021]{RN131}
Zhu, Y. and Goverde, R. M.~P. (2021).
\newblock Dynamic railway timetable rescheduling for multiple connected
  disruptions.
\newblock {\em Transportation Research Part C: Emerging Technologies},
  125:103080.

\bibitem[Zhu et~al., 2017]{zhu_bi-level_2017}
Zhu, Y., Mao, B., Bai, Y., and Chen, S. (2017).
\newblock A bi-level model for single-line rail timetable design with
  consideration of demand and capacity.
\newblock {\em Transportation Research Part C: Emerging Technologies},
  85:211--233.

\bibitem[Zhu et~al., 2020]{zhu_re_2020}
Zhu, Y., Wang, H., and Goverde, R.~M. (2020).
\newblock Reinforcement learning in railway timetable rescheduling.
\newblock In {\em 2020 {IEEE} 23rd International Conference on Intelligent
  Transportation Systems ({ITSC})}, pages 1--6. {IEEE}.

\bibitem[Zhu et~al., 2023]{zhu_reinforcement_2023}
Zhu, Y., Wang, P., and Corman, F. (2023).
\newblock Reinforcement {Learning} in {Railway} {Delay} {Management}.
\newblock preprint, SSRN.

\bibitem[Šemrov et~al., 2016]{SEMROV2016250}
Šemrov, D., Marsetič, R., Žura, M., Todorovski, L., and Srdic, A. (2016).
\newblock Reinforcement learning approach for train rescheduling on a
  single-track railway.
\newblock {\em Transportation Research Part B: Methodological}, 86:250--267.

\end{thebibliography}

\appendixpage

\setcounter{section}{0}
\renewcommand{\thesection}{A\arabic{section}}

\section{Estimating the waiting tolerance and target train from MD}\label{App-1}

In order to connect train information with MD, we, first, select data which leave by railway, and then,  match the leaving time with the railway timetable in July 2021. Considering the daily disturbance of departure time and latency of record, we set a buffer to group the data. Specifically, when a users' leaving time is greater than or equal to a departure time of train $i$ and less than the next departure time of train $j$, we deem this user as a passenger for train $i$. Furthermore, we sum the weight of all users of a train as its passenger demand. In addition, there are some departure times are near or equal. We use passengers' next holding point to determine which train they belong to. It is reasonable because most of trains with similar departure time are on different routes.

The timetable provided in Table \ref{Tab-Timetable} illustrates a segment of a railway timetable passing through Xi'an. The table contains important information regarding the trains such as their respective codes, arrival and departure times, routes, starting and ending stations, and an indication of whether the train was disrupted by the Zhengzhou flood.

\begin{table}[ht]
\caption{Segment of railway timetable passing Xi'an}
\label{Tab-Timetable}
\begin{tabular}{llrlrrl}
\toprule
Train Code & Arrive                       & \multicolumn{1}{l}{Departure} & Route & \multicolumn{1}{l}{Starting station} & \multicolumn{1}{l}{Ending station} & Disrupted \\
\midrule
K559       & \multicolumn{1}{r}{16:53:00} & 17:13:00                      & 4     & Yan'an                               & Shanghai                           & 1         \\
T232       &                              & 17:26:00                      & 7     & Xi'an                                & Beijing West                       & 1         \\
K5468      & \multicolumn{1}{r}{17:20:00} & 17:35:00                      & 0     & Baoji                                & Ankang                             & 0         \\
K1088      &                              & 17:46:00                      & 0     & Xi'an                                & Yinchuan                           & 0         \\
K1674      &                              & 18:06:00                      & 0     & Xi'an                                & Huhehaote                          & 0         \\
K4033      &                              & 18:12:00                      & 0     & Xi'an        & Guiyang    & 0         \\
K420       & \multicolumn{1}{r}{18:10:00} & 18:20:00                      & 2     & Lanzhou                              & Yangzhou                           & 1         \\
1148       & \multicolumn{1}{r}{18:17:00} & 18:29:00                      & 9     & Baoji                                & Lianyungang East                   & 1         \\
K4816      & \multicolumn{1}{r}{18:38:00} & 18:54:00                      & 0     & Baoji                                & Beijing                            & 0         \\
2096       &                              & 19:00:00                      & 0     & Xi'an                                & Taiyuan                            & 0         \\
Z20        &                              & 19:12:00                      & 7     & Xi'an                                & Beijing West                       & 1         \\
Z88        &                              & 19:18:00                      & 0     & Xi'an                                & Hangzhou                           & 0         \\
Z44        &                              & 19:27:00                      & 0     & Xi'an                                & Beijing West                       & 0         \\
Z106       & \multicolumn{1}{r}{19:25:00} & 19:33:00                      & 1     & Wulumuqi                             & Ji'nan                             & 1\\     
\bottomrule
\end{tabular}
\end{table}

\begin{figure}
    \centering
    \includegraphics[width=0.6\linewidth]{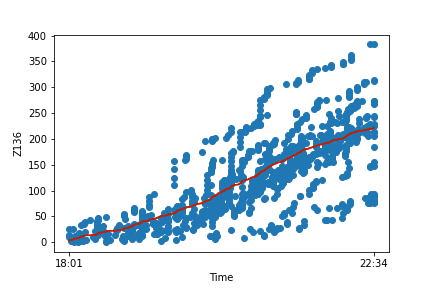}
    \caption{Example of accumulating process for a train on different days}
    \label{Fig-Arr}
\end{figure}

By traversing the timetable of study period on the normal days, demand of all trains and passengers' arriving process are detected from the MD. For each train, the passenger accumulating process on different days can be illustrated according to their arriving/entering time, for example, the process for train $Z136$ (departure at 22:34) on July 1-19 is given in Fig. \ref{Fig-Arr}. There are some noisy data due to variant daily demand. A signal filter method, Savitzky-Golay, is applied to clarify the mean curve of accumulation, which is red in the figure.

\begin{figure}[htbp]
\centering
\subfigure[Elbow method]{
\includegraphics[width=0.3\linewidth]{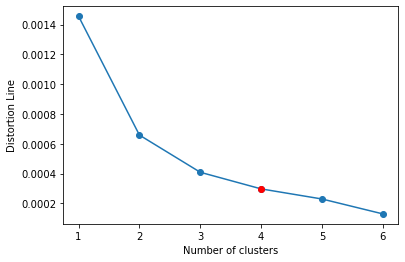}
}
\subfigure[Clustering result]{
\includegraphics[width=0.3\linewidth]{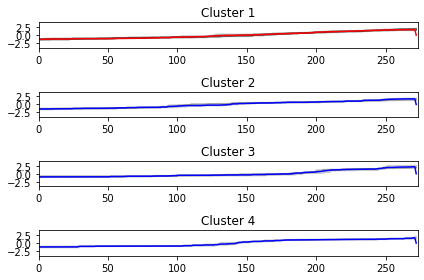}
}
\subfigure[Representing curve]{
\includegraphics[width=0.3\linewidth]{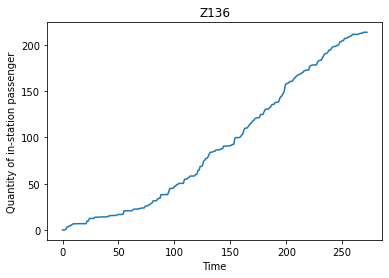}
}
\caption{Example of finding the representing curve for the accumulating process of a train}
\label{Fig-ArrRepre}
\end{figure}

In order to filter out the abnormal accumulating process, a time-series clustering method, Kshape, is applied to classify the typical process. It normalizes the scale of demand, and focuses on the time-depend tendency. An elbow method pre-determines the number of cluster. Then, the cluster to which the most daily accumulating processes belongs is picked as the typical accumulating process for a train. For example of train $Z136$ whose cluster number is 4 according to elbow point in Fig \ref{Fig-ArrRepre}(a), then, Fig. \ref{Fig-ArrRepre}(b) shows the accumulating process and centering curve of each cluster where the red one is the typical cluster. Fig. \ref{Fig-ArrRepre}(c) shows the representing curve by extending the centering curve of typical cluster to the average value in the dimension of demand.

\begin{figure}
    \centering
    \includegraphics[width=0.6\linewidth]{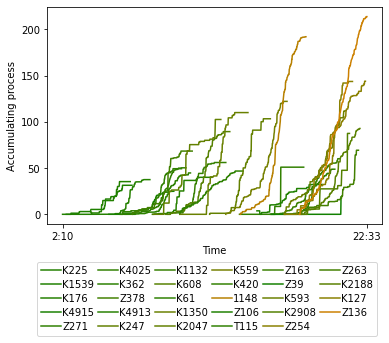}
    \caption{Example of the combination of accumulating process of several trains}
    \label{Fig-AllTrain}
\end{figure}

By combining the accumulating process of all disrupted trains, the in-station passenger quantity can be detected and predicted in real-time. Several trends are picked out in Fig. \ref{Fig-AllTrain} where the color distinguishes the total demand of the train. By vertically summing up the accumulated number for different trains, the in-station quantity at that instant can be found.

Section \ref{Method-Env} described a posterior method to determine a user's target train, based on the historical data. However, in reality, we cannot determine the train when a user enters the station. Therefore, a estimating approach is needed to bridge the gap between the user and train demand according to the information from MD.

The independent variables include arriving time (timestamp), localization (binary), weekday (integer), accumulating process (category) and the dependent variable is the train code (category). The data related to disrupted trains on Longhai railway from July 1 to 19 is filtered out and split into training and texting sets whose size is 9,187 and 2,297, respectively. Machine learning models from $sklearn$ classifier packages in Python 3.9 are used to predict the train code. Performance of different models, including Random Forest, MLP, KNN, SVC, Ridge, Lasso, Decision Tree, Extra Tree, AdaBoost, GradientBoosting, Bagging, are compared. Random Forest Classifier has the best accuracy of 0.74, and the solving efficiency meets the requirement of real-time.

The waiting tolerance is predicted by machine learning regression approach. The independent variables include arriving time (timestamp) and weekday (integer), and the dependent variable is the waiting time (timestamp). In order to analyse passenger's tolerance without advance-noticed disruption, we combined daily data of July 4, 11, 18, and 20 which are the days suffered heavy delay before or once after the Zhengzhou flood. The dataset is split into training and testing sets whose size is 9,534 and 2,384. Random Forest Regression has the best r2 score of about 0.87, and mean absolute error of about 23.

\begin{figure}[htbp]
\centering
\subfigure[Type 1: fast accumulating \& slow returning]{
\includegraphics[height=4cm]{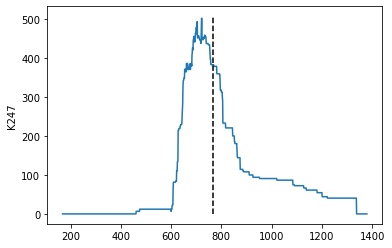}
}
\subfigure[Type 2: bell shape]{
\includegraphics[height=4cm]{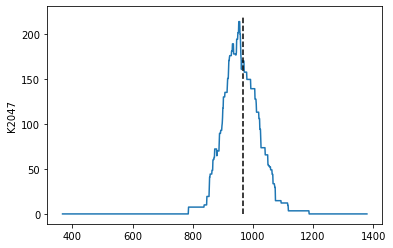}
}
\subfigure[Type 3: slow accumulating \& fast returning]{
\includegraphics[height=4cm]{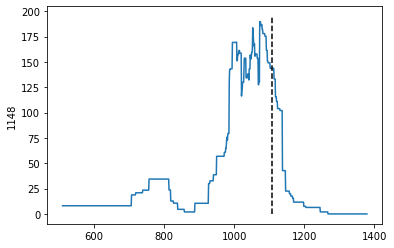}
}
\subfigure[Type 4: simultaneously accumulating \& returning]{
\includegraphics[height=4cm]{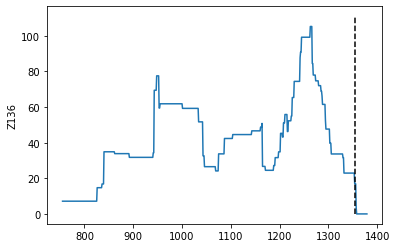}
}
\caption{Typical accumulating and returning process after unexpected disruption}
\label{Fig-Whole}
\end{figure}

Accordingly, with the estimating tools for passengers' target train code and their waiting tolerance, the arriving and returning process for each train can be known dynamically. In the disrupted period, the timetable is not reliable where most of trains are delayed. We take MD on July 20 as an example to estimate real-time demand for each train, accounting for passengers mobility. Four typical processes are found and shown in Fig. \ref{Fig-Whole}, which are fast accumulating \& slow returning, bell shape, slow accumulating \& fast returning, simultaneously accumulating \& returning. The dashed lines show the departure time on the timetable which has been cancelled on that day. Therefore, we address the complexity of dynamic demand by real-world data, and this study treats the passenger demand by real-world passenger mobility rather than model-based estimation.

\section{Case description through MD}\label{App-2}
\subsection{Identifying the potential passengers from MD}\label{App-2.1}

The dataset was divided into two categories, normal days and abnormal days, based on the presence of any disturbance or disruption during the day. Normal days were recorded from July 1-19, excluding July 4, 11, and 18 which experienced significant delays due to heavy rainfall. Abnormal days include the aforementioned three days and the period from July 20 to August 4, which experienced troubleshooting during the Zhengzhou flood. Within each dataset, passenger groups were classified into arrive-departure and arrive-return, resulting in four datasets: Normal-Return, Normal-Departure, Abnormal-Return, and Abnormal-Departure.

The dimensions related to each group within each dataset included entering time, leaving time, leaving mode, next stop, localization, visiting frequency of the station, and the visiting frequency of the next stop. Labels were assigned to all passenger groups in all four databases: 1 for departure and 0 for return. The random forest methods in sklearn in Python 3.9 was used for machine learning. The MD from July 1 to August 4 was split into a training set (226,332 data) and a testing set (56,583 data). After training, the r2 score for the testing set was approximately 0.92.

Using the model on the normal days data, the prediction accuracy of the user group's next movement (i.e. return or departure) was approximately 1, while the accuracy for abnormal days passenger groups was approximately 0.55 due to misunderstandings about passengers' next movements during abnormal days. This means some passengers were predicted to depart but ultimately returned. These passengers were considered potential passengers for the disrupted trains. Thus, these records were examined for their waiting tolerance to identify waiting tolerance patterns during heavy delays.

\subsection{Consequence of unexpected railway disruption}\label{App-2.2}

\begin{figure}[htbp]
  \centering
  \includegraphics[width=0.6\linewidth]{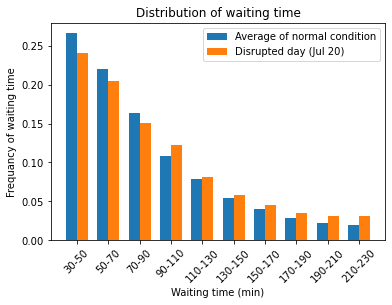}
  \caption{Distribution of waiting time}
  \label{Fig-waiting}
\end{figure}

To analyze the consequences of the disruption, a detailed study was conducted on the MD specifically for July 20th. Firstly, it was observed that passengers experienced prolonged waiting times due to train delays. A comparison of waiting times between normal days and the day of the disruption, as depicted in Figure \ref{Fig-waiting}, revealed important insights. The blue bars represent the distribution of average waiting times on normal days (July 1st to July 19th, excluding July 4th, 11th, and 18th), while the orange bars represent the waiting times on the disrupted day, July 20th. Based on Figure \ref{Fig-waiting}, it is evident that the frequency of shorter waiting times (30-90 minutes) on normal days was higher compared to the disrupted day. Conversely, a larger number of users experienced longer waiting times (90-240 minutes) during the disruption.

\begin{figure*}[htbp]
  \centering
  \includegraphics[width=0.6\linewidth]{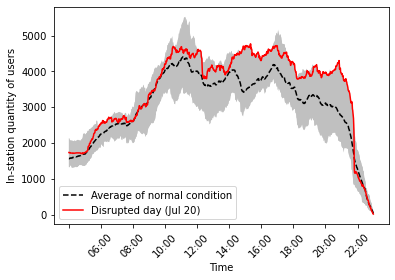}
  \caption{Dynamic of in-station users quantity}
  \label{Fig-instation}
\end{figure*}

Secondly, the analysis focused on the quantity of in-station users, which remained at a consistently high level for an extended duration compared to normal days. Figure \ref{Fig-instation} presents a comparison of in-station user quantities between the disrupted and normal days, containing the records of departing and returning users. The level of crowding within a station is dynamic, fluctuating with the arrival and departure of passengers. In normal conditions, there were two peaks observed, occurring around 10 o'clock in the morning and 4 o'clock in the afternoon. However, during the disruption, the peak persisted for a longer duration, and the overall quantity of in-station users remained at a higher level. This transformed the peak into a plateau, indicating a more severe and prolonged period of crowding. Consequently, passengers were subjected to heightened levels of congestion for an extended period of time when unexpected disruptions occurred.

\section{Case study settings}\label{App-3}
\subsection{Timetable of disrupted trains}\label{App-3.1}

Table \ref{Tab-33train} displays the timetable of the 33 disrupted trains that pass through Xi'an station. The table includes information such as the train code, arrival time, departure time, and the route to which each train belongs. For trains that depart from Xi'an, the arrival time is left empty. The disrupted trains are distributed across various routes as follows: 6 trains on route 1, 2 trains on route 2, 3 trains on route 3, 11 trains on route 4, 2 trains on route 5, 4 trains on route 6, 3 trains on route 7, 1 train on route 8, and 1 train on route 9.

Out of the 33 disrupted trains, 10 of them originate from Xi'an, while the remaining 23 trains pass through Xi'an. Among these trains, those arriving at Xi'an before 17:00 are considered available for rescheduling, while the trains that arrive after that time are disrupted due to the hub failure and are subsequently stopped en route. The mid rule is applied to separate the timetable. Trains above the mid rule are eligible for rescheduling, while those below it are not. 

By distinguishing between available and disrupted trains based on their arrival times at Xi'an, it becomes possible to determine which trains can be rescheduled and which ones are affected by the hub failure.

\begin{table}[h]
\caption{Timetable for 33 disrupted trains}
\label{Tab-33train}
\begin{tabular}{lrrl}
\toprule
\multicolumn{1}{c}{TrainCode} & \multicolumn{1}{c}{ArriveTime} & \multicolumn{1}{c}{StartTime} & \multicolumn{1}{c}{Route} \\
\midrule
K225                          & 6:33:00                        & 6:43:00                       & 6                         \\
K1539                         & 6:40:00                        & 6:50:00                       & 4                         \\
K176                          & 7:54:00                        & 8:02:00                       & 1                         \\
K4915                         & 9:28:00                        & 9:38:00                       & 4                         \\
Z271                          & 10:01:00                       & 10:09:00                      & 1                         \\
K4025                         & 10:17:00                       & 10:27:00                      & 1                         \\
K362                          & 10:32:00                       & 10:45:00                      & 4                         \\
Z378                          & 10:44:00                       & 10:52:00                      & 4                         \\
K4913                         & \multicolumn{1}{l}{}           & 12:12:00                      & 1                         \\
K247                          & 12:27:00                       & 12:47:00                      & 2                         \\
K1132                         & 12:41:00                       & 13:07:00                      & 1                         \\
K608                          & \multicolumn{1}{l}{}           & 13:23:00                      & 3                         \\
K61                           & \multicolumn{1}{l}{}           & 14:16:00                      & 8                         \\
K1350                         & \multicolumn{1}{l}{}           & 14:36:00                      & 6                         \\
K2047                         & \multicolumn{1}{l}{}           & 16:06:00                      & 5                         \\
Z94                           & \multicolumn{1}{l}{}           & 16:45:00                      & 4                         \\
K559                          & 16:53:00                       & 17:13:00                      & 4                         \\
\midrule
T232                          & \multicolumn{1}{l}{}           & 17:26:00                      & 7                         \\
K420                          & 18:10:00                       & 18:20:00                      & 2                         \\
1148                          & 18:17:00                       & 18:29:00                      & 9                         \\
Z20                           & \multicolumn{1}{l}{}           & 19:12:00                      & 7                         \\
Z106                          & 19:25:00                       & 19:33:00                      & 1                         \\
T115                          & 20:15:00                       & 20:25:00                      & 4                         \\
T57                           & 20:21:00                       & 20:38:00                      & 7                         \\
Z163                          & 20:34:00                       & 20:44:00                      & 4                         \\
Z39                           & 20:41:00                       & 20:50:00                      & 4                         \\
K593                          & 20:57:00                       & 21:08:00                      & 3                         \\
K2908                         & \multicolumn{1}{l}{}           & 21:24:00                      & 3                         \\
Z254                          & \multicolumn{1}{l}{}           & 21:36:00                      & 4                         \\
Z263                          & 21:46:00                       & 22:00:00                      & 6                         \\
K2188                         & 21:54:00                       & 22:06:00                      & 4                         \\
K127                          & 22:05:00                       & 22:28:00                      & 5                         \\
Z136                          & 22:26:00                       & 22:34:00                      & 6\\
\bottomrule
\end{tabular}
\end{table}

\subsection{Example of predefined stop patterns}\label{App-3.2}
Table \ref{Tab-RnSn} presents an example of the predefined stop pattern for route 4. This route accommodates a total of 11 trains throughout the day. The stations along the route are categorized into three sets: $p_1$, $p_2$, and $p_3$.

$p_1$ consists of 16 stations that appear more than 3 times across different trains on route 4. These stations have a higher frequency of appearance and are considered significant for the route.

$p_2$ consists of 21 stations that appear more than 2 times on route 4. Although not as frequently visited as those in $p_1$, these stations still play an important role in the overall operations of the route.

$p_3$ encompasses all 26 stations along route 4, including those in $p_1$ and $p_2$. This set represents the complete list of stations on the route.

The arrival times at downstream stations, originating from Xi'an, are provided in the corresponding columns of the table. These arrival times serve as references for scheduling and decision-making within the simulated environment.

\begin{table}[h]
\caption{Predefined stop patterns on Route 4}
\label{Tab-RnSn}
\begin{tabularx}{\linewidth}{cccccc}
\toprule
\multicolumn{2}{c}{S1}      & \multicolumn{2}{c}{S2}      & \multicolumn{2}{c}{S3}      \\
\midrule
Station   & Travel Time & Station   & Travel Time & Station   & Travel Time \\
\midrule
Xi'an          & Departure          & Xi'an          & Departure          & Xi'an          & Departure \\
Weinan         & +38       & Weinan         & +38       & Weinan         & +38       \\
Tongguan       & +104      & Tongguan       & +104      & Tongguan       & +104      \\
Sanmenxia West & +189      & Sanmenxia West & +189      & Lingbao        & +164      \\
Sanmenxia      & +214      & Sanmenxia      & +214      & Sanmenxia West & +192      \\
Luoyang        & +320      & Luoyang        & +320      & Sanmenxia      & +217      \\
Zhengzhou      & +418      & Zhengzhou      & +418      & Mianchi        & +261      \\
Shangqiu       & +560      & Kaifeng        & +464      & Luoyang        & +326      \\
Xuzhou         & +679      & Minquan        & +516      & Zhengzhou      & +424      \\
Bengbu         & +797      & Ningling       & +531      & Kaifeng        & +470      \\
Chuzhou North  & +894      & Shangqiu       & +569      & Lankao         & +502      \\
Nanjing        & +984      & Yucheng        & +610      & Minquan        & +525      \\
Zhenjiang      & +1031     & Dangshan       & +636      & Ningling       & +540      \\
Changzhou      & +1100     & Xuzhou         & +694      & Shangqiu       & +569      \\
Wuxi           & +1136     & Bengbu         & +813      & Yucheng        & +619      \\
Suzhou         & +1168     & Chuzhou North  & +909      & Dangshan       & +645      \\
Shanghai       & +1234     & Nanjing        & +999      & Xuzhou         & +703      \\
               &            & Zhenjiang      & +1046     & Bengbu         & +822      \\
               &            & Changzhou      & +1115     & Mingguang      & +846      \\
               &            & Wuxi           & +1151     & Chuzhou North  & +918      \\
               &            & Suzhou         & +1183     & Nanjing        & +1008     \\
               &            & Shanghai       & +1249     & Zhenjiang      & +1055     \\
               &            &                &            & Changzhou      & +1124     \\
               &            &                &            & Wuxi           & +1160     \\
               &            &                &            & Suzhou         & +1192     \\
               &            &                &            & Kunshan        & +1201     \\
               &            &                &            & Shanghai       & +1258\\
\bottomrule
\end{tabularx}
\end{table}

\section{GA pseudocode}\label{App-4}




\begin{algorithm}
\caption{Genetic Algorithm for One-off Rescheduling}
\label{alg:GA-Rescheduling}
\begin{algorithmic}[1]
\State Let $G$ be the number of iterations (generations)
\State Let $C_r$ be the crossover rate
\State Let $M_r$ be the mutation rate
\State Let $L$ be the chromosome length ($L = 240$)
\State Let $A$ be the set of scheduling actions with $|A| = 20$ and $0$ representing inaction
\State Let $N_{RS}$ and $N_{RS}^*$ be the numbers of available and restricted rolling stock
\State Initialize population $\mathcal{P}_0$ with random chromosomes of length $L$
\State Randomly allocate rolling stock to gene positions in chromosomes, set others to 0

\For{$g = 1$ to $G$}
    \For{each chromosome $c \in \mathcal{P}_{g-1}$}
        \State $R_s(c) \gets$ reward from time steps for chromosome $c$
        \State $R_E(c) \gets$ episode reward for chromosome $c$
        \State Fitness($c$) $\gets \sum_{s} R_s(c) + R_E(c)$
    \EndFor
    \State Select parents $\mathcal{P}_{\text{parents}}$ based on Fitness
    \State $\mathcal{P}' \gets$ Crossover($\mathcal{P}_{\text{parents}}, C_r$)
    \State $\mathcal{P}'' \gets$ Mutation($\mathcal{P}', M_r$)
    \For{each chromosome $c' \in \mathcal{P}''$}
        \If{CountNonZero($c'$) $>$ $N_{RS}$}
            \State $c' \gets$ ReduceNonZero($c'$, $N_{RS}$)
        \EndIf
        \If{NotFeasible($c'$, $N_{RS}^*$)}
            \State Discard $c'$
        \EndIf
    \EndFor
    \State $\mathcal{P}_{g} \gets \mathcal{P}''$
\EndFor
\end{algorithmic}
\end{algorithm}

The Genetic Algorithm (GA) for Rescheduling, as outlined in Algorithm \ref{alg:GA-Rescheduling}, operates within a framework where a population of chromosomes $\mathcal{P}$ evolves over $G$ generations to optimize scheduling decisions. Each chromosome $c_i$ in the population represents a sequence of decisions across a planning horizon with length $L$, where actions from a set $A$ including 20 rescheduling plans and the inaction ``0'' are chosen for each time step. Initially, the population $\mathcal{P}_0$ is generated with random allocation of rolling stock to certain gene positions, leaving the remainder set to 0. In each generation $g$, the algorithm evaluates the fitness of each chromosome based on the sum of step rewards $R_s(c)$ and an episode reward $R_E(c)$, effectively capturing both short-term and long-term planning efficacy. With the fitness calculated, a selection process yields a subset $\mathcal{P}_{\mathrm{parents}}$ from which new offspring $\mathcal{P}'$ are created via a crossover operation with rate $C_r$. The offspring then undergo mutation with rate $M_r$ to form $\mathcal{P}''$, ensuring genetic diversity and exploration of the search space. This new population is subjected to constraints ensuring that the number of non-zero decisions does not exceed available rolling stock and that the rescheduling plans are feasible given restricted rolling stock $N_{RS}^*$. Unsuitable chromosomes are discarded, and the remainder forms the updated population $\mathcal{P}_g$ for the next iteration. The process iterates until it completes $G$ generations, with the aim of converging to a set of scheduling decisions that maximize the fitness, balancing immediate rewards and overall episodic performance.

\end{document}